\documentclass[preprint]{aastex}
\usepackage{natbib}
\usepackage{graphicx}               
\usepackage{multicol}                 
\usepackage[bottom]{footmisc}  
\usepackage{makeidx} 

\usepackage{fancyhdr}
\newcommand{\dove}{Figure}

\newcommand{\lsim}{\mathrel{\mathop{\kern 0pt \rlap
  {\raise.2ex\hbox{$<$}}}
  \lower.9ex\hbox{\kern-.190em $\sim$}}}
\newcommand{\gsim}{\mathrel{\mathop{\kern 0pt \rlap
  {\raise.2ex\hbox{$>$}}}
  \lower.9ex\hbox{\kern-.190em $\sim$}}}

\newcommand{\pbar}{\bar{p}}

\newcommand{\beq}{\begin{equation}}
\newcommand{\eeq}{\end{equation}}
\newcommand{\bea}{\begin{eqnarray}}
\newcommand{\ena}{\end{eqnarray}}

\def\pbar{$\overline{p}$}
\def\nbar{$\overline{n}$}

\def\dbar{$\overline{d}$}
\def\eplus{$e^+$}
\def\efrac{$e^+/(e^+ + e^-)$}

\title{Astrophysics of Galactic charged Cosmic Rays}
\author{A.Castellina$^1$, F.~Donato$^2$}

\affil{(1) Istituto Nazionale di Astrofisica, IFSI,  via P.Giuria 1, I-10125 Torino, Italy} 
\affil{(2) Dipartimento di Fisica, Universit\`a di Torino, via P. Giuria 1, I-10125 Torino, Italy}

\begin{document}

\begin{abstract}
A review is given of the main properties of the charged component of galactic cosmic rays, particles detected at Earth with an energy
spanning from tens of MeV up to about $10^{19}$ eV.
After a short introduction to the topic and an historical overview, the properties of cosmic rays are discussed with respect to different energy ranges. 
The origin and the propagation of nuclei in the Galaxy are dealt with from a  theoretical point of view. The mechanisms leading to the
acceleration of nuclei by supernova remnants and to their subsequent diffusion through the inhomogeneities of the galactic magnetic field are discussed and some clue is
given on the predictions and observations of fluxes of antimatter, both from astrophysical sources and from dark matter annihilation in the galactic halo.\\
The experimental techniques and instrumentations employed for the detection of cosmic rays at Earth are described. Direct methods are viable up to $\simeq 10^{14}$ eV, by means of experiments flown on balloons or satellites, while above that energy, due to their very low flux, cosmic rays can be studied only indirectly by exploiting the particle cascades they produce in the atmosphere.\\
The possible physical interpretation of the  peculiar features observed in the energy spectrum of galactic cosmic rays, and in particular the so-called "knee" at about 4$\times 10^{15}$ eV, are discussed.
A section is devoted to the region between about $10^{18}$ and $10^{19}$ eV, which is believed to host the transition between galactic and extragalactic cosmic rays. 
The conclusion gives some perspectives on the cosmic ray astrophysics field. Thanks to a wealth of different experiments, this research area is living a very flourishing era. The activity is exciting both from the theoretical and the instrumental sides, and its interconnection with astronomy, astrophysics and particle physics experiences non-stop growth.
\end{abstract}

{\it Subject headings} : 
Cosmic rays -- Origin and Propagation -- Instrumentation: detectors -- Energy spectrum and Composition --  Galactic to extragalactic transition\\

\pagebreak
\tableofcontents

\section{Introduction}
\label{sec:1}
{\it "The subject of cosmic rays is unique in modern physics for the minuteness of the phenomena, the delicacy of the observations, 
the adventurous excursions of the observers, the subtlety of the analysis, the grandeur of the inferences" (K.K.Darrow)} 
\vspace{0.4cm}

\begin{figure}[h]
\begin{minipage}{0.42\linewidth}
 \centering
 \includegraphics[width=7.5cm,height=7cm]{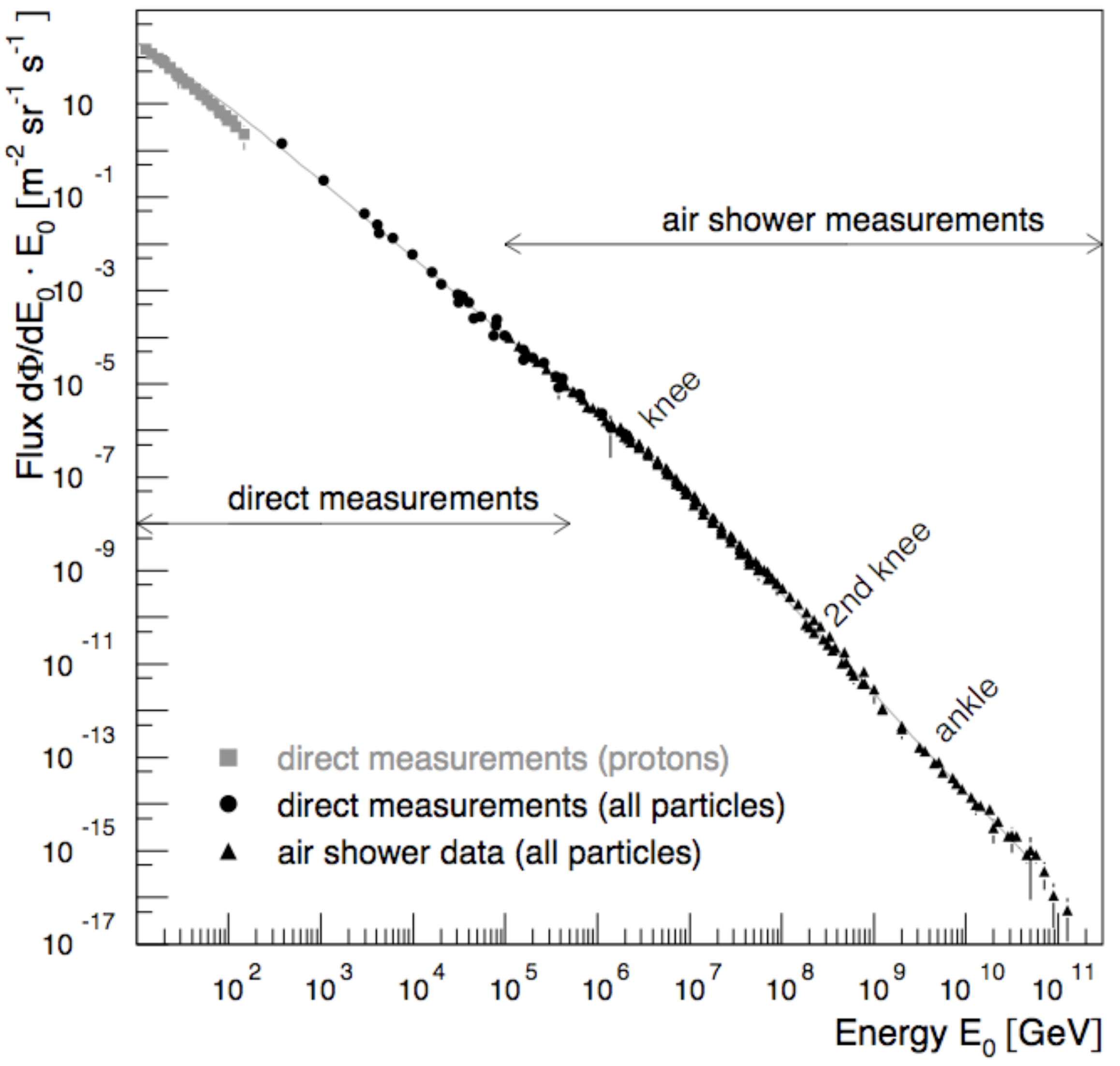}   
 \caption{\em {The all-particle energy spectrum. The changes of slope correspond to the indicated knee and ankle energies (modified from \citep{blu09}).}}  
 \label{fi:fig1}
 \end{minipage}\hfill
 \begin{minipage}{0.52\linewidth}
 \centering
 \includegraphics[width=8.5cm,height=7cm]{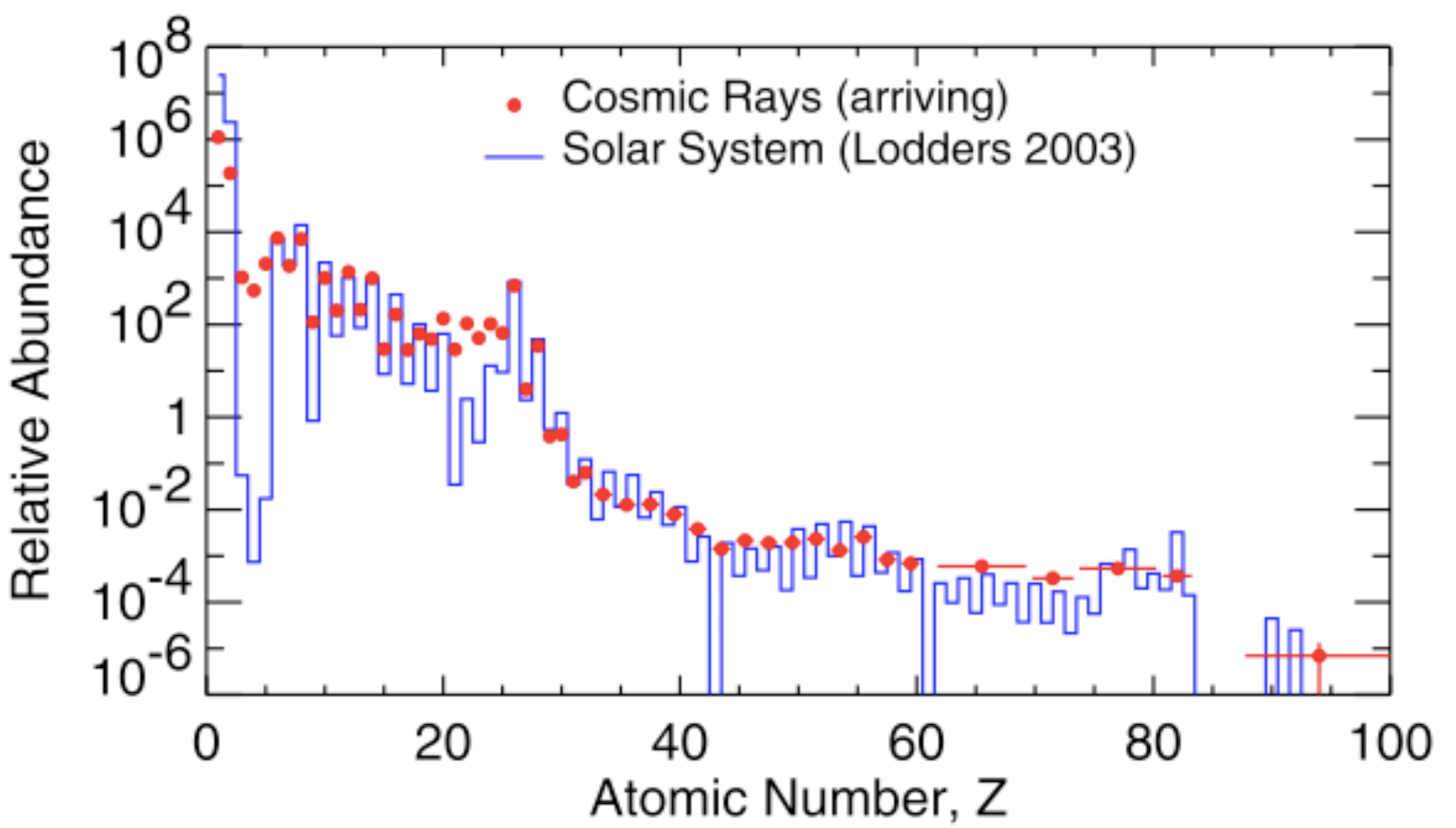}   
 \caption{\em {Elemental abundances in the arriving CRs for $E \leq 1$ GeV/n near the peak of 
their spectrum (relative to Carbon) compared to those found in the Solar System material \citep{cri07}.}}  
 \label{fi:solcr}
 \end{minipage}
\end{figure}

A large contribution to the knowledge and understanding of the  Galaxy is given  by the observation of the most energetic particles, the cosmic rays (CRs). These relativistic particles, reaching the Earth from the outer space, are either primary nuclei, arriving directly from the sources, or secondary products of the 
spallation processes  (i.e. fragmentation by nuclear destruction) taking place during the propagation from the sources through the interstellar medium (ISM). \\
CRs play an important role in the dynamics of the Galaxy: their energy density $\rho_{E} \simeq 1$ eV/cm$^{3}$ 
is  comparable to the energy density of the visible starlight $\rho_{S} \simeq 0.3$ eV/cm$^3$, the galactic magnetic fields $B^{2}/2 \mu_{0} \simeq 0.25$ eV/cm$^3$ (if $B \simeq 3 \mu G$) or the cosmic microwave background (CMB) radiation $\rho_{CMB} \simeq 0.25$ eV/cm$^3$. Such non thermal component is indeed strictly linked to radiation and magnetic fields.
Measuring CRs and their properties could be an effective tool to understand  the stellar nucleosynthesis  and the Supernovae evolution. These particles strongly influence the galactic chemical composition and evolution and their interactions with the cosmic radiation background, the interstellar radiation field and interstellar gas give rise to diffuse$\gamma$ ray emission. 
The challenge we are facing is that of identifying the  sources of CRs and the mechanism through which low energy particles (the seeds being just  single elements, or dust and grains) are accelerated to such high energy to be called CRs and of understanding their propagation through the galactic magnetic fields.   \\
The CR  extreme energies are by far beyond the reach of the most powerful man-made accelerators, so that they are of great interest also from the point of view of particle physics, probing the standard model of hadronic interactions and the laws of relativity in extreme domains.\\
The flux of CRs is seen at Earth varying about 32 orders of magnitude across an energy range spanning 14 orders of magnitude. It amounts to $\simeq 10^{4}$ m$^{-2}$s$^{-1}$ at $\simeq 10^6$ eV to less than $1$ km$^{-2}$ century$^{-1}$ at $\simeq 10^{19}$ eV. 
The all-particle energy spectrum is shown in Fig.\ref{fi:fig1}, where the main structures of an  otherwise almost pure power-law are  visible. Changes of slope are taking place at the  {\it knee},  at $\simeq 3-4$ PeV\footnote{energy is hereafter measured in units of 1 GeV, 1 TeV, 1 PeV, 1 EeV meaning $10^{9}, 10^{12}, 10^{15}, 10^{18}$ eV respectively},  the {\it second knee}, near $400$ PeV 
and the {\it ankle}, a broader feature around $3$ EeV. The investigation of these features is a valuable tool for the study of the CRs, their nuclear  composition and the energy above which an extragalactic component takes over.\\
Due to the wide energy range and the rapidly changing flux, it is obvious that different experimental techniques are needed in diverse energy regions. Direct measurements of the primary CRs can be performed up to $10^{14}$ eV. Above this threshold, the low fluxes due to the steeply falling spectrum force us to exploit indirect methods, detecting the extensive air showers generated by the interaction of CRs in the atmosphere. 

\subsection{Charged cosmic rays}

Galactic cosmic rays (GCR)  are supposed to be produced by nucleosynthesis processes in the stellar interior, where they are fractionated, ejected and then accelerated to CR energies and diffused throughout the whole Galaxy. When reaching the Solar System - where the slowest CRs experience  a spectral distortion by the solar wind - they can be observed both by space-based experiments or by detectors at ground.\\
Starting from the spectrum and composition of the CRs observed at Earth, and knowing the characteristics of propagation of these particles in the Galaxy,  which include diffusion and convection, energy losses, reacceleration and production of secondaries through spallation processes, it is possible to step back from  the observed spectrum at Earth to the source one. The similarity between the CR composition as measured directly with detectors on satellites or balloons, like BESS \citep{bes02}, HEAO-3 \citep{hea90} and ACE/CRIS \citep{cris,cri07}  and that found in the Solar System \citep{lod03} is shown in Fig.\ref{fi:solcr}. The overabundance of elements like Li, Be, B ($Z=3 \div 5$) and sub-Fe ($Z=21 \div 25$) in CRs can be explained by their origin: they are in fact secondaries produced in spallation processes during propagation in the ISM.\\
The problem of the origin of CRs is related to many different issues: what are the seeds out of which the sources accelerate particles, what accelerators have the power to boost them to such high energies, which part of the CR spectrum is of galactic origin. 
Charged particles are bent in the galactic magnetic fields, so there is no obvious way of tracking them back to their sources. The energy spectrum which is observed at Earth is folded with the source one through an energy dependent diffusion coefficient which shapes the effects of the galactic magnetic field in an effective way, and which can be deduced only phenomenologically.
These peculiarities are at variance with $\gamma$ rays, which point directly to the sources: the accelerated ions around a source are expected to interact with the ambient matter and produce high energy $\gamma$s which can be detected at Earth. \\
On the basis of energetic arguments, Supernovae Remnants  (SNRs) have long been considered the most probable sources of GCRs \citep{gs64}. 
In fact, the average kinetic energy of a Supernova ejecta is $L_{kin} \simeq 1.6 \times 10^{51}$ erg (for  10 solar masses traveling at $\simeq 4000$ km/s). Assuming a supernova rate of explosion in our Galaxy of  the order of $30 \ yr^{-1}$ and an efficiency for converting the kinetic energy into relativistic particles of $\simeq 10 \%$, SNRs can provide the $\simeq 10^{41} \ erg/s$  power required to keep the CR energy density. The energetic argument is however not conclusive, since other potential sources could meet the requirement, like stars with powerful winds or pulsars. \\
Based on the first theory of CR acceleration developped by E.Fermi \citep{fer49}, the most accredited mechanism to convert from  the kinetic motion of the plasma to kinetic energy of charged particles is the diffusive acceleration in presence of shock waves (DSA) powered by supernova explosions propagating from the remnant to the ISM \citep{ber99}. Traversing the boundary between the unshocked upstream and the shocked downstream region back and forth, charged particles gain each time an energy $\Delta E \propto E$; the acceleration spectrum follows a power-law in momentum, $Q(E) \propto p^{-\alpha}$, with $\alpha$  located between 2.0 and 2.5.
Taking into account  the diffusion of CRs in the Galaxy, with a diffusion coefficient $K$ expected to be proportional to the rigidity of the particle ($R=pc/Ze$, where Z is the charge and c is the speed of light) as $K \propto R^{\delta}$,  this will eventually lead to the observed spectrum at Earth $N_{obs}(E) \propto E^{-\gamma}$, with $\gamma \simeq \alpha+\delta$, up to a maximum energy 
\begin{equation}
E_{max} \simeq Ze(B/\mu G)(L/pc) \beta_{shock} \ \ PeV
\label{eq:eq1}
\end{equation}
where Ze is the particle charge, B is the galactic magnetic field strength, L is the linear dimension of the acceleration site and $\beta_{shock}$ is the shock velocity \citep{hill84}. \\
Direct measurements of the CRs performed by means of balloon or satellite-borne experiments can help in  the study of both diffusion and acceleration spectrum. The experimental  measurement of the secondary to primary ratio, i.e. the boron--to--carbon flux ratio B/C, gives the most direct  information on the slope of the diffusion coefficient, $\delta$ (see discussion in Sect. \ref{sec:propa1}).  A careful measurement of the fluxes of individual elements can determine the spectrum slope $\gamma$ for light nuclei with a negligible error, bringing  information on  the acceleration slope power index $\alpha$ after having taken into account the uncertainty due to $\delta$.\\
Besides the measurement of the all particle energy spectrum, remarkable achievements  of extensive air shower detectors include the determination of the energy spectra of single or groups of elements,  showing  cutoff energies at constant rigidity $E/Z$, even if spectral steepenings at constant energy/nucleon $E/A$, predicted by some of the models, cannot yet be disproven.\\
The interpretation of the knee feature in the energy spectrum is another important clue to understand the origin of the GCRs. 
Different models attribute the knee either to a limit on the maximum energy attainable during acceleration in the sources or to the leakage of particles out of the Galaxy during their propagation. Other ideas point to interactions with background particles like massive neutrinos or photodisintegration of nuclei in the fields of soft photons, or finally to new properties of hadronic interactions taking over at high energy.\\
The region between about $10^{18}$ and $10^{19}$ eV is of particular interest, as it is supposed to host the transition from galactic to extragalactic cosmic rays  (EGCRs):
CR nuclei of charge Z and energy E cannot be confined in the Galaxy if their Larmor radius $r_{L} \simeq E_{PeV}/(Z \ B_{\mu G})$ pc  becomes larger than the trasversal dimension of the galactic disk ($\simeq 100$ pc).  \\
For a proton, the confinement becomes impossible above about 1 EeV; a factor 26 higher energy would be required for iron.\\
The transition is described by the current models in terms of  source emissivity and energy spectrum  of particles at the source; the source density $n$ is assumed to be a function of the red shift $z$, with a source evolution parameter {\it m} such that $dn/dz \propto (1+z)^m$. \\
In the "{\it dip}" model \citep{ber06}, the transition is due to $e^{+}e^{-}$ pair production by extragalactic  protons on the photons of the CMB and takes place at the 2-nd knee. Only a very low contamination ($\leq 10 \div 15 \%$) by heavier nuclei can be allowed in an otherwise pure proton beam. The "{\it mixed}" model \citep{alla08}, on the contrary, puts the transition around $\simeq 3$ EeV; the nuclear composition is mixed as in the galactic component and the dip is filled by the contribution of elements other than protons. The more traditional  "{\it ankle}" model \citep{hil05} considers the ankle as the intersection between a very steep galactic component and a flat extragalactic  one.
Composition and anisotropy of high energy CRs are the most useful tools to disk riminate among the models.\\
\begin{figure}[h]
\begin{minipage}{0.47\linewidth}
 \centering
 \includegraphics[width=8.5cm,height=8.5cm]{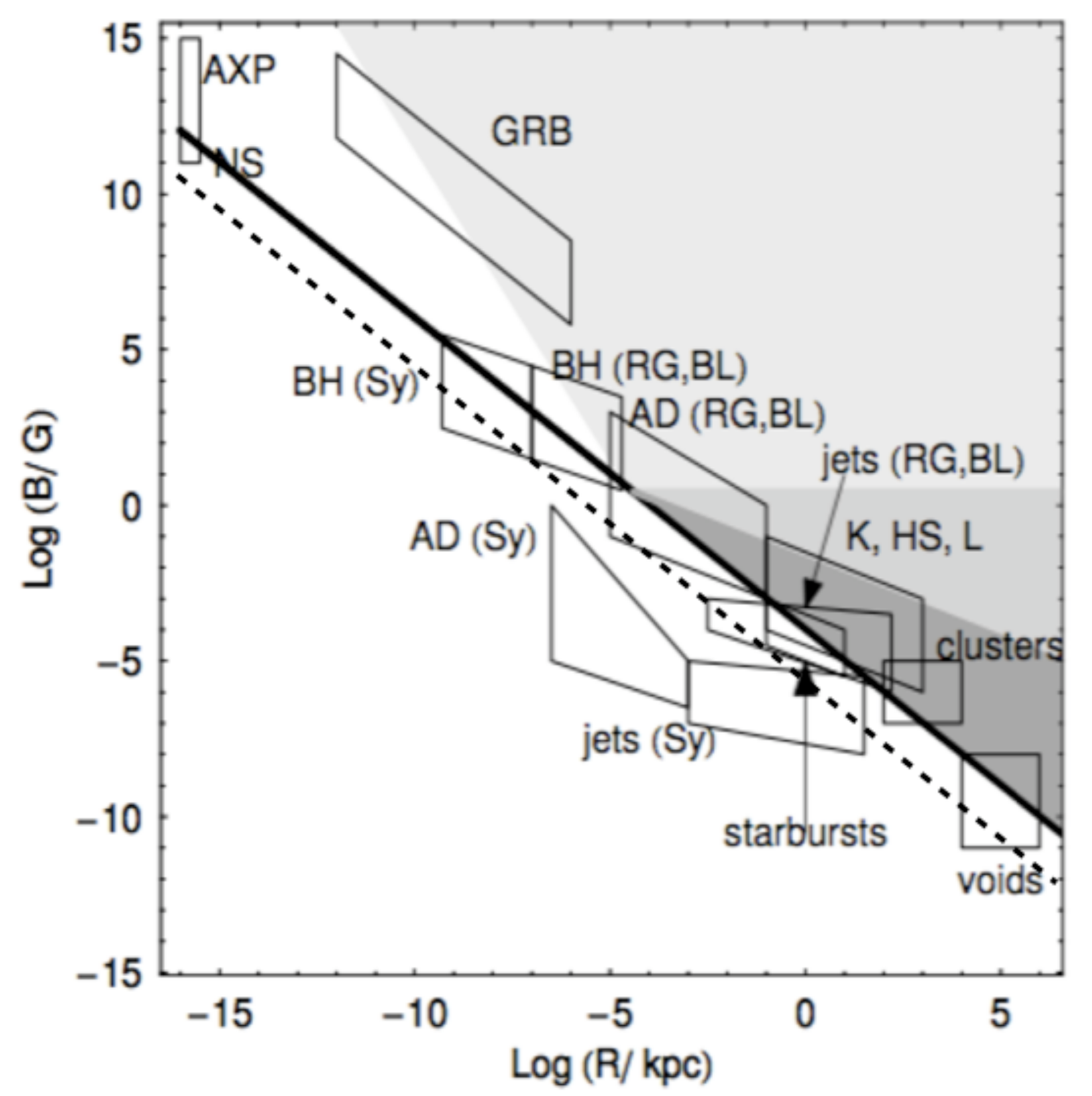} 
 \caption{\em {The magnetic field and size of potential CR accelerators \citep{pti10}. Full  and dashed lines: the lower boundary allowed by the Hillas criterium  for $10^{20}$ eV protons and iron nuclei.}}  
 \label{fi:fig2}
 \end{minipage}\hfill
 \begin{minipage}{0.47\linewidth}
 \centering
 \includegraphics[width=8.5cm,height=8.5cm]{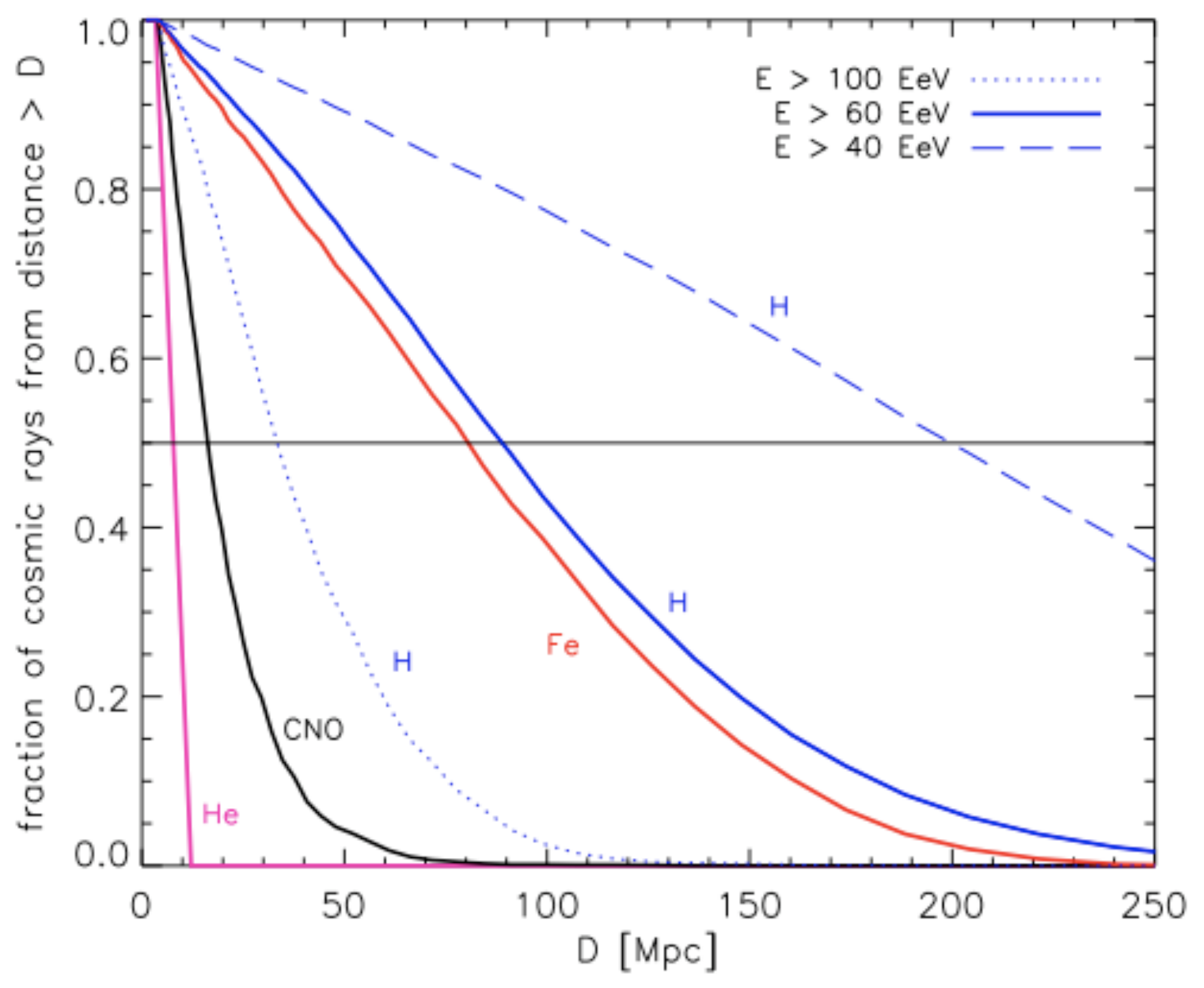} 
\caption{\em { GZK horizons of p, He, CNO and Fe above different energies \citep{oli11}. Light nuclei are 
promptly dissociated by CMB photons, while p and Fe may reach us from distances $\lesssim100$ Mpc.}}  
 \label{fi:uhe2}
 \end{minipage}
\end{figure}
Where is the natural end of the CR spectrum? As shown in Fig.\ref{fi:fig2}, only very few astrophysical objects are thought to be able to accelerate particles to ultra high energies (UHE), above $\simeq$ few $10^{19}$ eV : neutron stars (NS),$\gamma$ ray bursts (GRB), radio Galaxies. The acceleration sites in radio Galaxies (indicated as Seyfert-Sy, blazars-BL, radio-loud FR-I and FR-II) include both the region close to the black hole (BH) and the accretion disk (AD) and extended structures like knots (K) and hot spots (HS) in jets and lobes (L).The shaded areas are allowed by the radiation-loss constraints (shown for protons only).
A given energy  can be obtained either in a large region with low magnetic field or in a compact high field source (see Eq.\ref{eq:eq1}).\\
As recognized soon after the disk overy of the CMB, CRs of ultra high energy undergo interactions with the  intergalactic radiation fields, giving rise to an attenuation of their flux, the well known GZK effect \citep{gre66,zk66}. \\
For protons, the most important interactions with the microwave background are the photopion production ($p + \gamma_{CMB} \rightarrow N + \pi$) and the production of  $e^{+} + e^{-}$ pairs ($p + \gamma_{CMB} \rightarrow p + e^{+} + e^{-}$ ). The energy thresholds for these two processes are respectively given by $\simeq 6 \times 10^{19}$ and $\simeq 4 \times 10^{17}$ eV, but the energy loss per interaction  for pion production is much higher (20\% compared to 0.1\% for the pair production). For heavier nuclei, the most important energy losses are due to photodisintegration $(A + \gamma_{CMB} \rightarrow (A-1) + N)$ and pair production $(A +  \gamma_{CMB} \rightarrow A+ e^{+} + e^{-})$ on the photons of the CMB.\\
The fraction of CRs that arrive at Earth from a given distance for nuclei above different energies is shown in Fig.\ref{fi:uhe2}, where the  
GZK horizon is defined as the distance from which 50\% of primaries originate.\\
A discussion of the extragalactic spectrum features, above the transition, is outside the scope of this review. It is however important to remind that at these extreme energies a clear  evidence for a flux suppression has been measured both by the HiRes \citep{abb08} and Pierre Auger \citep{aug08} collaborations, consistent with that expected due to the GZK effect. However, the same effect could also be  related to the maximum energy of sources. 
The topic is discussed at length in the reviews of \citep{nag09,oli11}. 

\subsection{Brief history: from CRs to astroparticle physics}
\index{Brief history}
About a century ago, CRs were identified as being a source of radiation on Earth. The proof came from two independent experiments. The Italian physicist Domenico Pacini observed the radiation strength to decrease when going from the ground to few meters underwater (both in a lake and in the sea) \citep{pac12}.  Victor Hess during several ascents with hydrogen-filled balloons up to altitudes of 5 km \citep{hess} measured the ionization rate of air as a function of altitude. He explained the measured increase of ionizing radiation with increasing height with the presence of a radiation continuously penetrating into the Earth atmosphere from outer space; for this disk overy he was awarded the Nobel Price in 1936. 
Hess measurements were further extended up to 9 km by the subsequent ascents of W. Kolhorster \citep{kolh}. \\ 
The first hints for a corpuscular nature of the cosmic radiation came exploiting the newly invented Geiger-Muller counters in the experiment of W. Bothe and W. Kolhorster \citep{bk} and from measurements by J. Clay, who showed that the intensity of CRs depends on the magnetic latitude of the observer \citep{clay} and they thus had to be charged particles. \\
Using a cloud chamber, in 1927 D.~Skobelzyn observed the first tracks left by CRs. In 1934 Bruno Rossi reported an observation of near-simultaneous disk harges of two Geiger counters widely separated in a horizontal plane, noting that "It seems that extended showers of particles every now and then reach our detectors...determining temporal coincidences among counters even quite far away from each other...".  
By 1938,  using Wilson chambers and Geiger tubes spread over a wide area and working in coincidence,  Pierre Auger showed that very high energy CRs trigger extensive air showers in the Earth atmosphere, sharing the primary energy among billions of lower energy particles that reach the ground together \citep{kolh2,auma}. On the basis of his measurements, Auger concluded that he had observed showers with energies of $10^{15}$ eV, incredibly higher compared to any previously measured before. \\
From the 1930s to the 1950s, before man-made accelerators reached very high energies, CRs served as a source of particles for high energy physics investigations, and led to the disk overy of a wealth of subatomic particles, such as the muon, the pion, hyperons etc.
The field benefited by many different technical improvements, from Wilson chambers to Geiger-Muller tubes, to photographic emulsions. The development of photomultipliers allowed the use of scintillators and Cherenkov detectors. Experimental measurements were performed in very different environments, with balloons in the highest atmosphere, high altitude observatories and more recently underground laboratories.\\
Since the mid 1940s large apparatuses were used to detect the extensive air showers. In 1958, G.V. Kulikov and G.B. Khristiansen measured the integral electron number spectrum in air showers using an array of hodoscope counters \citep{kk}; this brought to the first detection of the "knee"  around few PeV.\\
In 1962, using an air shower array in Volcano Ranch, New Mexico, an event with an energy of tens of Joules (about $10^{20}$ eV) was observed \citep{linsley}.
In 1966, an abrupt steepening of the CR spectrum above $10^{20}$ eV was predicted \citep{gre66,zk66} as a result of the interactions of the CRs with the  newly disk overed CMB radiation, since then called the "GZK cut-off".
In the subsequent years, the ultra high energy events have been studied either by means of scintillator or water detectors in ground arrays at Haverah Park \citep{edg73}, SUGAR \citep{bel74}, Yakutsk \citep{afa03}, Akeno \citep{nag92}, AGASA \citep{chi92} or by means of a new technique exploiting air fluorescence in atmosphere  with Fly's Eye \citep{bal85} and HiRes \citep{bir93}. Most recently, both techniques have been employed in the Pierre Auger Observatory \citep{abr04}.\\
Since 1927, when  R.~Millikan introduced the term "CRs", the main focus of this research has been directed towards the astrophysical investigations of their origin, acceleration and propagation, what role they play in the dynamics of the Galaxy, and what their composition tells us about matter from outside the solar system.  Starting at the end of the 1980s, the new interdisk iplinary field of "Astroparticle Physics" was born, in which astrophysics, cosmology, CRs and particle physics together contribute to shed light on the nature and structure of the matter in the Universe.

\section{From 100 MeV/n to 100 TeV/n}

\label{sec:propa}
\subsection{The diffusion equation}
\label{sec:propa1}
The spectrum of GCRs - spanning from tens of MeV/nucleon up to $\gsim 10^{18}$ eV --
is fundamentally shaped by acceleration and diffusion, at least for 
energies  $\gsim 10^2-10^3$ GeV/n. 
Several other phenomena - for instance convection, reacceleration, nuclear fragmentation, electromagnetic losses, solar modulation - compete at lower energies, where their effect is 
often degenerate and prevents unambiguous interpretation of the wealth of data collected in the lower tail of the galactic spectrum.
Nevertheless,  phenomenological models able to reproduce data on a wide energy range can be built (the milestone in the field literature being \citep{1990acr..book.....B}).  
The most realistic propagation models are the diffusion
ones, even if the so--called leaky box model has been often preferred in the past for its simplicity.
In the leaky box model, the densities of sources $q^j$, of interstellar matter {\it n} and 
of CRs $N^j$ are assumed to be homogeneous in a finite propagation volume
delimited by a surface. In addition, each nucleus is supposed to escape from
the leaky box with a probability per unit of time $1/\tau_{\rm esc}$.
In a steady state regime, the relevant densities, for a given nucleus {\it j}, obey the following equation:
\begin{equation}
          \frac{N^j}{\tau_{\rm esc}}+nv\sigma^j N^j=q^j+\sum_{{\rm heavier}~k}
          nv\sigma^{kj}N^k\;,
\label{LB_eq}
\end{equation}
where {\it v} is the nucleus velocity and $\sigma^j$ is its destruction cross section. 
The leaky box model has been successful to explain most
of the observed fluxes for stable species by the single function $\tau_{\rm esc}(E)$.
This function is usually  adjusted to the data,
its physical interpretation being found afterwards, or
extracted  directly from more complete propagation
equations~\citep{2001ApJ...547..264J}.\\
At variance, diffusive models account for spatial dependence of 
sources, CR densities and in principle of the interstellar matter. 
The Galaxy is shaped as a thin gaseous disk where all the astrophysical sources are located, 
embedded in a thick diffusive magnetic halo. 
Diffusive models, besides being more realistic and closer to a physical interpretation for
each component, have proven to be successful in reproducing the nuclear, antiproton and radioactive isotopes data. They also allow to
treat contributions from dark matter (or other exotic) sources
located in the diffusive halo.\\
The relevant transport equation for a charged particle wandering through the magnetic inhomogeneities of the galactic magnetic field
writes in terms of the differential density $N(E,\vec{r})$ as a function of the total energy $E$ and
the position $\vec{r}$ in the Galaxy.  Assuming steady-state ($\partial N/\partial t=0$),  the transport equation for a given nucleus (the subscript {\it j} is omitted) can be written in a compact form as 
\begin{equation}
(-\vec{\nabla} \cdot (K\vec{\nabla}) + \vec{\nabla}\cdot\vec{V_C} +
     \Gamma_{\rm rad} + \Gamma_{\rm inel})
 N + \frac{\partial}{\partial E}\left( b N - c \frac{\partial N}{\partial E} \right) = {\cal S}\;.
\label{eq:CR}
\end{equation}
The first bracket in the l.h.s. accounts for: i) spatial diffusion
$K(\vec{r},E)$, ii) convection with speed $\vec{V_C}(\vec{r})$, 
iii) the (possible, for some isotopes with half lifetime $\tau_0$) radioactive decay rate
$\Gamma_{\rm rad}(E)= 1/(\gamma\tau_0)$ ($\gamma$ here is the Lorentz factor), 
iv) the destruction rate 
$\Gamma_{inel}(\vec{r},E)=\sum_{ISM} n_{\rm ISM}(\vec{r}) v \sigma_{\rm inel}(E)$ due to collisions with the ISM.
In this last expression $n_{\rm ISM}(\vec{r})$ is the density of the ISM in the various locations of the Galaxy and in its different H and He components, 
and $\sigma_{\rm inel}(E)$ is the destruction (inelastic) cross section for a given nucleus. 
The coefficients $b$ and $c$ are respectively first and
second order gains/losses in energy.
The source term ${\cal S}$ includes  primary sources of CRs (e.g. supernovae), secondary  sources due to the fragmentation of heavier nuclei, and secondary decay-induced sources. \\
In the following, each of the components of Eq.\ref{eq:CR} is reviewed.

\begin{figure}[h]
\begin{minipage}{0.47\linewidth}
 \centering
\includegraphics[width=8.cm,height=8cm]{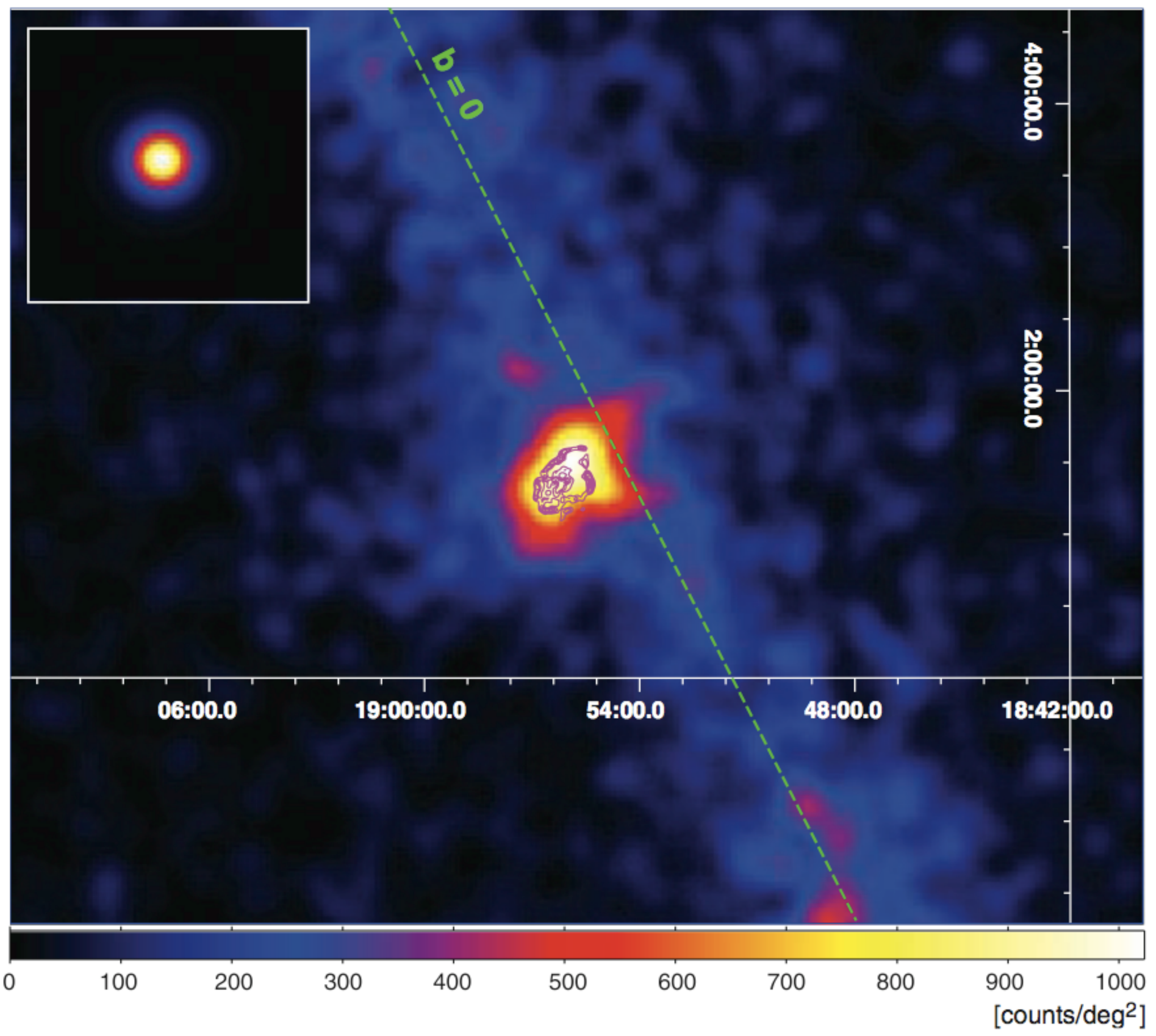} 
 \caption{Fermi-LAT image (2-10 GeV) of the region where the SNR W44 is located.
The color scale indicates counts per solid angle on a linear scale. The green line
corresponds to the galactic plane. The radio image of W44 as seen in 20-cm 
wavelength by the Very Large array VLA is overlaid as the magenta contour \citep{Abdo:2009zz}.}
 \label{fi:W44}
 \end{minipage}\hfill
 \begin{minipage}{0.47\linewidth}
 \centering
\includegraphics[width=8.cm,height=8cm,angle=0]{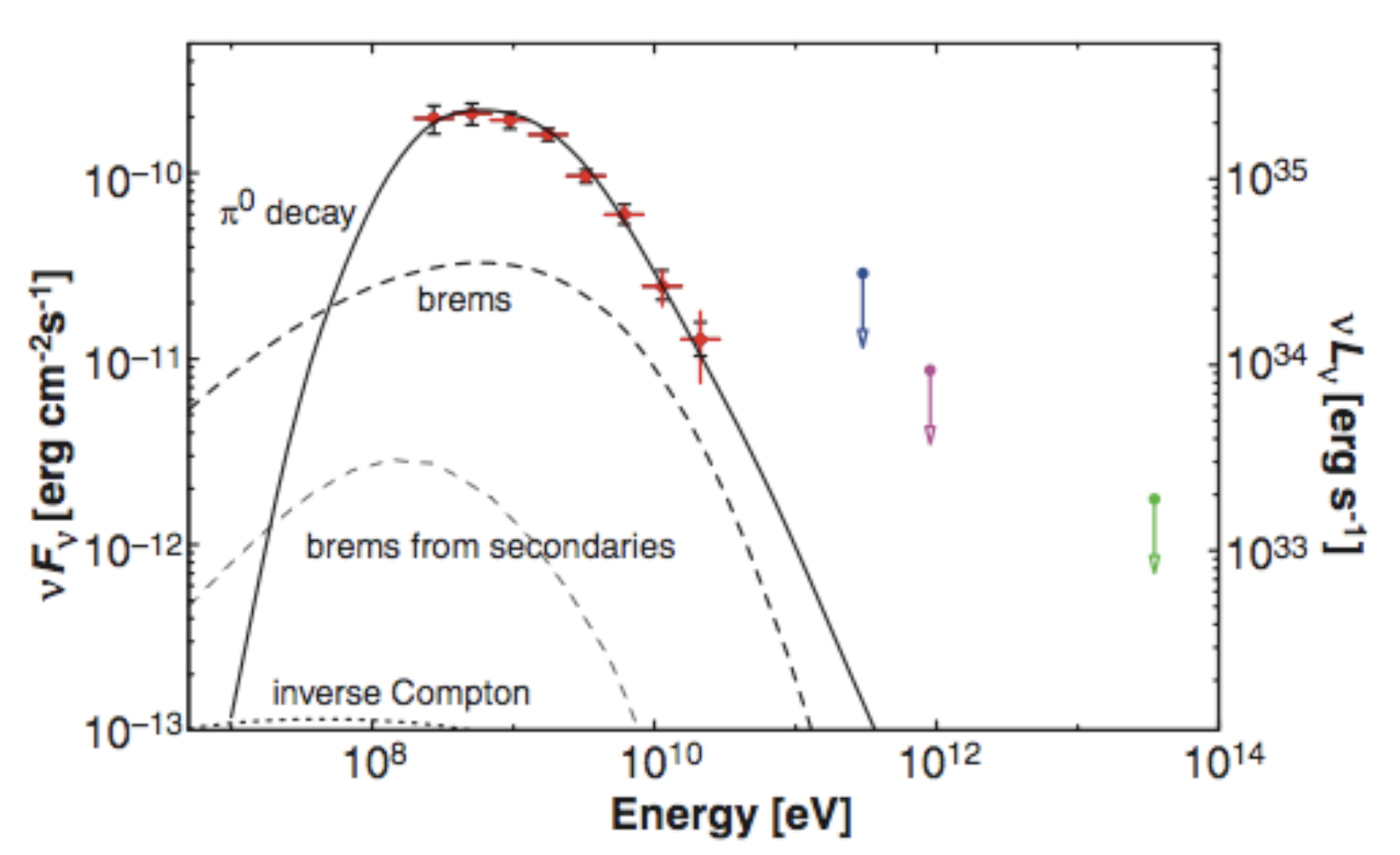} 
 \caption{Fermi LAT spectral energy distributions of SNR W44. 
Each curve corresponds to contributions from: $\pi^0$ decay (solid),
electron bremsstrahlung (dashed), inverse Compton scattering (dots), 
and bremsstrahlung from secondary electrons and positrons \citep{Abdo:2009zz}.
Colored arrows refer to upper limits from Whipple, Hegra and Milagro experiments.}  
 \label{fi:W44_s}
 \end{minipage}
\end{figure}

{\bf Acceleration.}
As we already mentioned, the acceleration spectrum of primary GCRs is believed to be determined by supernovae. 
The acceleration process takes place in the 
SNR phase and the most plausible mechanism is  
diffusive shock acceleration. 
Indeed, if the acceleration would take place in the explosion 
itself, the adiabatic losses suffered by charged particles would require 
an enormous energy amount to reach the observed energy levels
\citep{1982MNRAS.198..833D,2001astro.ph..6046O}.
   The  spectrum of accelerated particles escaping the remnant follows  a
power-law in momentum, $Q(E)\propto p^{-\alpha}$, with  $\alpha$  located between 2.0 and 2.5.
Even if a precise value for $\alpha$ cannot be predicted, the effective spectral index as derived by several indirect observational tests and by 
numerical studies is close to $\alpha \sim$ 2.0-2.1
(\citep{2001astro.ph..6046O} and refs. therein, \citep{1994APh.....2..215B}).  \\
The diffusive SNR shock acceleration model has been improved by the nonlinear reaction effects of the
accelerated particles on the shock structure \citep{1997ApJ...487..182M,2000ApJ...540..923B}. A complete review on the topic can be found in \citep{2001RPPh...64..429M}.\\
The structure of the shock is modified by the accelerated
particles  pressure acting on the background plasma, though in a collisionless
manner. Particles with different momenta experience different effective accelerations.
One consequence of this effect is that the spectrum {\it inside}
the SNR is no more a pure power law but is concave as a function of momentum
\citep{2006MNRAS.371.1251A}. 
Additionally, it was noticed in \citep{1983A&A...118..223L,1983A&A...125..249L}
that the maximum energy reachable by particles in a SNR is by far 
lower than the knee energy. A possible way out is to assume Bohm-like
diffusion coefficient -- which scales as the inverse of the magnetic field 
instead of the inverse squared field -- and a strongly non-linear magnetic field amplification.
The latter is due to the reaction effects of the accelerated particles on the shock structure.
In this scheme, the maximum momentum a proton can acquire is about 
few $10^6$ GeV, thus satisfying the knee energy reach \citep{2007MNRAS.375.1471B}. \\
A convincing evidence of particle acceleration in young SNRs 
would be the observation of high energy $\gamma$'s 
with a clear hadronic origin:  due to their hard energy spectra, 
the relativistic particles produced at the 
source with high acceleration efficiency can interact with the ISM, producing neutral and 
charged pions which in turn generate photons and neutrinos 
\citep{1994A&A...287..959D,2000ApJ...540..923B}. 
This signal must however by tagged against a 
competing emission process for $\gamma$'s from inverse Compton scattering by high energy electrons. 
To get an unambiguous determination of the origin of the radiation, $\gamma$ ray telescopes must provide 
spatial resolved spectral measurements and correlate their studies with other wavelength data (a deeper 
discussion can be found in the contribution by F.Aharonian, this book). As an example, RX J1713.7-3946 
\citep{eno02,aha04,aha06,aha10,2011ApJ...734...28A} is a source of  TeV $\gamma$ emission  corresponding in morphology to the 
non thermal X-ray emission seen by the ASCA satellite \citep{uch03} and to  the CO observations by the 
NANTEN sub-mm telescope \citep{fuk03}. 
Similarly, gamma observations of the SNR W44 by the Fermi Large Area Telescope  \citep{Abdo:2009zz}
are in remarkable agreement with the morphology of the remnant in the radio image
 in the 20 cm wavelength, as displayed in Fig. \ref{fi:W44}. 
The contribution from inverse Compton scattering of electrons cannot yet
been excluded, but it seems unlikely that it could dominate in the GeV band. 
These results are visible in Fig. \ref{fi:W44_s}, where contributions from 
$\pi^0$ decay are displayed along with electron bremsstrahlung, 
inverse Compton scattering and bremsstrahlung from secondary electrons 
and positrons. 
A different treatment should be deserved for the acceleration and propagation
of UHECRs, which are commonly believed to be of 
extragalactic origin (\citet{hill84,2005MPLA...20.3055B} and refs. therein). This topic is discussed at length in Sec.\ref{sec5}.

\begin{figure}[h]
\begin{minipage}{0.47\linewidth}
\centering
\includegraphics[width=8.cm,height=8cm]{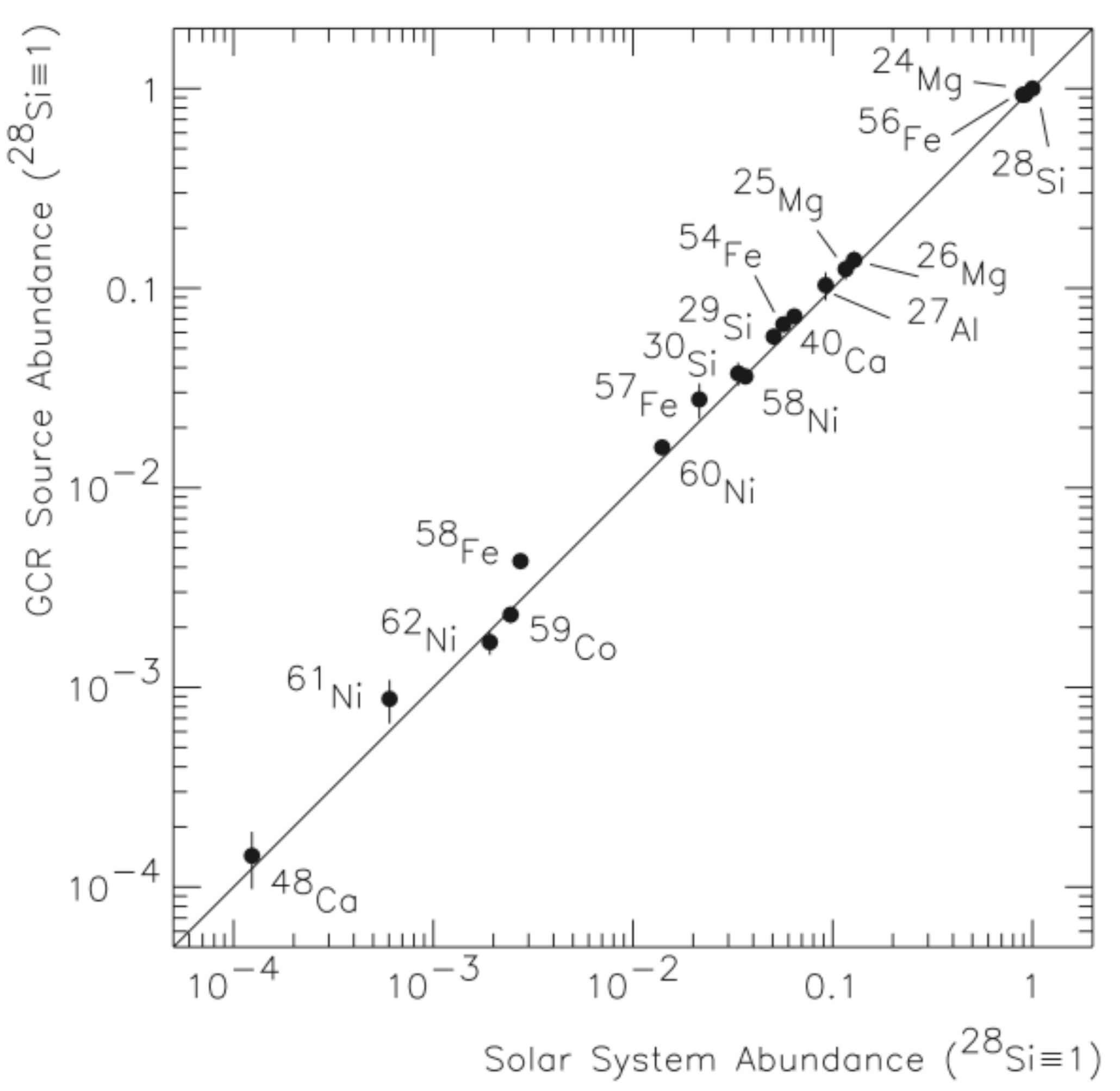}   
\caption{\em {Abundances of refractory nuclides in GCR and Solar
System (propagation modeled by a leaky-box model) \citep{cri03}.}}
\label{fi:abun}
\end{minipage}\hfill
\begin{minipage}{0.47\linewidth}
\centering
\includegraphics[width=8.cm,height=8cm]{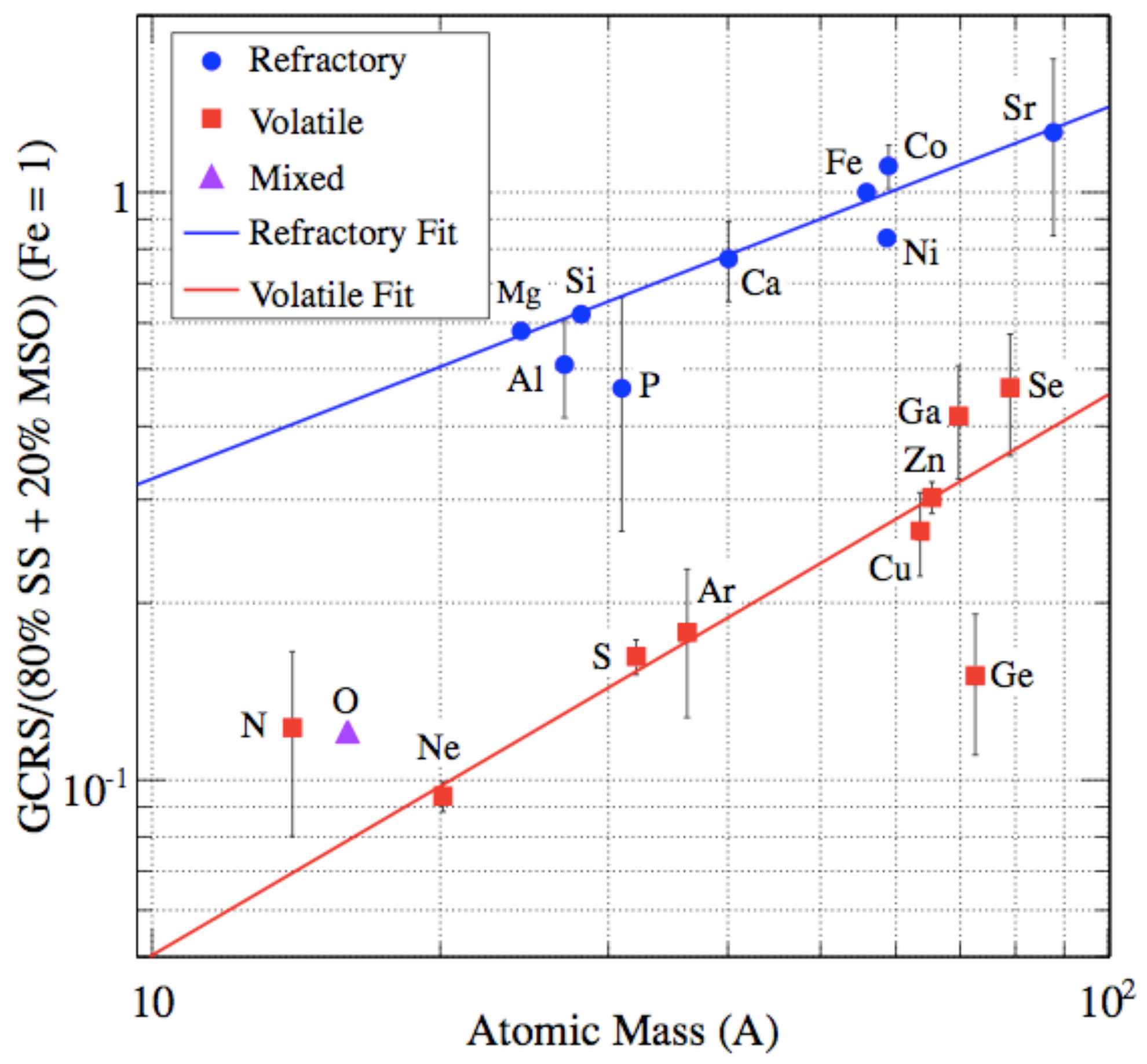}   
\caption{\em {Ratio of GCR abundances to a mixture of $80\%$ solar system
and $20\%$ massive star outflow vs atomic mass A \citep{lin09}.}}
\label{fi:tiger}
\end{minipage}
\end{figure}

{\bf Source composition}
The composition of CRs at their source can be inferred either as the solar
System composition convoluted with the first ionization potential (FIP)
\citep{cas78}, or with the volatility \citep{mey97}, a parameter indicating 
the condensation temperature of an element (volatile elements are 
the light ones while refractory elements are mostly metals, having high melting
points). 
In the first case, it has been suggested that the seed material forming
the CRs be in coronal mass emissions from Sun-like stars, where
low FIP elements are overabundant.
Most interestingly, in the
latter model the seeds are instead refractory elements condensed in
grains, which are accelerated to relativistic energies (thus becoming
"CRs") in supernova shocks with an efficiency proportional to
their high charge-to-mass ratio.
Fig.\ref{fi:abun}, taken from \citep{cri07}, shows the CR abundances in
comparison with the Solar System ones \citep{solar}. The first ones have been
obtained  using a leaky box model for
interstellar propagation with a solar modulation one, and normalized to
the energy spectra of
individual isotopes from CRIS data \citep{cri07}. These elements come from
different
nucleosynthesis processes and stars with various initial masses: the striking
similarity (within $20 \%$ taking into account the systematics) between
the two populations  can be understood only if the two samples are coming
from a similar, well mixed material.
\\
The largest differences are instead observed for the $^{22}Ne/^{20}Ne$,
and at a lower extent for  $^{58}Fe/^{56}Fe$.
The isotopic composition of Ne, Fe and other species has been measured by
the CRIS spectrometer \citep{cri07}.
The ratio $^{22}Ne/^{20}Ne$ so  found is consistent with GCRs sources with $\simeq
80\%$ solar system composition plus a $\simeq 20\%$  of material from
Wolf-Rayet (WR) stars. These kind of stars are evolutionary products of the hot and massive 
OB stars, which are short living and highly radiating objects, loosely organized in groups called OB associations. As a consequence,
any model of origin of GCRs must include an efficient mechanism to inject
the WR material in the accelerator.
Measurements of the $^{59}Ni$, an isotope which decays only by electron
capture, have shown that it has almost completely decayed to $^{59}Co$,
allowing to set up a limit of $\simeq 10^5$ y for the time that has to
elapse from the production of the material and its acceleration to cosmic
ray energies \citep{wie99}.  Models of superbubbles environments
\citep{hig98}  host Supernova events every 0.3-3.5 My, thus allowing enough
time for Ni to decay.
With 50 days of data, the TIGER experiment \citep{lin09} measured a clear
ordering of abundances of GCRs when compared to a mixture of $80\%$ solar
system and $20\%$ star outflow, thus  strongly confirming that GCRs
originate in OB associations. As shown in  Fig.\ref{fi:tiger}, this result
also confirms the preferred acceleration of refractory elements as
compared to volatile ones and derives a mass dependence of the relative
abundances for both refractory ($\propto A^{2/3}$) and volatile ($\propto
A$) elements.  This results has been confirmed also by measurements at higher energies, up to almost 4 TeV \citep{2010ApJ...715.1400A}.

{\bf Spatial diffusion.}
The diffusion on the irregularities of the galactic magnetic
field explains the highly isotropic distribution 
of CRs and their confinement in the Galaxy. 
The scatter of charged particles on the random 
inhomogeneities of the magnetic field $B$ is usually treated in the 
quasi-linear approximation, according to which the average and fluctuating 
fields are separated and it is assumed that fluctuations $\delta B << B$ are 
small (B is the regular field). 
The interaction between the waves and the cosmic particles
is of resonant type and is maximized when the irregularities on the magnetic field
have the wave vector component parallel to the average magnetic field 
$K_{//}= \pm s/(r_g \mu)$, where the integer $s$ is the order of cyclotron 
resonances, $r_g=R/B$ is the gyroradius, $R$ is the particle rigidity, 
 and $\mu$ is the particle pitch angle. 
According to \citep{1993A&A...268..726P}
 $$ K_{||}(\vec{r},R)\simeq \frac{1}{3} v r_g/P$$
where P is the integral of the normalized power
spectrum of the turbulent hydromagnetic fluctuations 
$P=\int \Delta B(k)^2/B^2$ . It is generally assumed $B(k)^2\propto k^{-a}$, 
with different predictions for $a$. The Kolmogorov turbulence spectrum 
is identified by $a$=5/3, which gives $K_{||}\propto R^{1/3}$. 
An exponent $a$=3/2 is typical for the Kraichnan turbulent spectrum, corresponding
to $K_{||}\propto R^{1/2}$, which is closer to the trend of
the diffusion coefficient derived from pure diffusion interpretation of 
high energy CR data. 
The most direct way to deduce phenomenologically  the diffusion coefficient is 
the comparison with data on secondary--to--primary nuclei fluxes. The best 
data are for the ratio of boron--to-carbon (B/C). The boron is a purely secondary nucleus, produced mostly by the fragmentation of carbon, nitrogen and oxygen nuclei, while carbon
is a primary species directly produced in the SNR.  
In the case of pure diffusion (neglecting energy losses and gains, convection and nuclear destructions) B/C $\propto K(E) \propto R^{-\delta}$. 
The main difference between the two diffusive schemes mentioned above is in that the Kraichnan
model predicts a softer (steeper) observed spectrum $\propto R^{-0.5}$, and closer
to the observation of the secondary--to--primary ratio (B/C above all) than 
the Kolmogorov spectrum, which would predict a B/C $\propto R^{-0.3}$ at $\geq 20-30$ GeV.
In the quasi-linear theory the perpendicular diffusion coefficient is 
suppressed 
with respect to the parallel one, meaning that diffusion takes place 
mostly along the regular field direction. 
The quasi-linear theory leads to a rigidity power law for the diffusion coefficient, which  is usually assumed to have the form: 
\begin{equation}
K(E)=K_0 \beta^{\eta} R^\delta
\label{diff_coeff}
\end{equation}
($\beta$ is the Lorentz factor).  
$K_0$ is linked to the level of the hydromagnetic turbulence and
$\delta$ to the density spectrum of these irregularities at different
wavelength. $\eta$ is usually set to 1, while a different value parameterizes
very low energy deviations.  
The lack of information on the magnetic field irregularities prevents 
us from a  precise determination of the diffusion coefficient, which is 
instead possible only from interpretation of CR data. 
The $\delta$=1/3 case, together with some amount of reacceleration, can
reproduce the peculiar boron--to--carbon (B/C) peak observed at $\sim$ 1 GeV/n 
but it tends to overestimate the data at higher energies, for which 
$\delta$=0.5-0.7 seems preferred.  
On the other hand, a diffusion coefficient $\delta \sim 0.6$, that is a short confinement time, would imply a high degree of 
anisotropy in the very high energy region above $10^{14}$ eV, at variance with the results of the experimental measurements (see Sect.\ref{solu1} and \ref{anis}).

{\bf Reacceleration.} Diffusive, or continuous, 
reacceleration is a second order Fermi acceleration process 
due to the scattering of charged particles on the magnetic turbulence in the
interstellar hydrodynamical plasma. The diffusive reacceleration coefficient is related to the
velocity of  such disturbances, called the Alfven velocity $V_A$, and is naturally connected to the space diffusion coefficient through the relation
\citep{1994ApJ...431..705S}:
\begin{equation}
K_{pp}\times K= \frac{4}{3}\;V_A^2\;\frac{p^2}{\delta\,(4-\delta^2)\,(4-\delta)}.
\label{eq:Va}
\end{equation}
where p is the particle momentum and  $\delta$ is the power index of the diffusion coefficient K
as in Eq.~\ref{diff_coeff}. The scattering centers drift in the Milky Way with
a velocity $V_{A} \sim$ 20 to 100 km s$^{-1}$.
Diffusive reacceleration contributes
significantly in shaping the B/C ratio at kinetic energy per nucleon E around
1 GeV/n. At energies greater than few tens of GeV/n the
energy change due to reacceleration is negligible and the relevant effect 
can be safely neglected in the propagation equation. 

{\bf Convection.}
The stellar activity and the energetic phenomena associated to the late stage
of stellar evolution may push the ISM and the
associated magnetic field out of the galactic plane.
The net effect is likely a convective wind directed
outwards from the galactic plane, which adds a convective term 
to the diffusion equation (Eq.\ref{eq:CR})  \citep{1965ApJ...142..584P,1975ApJ...196..107I}.
The properties of this wind may be studied by hydrodynamical~\citep{1971ApJ...165..381J,1989A&A...225...37Z} and
magneto-hydrodynamical (MHD)~\citep{1985A&A...151..151S} methods. 
Convection may dominate at low energies, 
depending on its intensity $V_C$  and if $\delta$ is large, and compete with diffusion up to  few tens of GeV/n, 
but its role becomes negligible at higher energies. 

{\bf Energy losses.} 
Nuclei lose energy via ionization in the ISM 
neutral matter ($90$\% H and $10$\% He), 
and Coulomb interaction in a completely ionized plasma,
dominated by scattering off the thermal electrons.
The expressions for the corresponding energy loss rates can be found
in Refs. \citep{1994A&A...286..983M} and \citep{1998ApJ...509..212S}.

{\bf Nuclear fragmentation.} 
Charged CR nuclei heavier than protons suffer catastrophic losses due to
nuclear destruction on the interstellar H and He. These reactions become irrelevant on the
CR spectrum  at energies $\gsim$ 1 TeV/n, depending on the nuclear species. 
Inelastic interactions  lower and flatten the flux of the 
broken nucleus, while giving rise to the secondary nuclei produced in the inelastic collision. 
Fragmentation cross sections can be derived  from the semi-empirical formulation 
of \citep{1990PhRvC..41..566W}  (see also \citet{2001ApJ...555..585M} and references therein). 

\subsection{Experimental methods for direct measurements}
Direct measurements of the GCRs are possible up to $\simeq 10^{14}$ eV by means of experiments flown on balloons or satellites.\\ 
$\bullet$ {\it Balloon-borne experiments} can be assembled  with a moderate budget, and in general allow multiple measurements, in that their payload can be eventually recovered, repaired and flown again. 
On the other hand, they can provide only a limited exposure of few days of flight, due to the limited resources on-board and to the winds, which direction and velocity can drive balloons far from the launch site or above populated areas.\\
This limit is constantly being pushed forward with the recent advent of the so-called Long Duration Balloon Flights, where balloons with volume $\gsim 10^6$ m$^3$ and suspended weight $\simeq 3-4$ ton can fly for tens of days. The CREAM experiment \citep{cre04} reached the record flight duration of 42 days in 2005 in Antarctica. 
An intense activity of research and development is ongoing to produce Ultra Long Duration Balloons, pumpkin-shaped super-pressure  balloons using a very thin closed plastic skin, able to stay aloft up to 100 days. Prototypes were successfully flown in Antarctica in the last couple of years.\\
Two different classes of experiments on balloons have been realized so far, the first searching for antimatter in CRs \citep{bar97,boe01,bes02,yos04} and the second aiming at the measure of the CR primary composition up to 100 TeV \citep{cre04,ati09,2011arXiv1108.4838O}. \\
$\bullet$ {\it Satellite experiments} have longer exposure and can avoid the background related to the  residual atmosphere above the balloons, but their cost is very high. They can fly at different orbits: polar ones, to study low energy CRs at high latitudes, or equatorial ones if they want to detect gamma radiation, to be screened from CRs. Research activity on charged CRs is performed both to search for antimatter and to study chemical abundances \citep{pam09}.
Typical detectors employ various combinations of different instruments, which aim is that of measuring the magnitude of the incoming particle charge and its energy. The goal of measuring the single primary elements, both in spectrum  and abundance, requires adequate exposures:  e.g., instruments aiming at reaching the knee energies with minimal statistics of about 10 events above $10^{15}$ eV need exposures from $\simeq 1$ to $20$ m$^2$sr y for H and heavy elements respectively.\\
$\bullet$ {\it Space-based experiments} can be hosted on board space stations, with stricter requirements and higher costs. The AMS detector has been designed to operate as an external module on the International Space Station, with the aim of  searching for antimatter and dark matter while performing precision measurements of CRs composition and flux \citep{zuc09}.  The Jem-EUSO detector is planned to be located on the Japanese module of the Space Station to study  the extreme high-energy phenomena in the Universe, at $E>10^{20}$ eV \citep{tak09}. \\
\vspace{0.4cm}

\centerline{\it The measure of energy}
In magnetic spectrometers, the rigidity $R$ and the sign of the charge of the crossing particles can be determined. Measuring the rigidity R as a function of the radius of curvature of the particle, $R=B \ r_{curv}$ in a magnetic field $B$, and the charge  from the ionization energy lost in the tracker, the particle momentum can be obtained. The performance of the spectrometer is characterized by the distribution of the maximum detectable rigidities (MDR) for all trajectories, which is generally defined such that the error in the measure of R be
$$\frac{\delta R}{R}=\frac{R}{MDR}$$
The sign of the charge of a particle can be reliably determined for rigidities up to $\frac{1}{3} \ MDR$.\\
Calorimeters or transition radiation detectors are used to derive the particle energy. The first are generally employed for the measure of low-charge particles such as protons and helium nuclei.
Due to the limitations in thickness and weight inherent to their location on balloons or satellites, they cannot be used to study higher Z nuclei.
If $E_{1}$ is the energy required to produce a detectable "quantum" of energy in the detector (a photon for a scintillator, a positron-electron pair for a semiconductor device, an electron-ion pair for a gas array, etc.) and $E_0$ is the initial energy, then $N=\epsilon E_{0}/E_{1}$ is the actual number of recorded quanta ($\epsilon$ being the collection and detection
efficiency of the sensitive element). The energy resolution is dominated by poissonian fluctuations and improves with increasing primary energy, as well as in detectors with smaller $E_{1}$ (e.g., $E_{1} \simeq 1$ eV for semiconductors, while $E_{1} \simeq 100$ eV for scintillators). Typically, thick targets with sufficient interaction lengths to force the interaction of hadrons are placed in front of an electromagnetic calorimeter, thick enough in radiation lengths to fully absorb the secondary cascade. Calorimeters can be calibrated at accelerator beams with protons and heavy ions.\\
For nuclei with $Z > 3$, transition radiation detectors (TRD) can measure the Lorentz factor $\gamma_{L}=E/m$, which together with the knowledge of the particle mass can provide an energy measurement.  This radiation  is emitted in the X-ray region when a particle traverses the boundary between two media with different dielectric properties and is proportional to $\gamma_{L}$. It is observable above $\gamma_{L} \gtrsim 400$, where all other methods already give saturated signals, up to $\gamma_{L} \simeq 50000$. The upper limit is a consequence of the interference in emission from the multiple layers of material which constitute the radiator; in fact, multiple foils are needed to enhance the probability of emission, which for a single charge is of the order of 1/137. \\
The TRD has the advantage of having a relatively low mass, thus larger collecting areas can be set on balloons or satellites despite the weight limitations. On the other hand, no measurement  for $Z < 3$ is possible, due to the low photon number and consequently the too large fluctuations.  When both TRD and calorimeters are used, a fraction of the incoming particles  can cross both of them, thus providing a cross-calibration of the energy determination, at least for $Z>2$, with completely different systematic uncertainties. 
Cherenkov counters and proportional tubes are also used, offering together with the TRD a response proportional to the $Z^2$;  differently from magnetic spectrometers, the misidentification of the charge can heavily influence the determination of the Lorentz factor and hence of the energy.\\
At higher energies, emulsion chambers are used to study the angle and energy of the electromagnetic cascades from the decays of the neutral pions produced in the first interaction of the primary in the detector \citep{jac98,run05}. 
Good accuracies are obtained in the measurement of the energy released in the electromagnetic component in such interaction (about $15\%$); the largest source of uncertainty is represented by the fluctuations in the fraction of primary energy going into $\pi^0$, which decreases for heavier primary nuclei.\\
\vspace{0.4cm}

\centerline{\it The measure of the charge}
 The measure of the charge is generally based on the use of scintillators or solid state detectors. The ionization energy lost by the particle is proportional to $Z^2$.  Combining the measure of the kinetic energy and of the energy loss in a thin detector, the atomic number of the particle can be obtained. 
For anti-matter studies, the rigidity, charge and charge sign of a particle can be measured by means of magnetic spectrometers, consisting of a magnet (either permanent or superconducting) and a tracker. 
In Cherenkov counters, the signal of a particle with charge Z and velocity $\beta$ (in units of c)  is $\propto Z^{2}(1-1/\beta^{2}n^{2})$, with $n$=refractive index of the medium; its response abruptly drops below a threshold Lorentz factor given by $\beta=1/n$.\\
Different detectors can be employed together to measure the charge in a redundant way. For example,  four independent instruments are used in the CREAM detector \citep{cre04}: a timing-based scintillator charge detector, a plastic Cherenkov detector, a scintillating fiber hodoscope and a silicon charge detector. The use of detector segmentation or timing technique allows  backsplash particles produced for example in a calorimeter in the lower part of the payload to be tagged.\\
The required resolution in charge must reach 0.2 charge units or better to resolve different elements (e.g.  boron from carbon, for which the flux ratio can be as low as $1\%$).  \\
At energies above 1 TeV, emulsion chambers are also used, where the particle charge is measured through the ionization produced in stacks of nuclear or X-ray emulsion plates; the "darkness" of the spots produced when a charged particle crosses the sensitive layers is proportional to the nuclear charge. 

\subsection{Diffusive models and comparison with data.}
\label{solu1}
The solution of the master diffusion equation (Eq.\ref{eq:CR})
 has been deeply investigated and several different techniques proposed in the
  literature can lead to similar fluxes at the Earth, at least for stable nuclei
\citep{Strong:2007nh,Maurin:2002ua,2001ApJ...547..264J}. \\ 
In diffusive models \citep{1980Ap&SS..68..295G,1992ApJ...390...96W}, CRs diffuse in a region called diffusive halo,
pictured as a thick disk  high somehow between 1 and 15 kpc, which matches the circular structure of the Milk Way. 
 The galactic disk  of stars and gas,
where primary CRs are accelerated, lies in the middle. It extends
radially 20 kpc from the center and has a half--thickness $h$ of 100 pc.

\begin{figure}[h]
\begin{minipage}{0.47\linewidth}
 \centering
\includegraphics[width=8.cm,height=8cm]{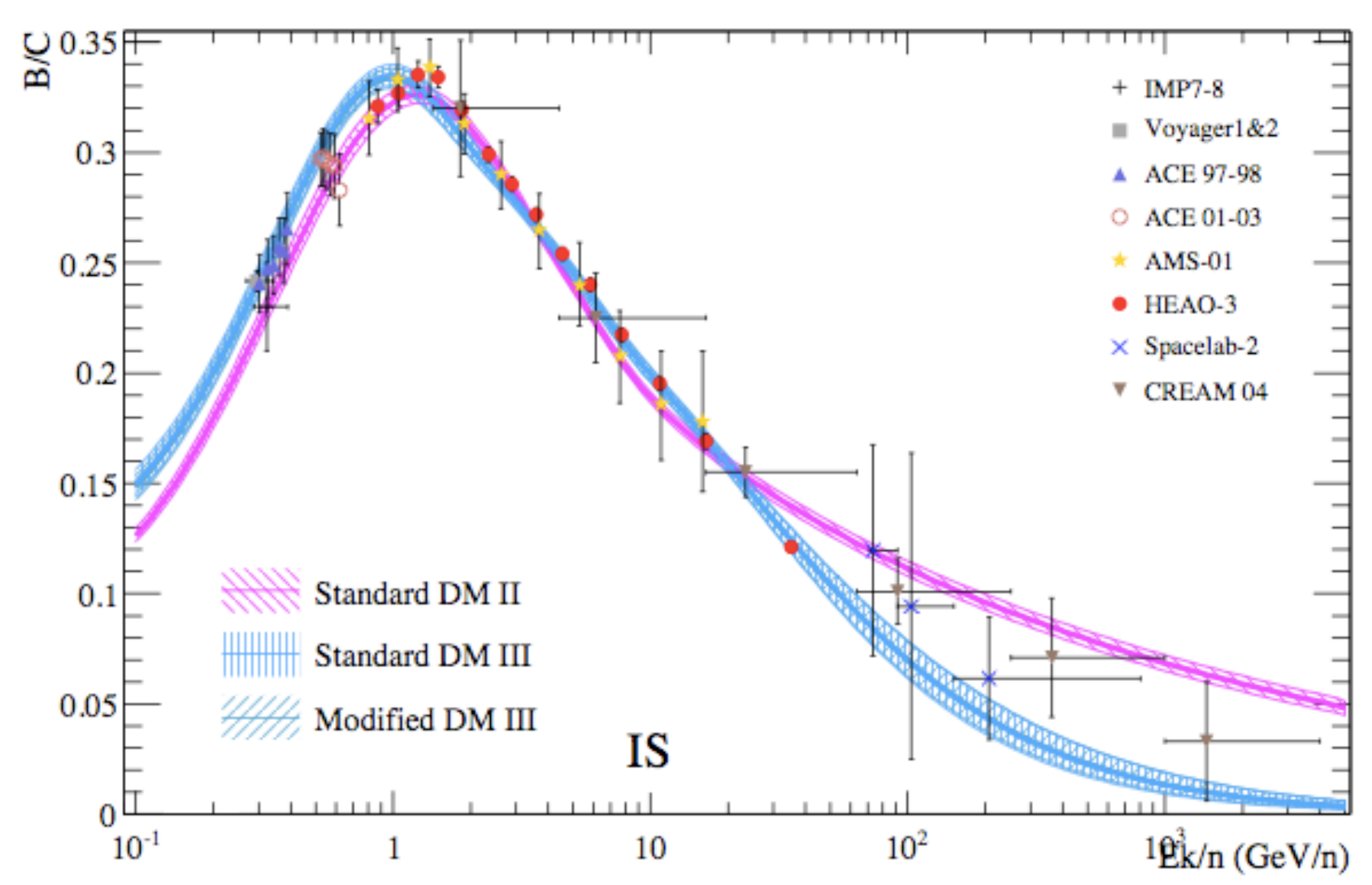}  
 \caption{68\% CL envelopes (shaded areas) and best-fit (thick lines) for 
 diffusive/reacceleration models  ($r_h = 0$, red) and for diffusive/reacceleration/convection
 models (blue) tested on B/C data. From \citep{Putze:2010zn}.}
 \label{fi:BC}
 \end{minipage}\hfill
 \begin{minipage}{0.47\linewidth}
 \centering
\includegraphics[width=8.cm,height=8cm]{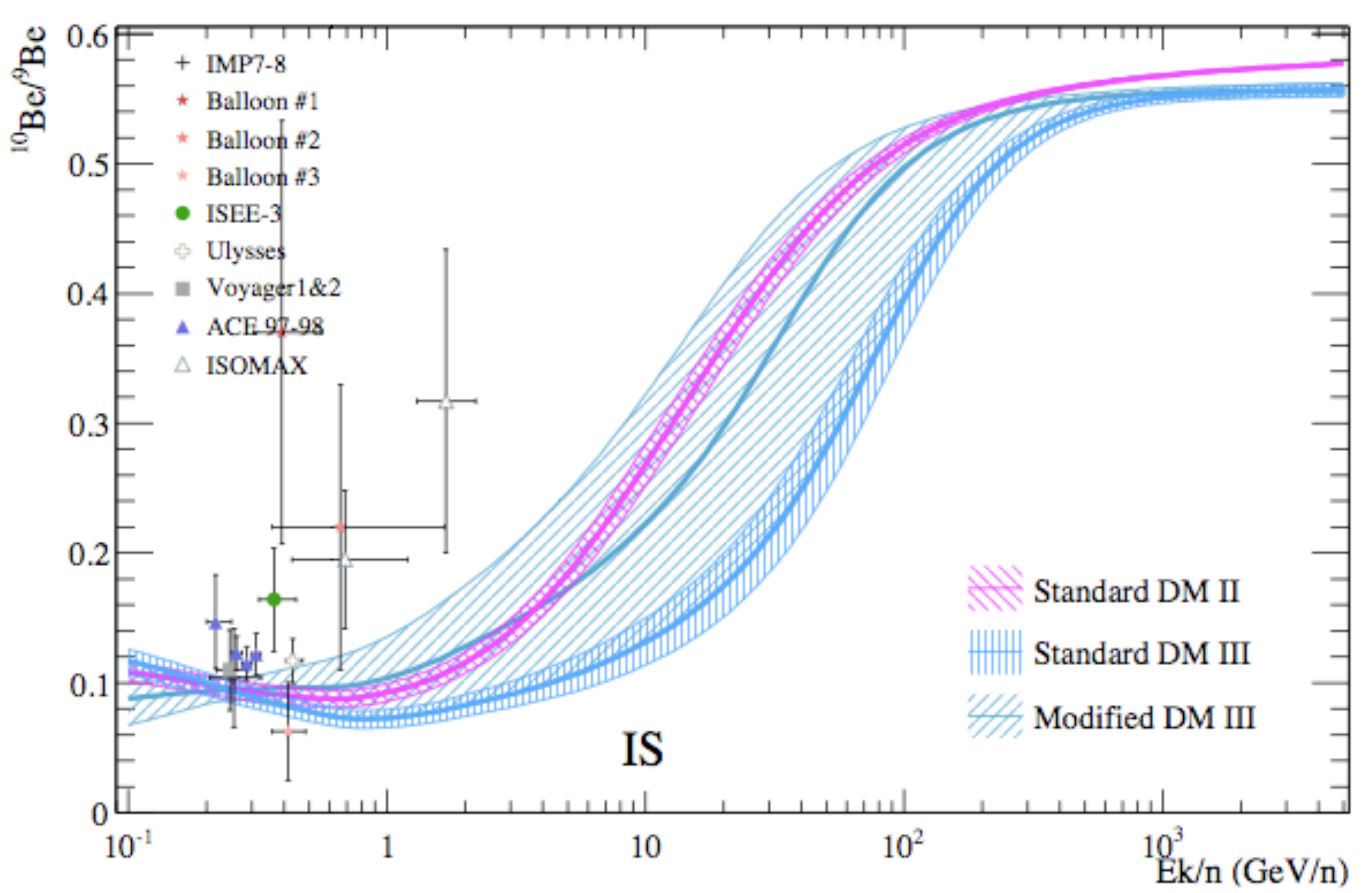}  
 \caption{68\% CL envelopes (shaded areas) and best-fit (thick lines) for 
 diffusive/reacceleration models  ($r_h = 0$, red) and for diffusive/reacceleration/convection
 models (blue) on $^{10}$Be/$^9$Be. From \citep{Putze:2010zn}.}  
 \label{fi:BC_rad}
 \end{minipage}
\end{figure}

Fully numerical solution is the one adopted by the GALPROP  
code \citep{1998ApJ...509..212S,2011ApJ...729..106T}, 
which is particularly useful if aiming at the production of 
gamma rays from charged CRs. Gamma rays do not diffuse and 
are therefore observed at their source, so that a full spatial treatment
is required \citep{Strong:2004de}. 
Comparable results for the propagation of stable primary and secondary
nuclei have been obtained with the similar Dragon code \citep{2008JCAP...10..018E}
and with the modified weighted slab
technique~\citep{2001ApJ...547..264J}.
\\
The Bessel expansion method, based on the cylindrical symmetry of the diffusive halo and 
on approximate values for the ISM (not relevant for charged CR propagation)
allows a two dimensional fully analytical model
\citep{2001ApJ...555..585M,Maurin:2002ua,Maurin:2002hw,2002A&A...381..539D}. 
Numerical solution is required only for the diffusion in energy space. 
If the interest is concentrated on charged particles, the 
complete spatial solution is often redundant, since they diffuse 
for long time before being detected and they do not backtrace their 
source. 
\\
The free parameters of diffusive models  - 
typically the normalization and power spectrum index of the diffusive coefficient  $K_{0}$  and $\delta$, 
the thickness L of the diffusive halo, convective and Alfv\'en velocities $V_{C}$ and $V_{A}$ - are fixed by 
studying the B/C ratio which
is quite sensitive to CR transport and which may be used efficiently as a constraint.
 These parameters were self-consistently constrained in 
\citep{2001ApJ...555..585M, Putze:2010zn}.
When only B/C is considered, models with diffusion, convection and reacceleration are preferred 
 over the diffusion and reacceleration case.
The former points towards $\delta\sim 0.7-0.8$, while the latter points towards $\delta\sim 0.3$. 
This result does not depend on the halo size  $L$.  Moreover,  a B/C analysis based on 
AMS-01 data (instead of HEAO-3) indicates that the presence of
convection and reacceleration is required and points to a diffusion slope
$\delta\sim 0.5$, closer to theoretical expectations. 
As regards the normalization of the diffusion coefficient $K_0$, it is worth
noting that $B/C$ measurements actually constrain $K_0/L$, not $K_0$ alone
\citep{2001ApJ...555..585M,2011ApJ...729..106T}. 
However, no definite answer is given to the precise values of the propagation 
parameters -- $\delta$ being the most relevant one -- and a high degeneracy 
among the parameters is observed. 
Data from PAMELA and from AMS02 are awaited to
help solving the long-standing issue on the value for $\delta$.
\\
One possible improvement to the understanding of the propagation physics 
would be the simultaneous study of stable and radioactive species -- usually believed to be
powerful tracers of the diffusive halo size. In fact, the combined analysis 
 does not yet allow to fully break this degeneracy~\citep{2002A&A...381..539D}.
Secondary radioactive species, such as $^{10}$Be, originate from the spallation of their heavier progenitors with the ISM and have half-lives $\tau_0$ ranging 
from 0.307 My for $^{36}$Cl to 1.51 My for $^{10}$Be.
The quantity $l_{\rm rad}=\sqrt{K \gamma \tau_0}$ is the typical distance on which a
radioactive nucleus diffuses before decaying. Using $K\approx 10^{28}$~cm$^2$~s$^{-1}$  and
$\tau\approx 1$~My, the diffusion length is $l_{\rm rad}\approx 200$ pc.
Therefore,  any under-dense region on a scale $r_h\sim 100$~pc around the Sun 
(as the observed Local Bubble) leads to an
exponential attenuation of the flux of radioactive nuclei. The attenuation is
energy and species dependent \citep{2002A&A...381..539D}.
In diffusion/convection/reacceleration models, $L\sim8$~kpc with $r_h\sim120$~pc \citep{Putze:2010zn}, 
and in diffusion/reacceleration models  $L\sim4$~kpc and no underdense region
is necessary to explain the data \citep{Putze:2010zn,2011ApJ...729..106T}.
The halo size comes out as an increasing function of the diffusion slope $\delta$.
 A striking feature is that in many models $r_h$ points to $\sim100$~pc, a value supported by direct astronomical
observations of the local ISM (see discussions in \citep{Putze:2010zn,2002A&A...381..539D} and refs. therein).
In Figs. \ref{fi:BC} and \ref{fi:BC_rad} 
 we display results obtained on a fit to B/C and $^{10}$Be/$^{9}$Be data. 
The free parameters of the model, namely L, $K_0, \delta, V_C, V_A$ have been varied in very large ranges. 
The results demonstrate the above considerations on the degeneracy among the free parameters 
of the diffusive models. A good fit is possible up to tens of GeV/n, where the data are good as well. 
The figure relevant to the $^{10}$Be/$^9$Be ratio also indicates the paucity of 
data for radioactive isotopes (the same is even more evident for all other radioactive isotopes). 

\noindent
A complementary view to CR propagation is given by the anisotropy considerations. 
A high isotropy is peculiar for GCRs. Indeed, the global leakage of particles 
from the Galaxy and the contribution of local sources can lead to an anisotropy, but the regular and
stochastic  magnetic fields tend to isotropize the angular distribution of CRs. 
Only at high energies it is possible to detect some deviation from pure isotropy 
(see also discussion in Sect. 5.1). 
The amplitude of the anisotropy, namely the gradient of the density of CRs, 
$\delta_{AN}$ may be written as 
$\delta_{AN} = -[3 K(E)] \nabla N/vN $ in the case of pure isotropic diffusion, where  
$K(E)$ is the diffusion tensor \citep{1990acr..book.....B} and $v$ is the velocity of the CR. 
The calculations indicate that a diffusion coefficient K(E) $\propto R^{0.3}$ (the 
Kolmogorov spectrum) is compatible with data on CR anisotropy within a factor 
of about 3, while a stronger dependence on energy as K(E) $\propto R^{0.6}$ (which 
is closer to observed absolute nuclei fluxes) tends to over-predict anisotropy 
at $E \geq 10^{14}$ eV \citep{2011arXiv1105.4529B}.

\subsection{Antimatter in CRs}
The presence of a small amount of antimatter in CRs 
is predicted from spallation reactions of incoming protons and helium nuclei
on the ISM (contributions from CRs with higher Z being negligible). 
The spallation products of these inelastic scatterings account for quarks and gluons,  
which immediately hadronize. In particular, 
the hadronization process includes the production of a small
amount of antiprotons, antideuterons and positrons (the latter 
induced by decay processes). 
Antiprotons and positrons in CRs have been measured in recent years with 
increasing accuracy and in an energy window spanning from 
few hundreds of MeV to hundreds of GeV.
For antideuterons, only upper limits have been set.\\
The search for cosmic antimatter is a further test of the 
propagation model. The study of light antimatter, due to its tiny flux, 
 is optimal to search for contributions with spectral peculiarities, 
such as non-thermal production or annihilation from 
dark matter particles in the galactic dark halo. 
The latter are usually predicted to be Weakly Interacting Massive Particles (WIMPs).\\
The indirect dark matter  detection is based on the search for anomalous
components due to the annihilation of dark matter pairs in the galactic halo, 
in addition to the standard astrophysical production 
\citep{Salati:2010rc,2005PhR...405..279B,2004PhRvD..70a5005B}. 
These contributions are potentially detectable as spectral distortions
in various cosmic antimatter fluxes, as well as $\gamma$ rays and $\nu$'s
\beq
\chi + \chi \to q \bar{q} , W^{+} W^{-} , \ldots \to
\bar{p} , \bar{D} , e^{+} \, \gamma \, \& \, \nu's \;\; .
\eeq
The search for exotic signals in cosmic antimatter is motivated by 
the very low astrophysical counterpart, acting as a background. 
Detection of the dark matter annihilation products has motivated the spectacular
development of several new experimental studies.\\
\vspace{0.4cm}

\begin{figure}[h]
\begin{minipage}{0.47\linewidth}
 \centering
\includegraphics[width=8.cm,height=8cm]{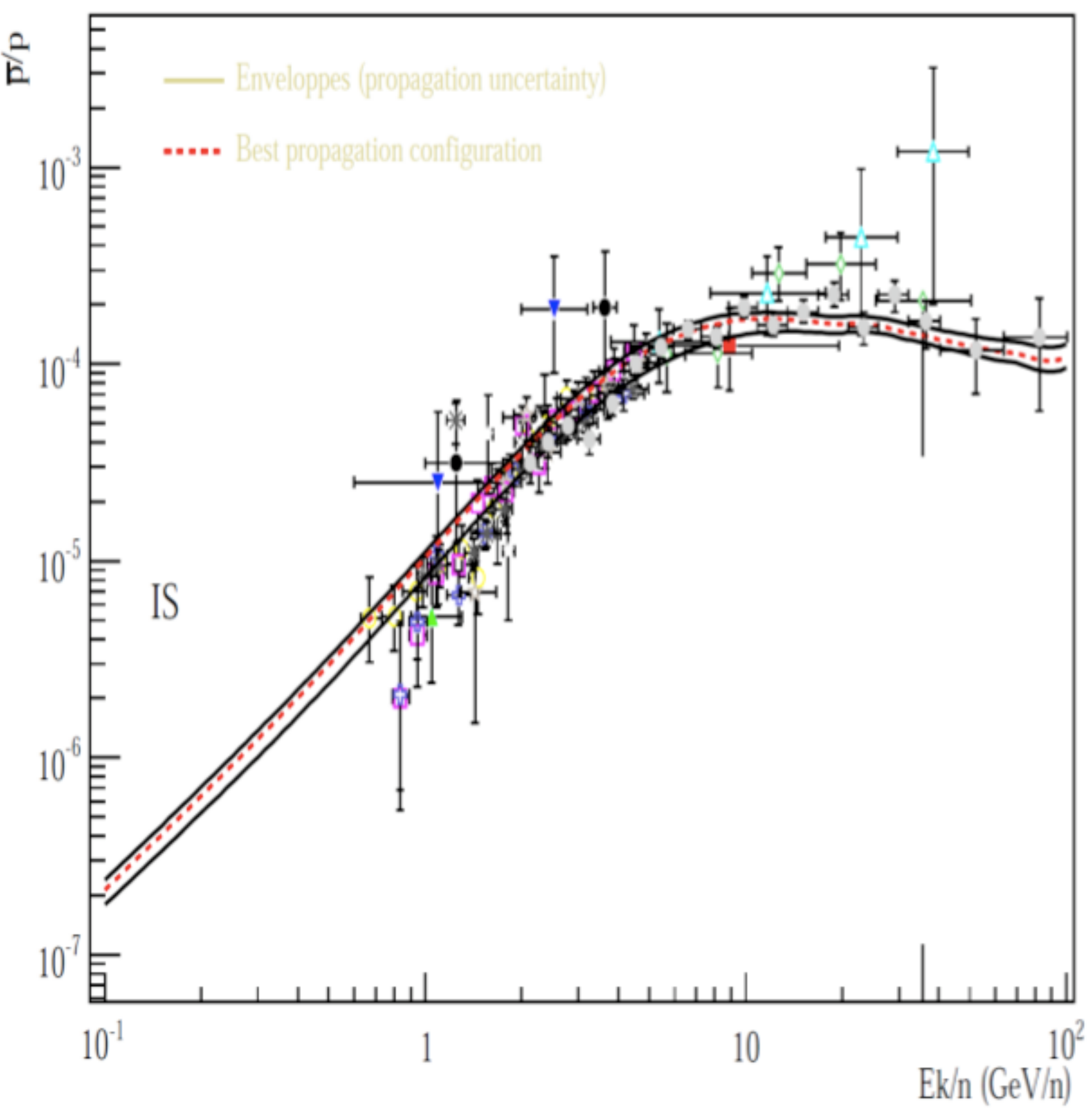}  
 \caption{Propagation uncertainty envelopes of the IS \pbar$/p$ ratio and two
parameterizations of the production cross section \citep{Donato:2008jk}.}
 \label{fi:pbarp}
\end{minipage}\hfill
 \begin{minipage}{0.47\linewidth}
 \centering
 \includegraphics[width=8.cm,height=8.cm]{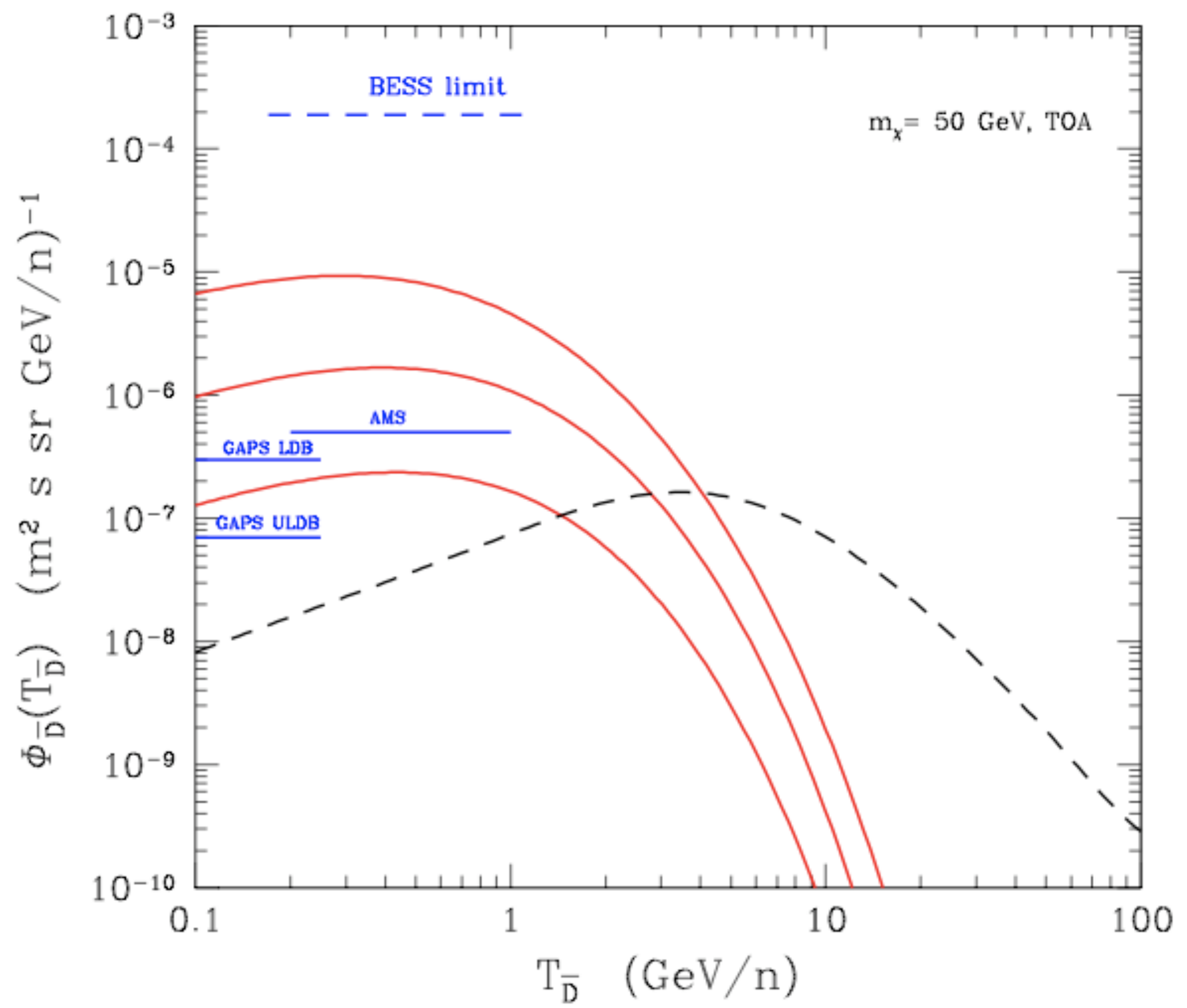} 
 \caption{Antideuteron flux for dark matter sources
(solid) and secondary contribution (dashed).
Horizontal solid lines show the estimated sensitivities for
next generation experiments 
\citep{Donato:2008yx}.}
\label{futuro_dbar} 
 \end{minipage}
\end{figure}

\centerline{\it Antiprotons}
Antiprotons may be produced by spallation of high--energy primary nuclei 
impinging on the atoms of the ISM inside the galactic disk  \citep{Bergstrom:1999jc,Donato:2001ms}. 
They represent the background when searching for small peculiar contributions 
such as  signals from dark matter annihilation.\\
The secondary antiproton flux has been predicted with small theoretical 
uncertainties \citep{Donato:2001ms,Donato:2008jk} and reproduces astonishingly 
well the data from 200 MeV up to at 100 GeV
\citep{2000PhRvL..84.1078O,2001APh....16..121M,2002PhRvL..88e1101A,2002PhR...366..331A,
2005ICRC....3...13H,2005AdSpR..35..135M,2008PhLB..670..103B,Adriani:2008zq,2010PhRvL.105l1101A}. 
This can be seen in Fig. \ref{fi:pbarp}, where we plot,  along with the demodulated \pbar$/p$ data
(data corrected by the effect of the solar wind), 
the curves bounding the propagation  uncertainty on the \pbar\ calculation.
We also illustrate the uncertainty related to the production cross sections. 
From a bare eye inspection, it is evident that the secondary contribution alone explains
the PAMELA data (grey filled circles, \cite{Adriani:2008zq,2010PhRvL.105l1101A}) on the whole energetic range. 
It is not necessary to invoke an additional
 component (such as, for instance, 
 from dark matter annihilation in the galactic halo -- see below) to the standard astrophysical one.

The annihilation of dark matter candidate particles throughout the whole Milky Way volume
may generate primary antiprotons. In this case
the  WIMP annihilations take place all over the diffusive halo.
The antiproton signal from annihilating dark matter particles leads to a primary component
directly produced throughout the diffusive halo \citep{Donato:2003xg}. 
The variation of the astrophysical 
parameters induces a much larger theoretical 
uncertainty on the primary than on the
secondary flux: in the first case, the uncertainty reaches
two orders of magnitude for energies $T_{\bar{p}}\lsim
1$~GeV, while in the second case it never exceeds 25\% (see Fig. \ref{fi:pbarp}
for the secondary population; for the primary dark matter component, 
see the discussion on Fig. \ref{futuro_dbar}).
The reason is in the location of the sources: the primary 
flux due to dark matter annihilation originates in the whole 
diffusive halo and is very sensitive to the halo size 
(which is varied between 1 and 15 kpc) and the convective velocity. 
Therefore, it is of the utmost importance to constrain the 
propagation parameters in order to evaluate any possible exotic contribution
to the \pbar\ flux from sources in the whole halo. 
\vspace{0.8cm}

\centerline{\it Antideuterons}
It was shown that the  antideuteron spectra deriving from dark matter annihilation 
are expected to be much flatter than the  secondary astrophysical component at low
kinetic energies, $T_{\overline{d}}\lsim$ 2-3 GeV/n,
thus offering a potentially very clear indirect detection channel \citep{Donato:1999gy,Donato:2008yx}. 
Antideuterons have not been measured so far, and only an 
 upper limit has been obtained in \citep{2005PhRvL..95h1101F}.
\\
The secondary \dbar\ flux is the sum of the six contributions corresponding to $p$, He
and \pbar\ CR fluxes impinging on H and He IS gas (other reactions are
negligible) \citep{1997PhLB..409..313C,Duperray:2005si}.
 The production cross sections for these specific processes
are those given in \cite{Duperray:2005si}.
The solution to the propagation equation has the same expression as for secondary
antiprotons \citep{Donato:2008yx}.
\\
The production of cosmic antideuterons is based on the fusion process of a \pbar\
and \nbar\ pair -- being the pair produced from CRs collisions or from dark matter 
annihilation in the dark galactic halo. 
One of the simplest but powerful treatment of the fusion of two or
more nucleons is based on the so--called coalescence model which, despite its
simplicity, is able to reproduce remarkably well the {available} data on light nuclei and
antinuclei production in different kinds of collisions (\cite{Duperray:2005si,Donato:1999gy,Donato:2008yx} and refs. therein).

We present in Fig.~\ref{futuro_dbar} a possible experimental 
 scenario.  The secondary \dbar\ flux is plotted  alongside the
primary flux from dark matter particle ($m_\chi=$50 GeV) annihilating in the halo.
The three curves bound the propagation uncertainties which -- as for the 
case of \pbar\ from galactic dark matter, grossly rescaled by a factor $10^{3}$ 
 -- span almost two orders of magnitude 
on the whole energetic range. 
The present BESS upper limit  \citep{2005PhRvL..95h1101F}  is at a
level of 2$\cdot 10^{-4}$ (m$^2$ s sr GeV/n)$^{-1}$. We also plot the
estimated sensitivities of the gaseous antiparticle spectrometer GAPS
on a long duration balloon flight (LDB) and  an ultra--long duration balloon
mission (ULDB) \citep{2010AdSpR..46.1349A} and of AMS--02 for three years of data taking. 
The perspectives to explore a part of  the parameter space
where dark matter annihilations are mostly expected (i.e. the low--energy tail) are  promising. 
\begin{figure}[!h]
\begin{minipage}{0.47\linewidth}
 \centering
\includegraphics[width=8.cm,height=8cm]{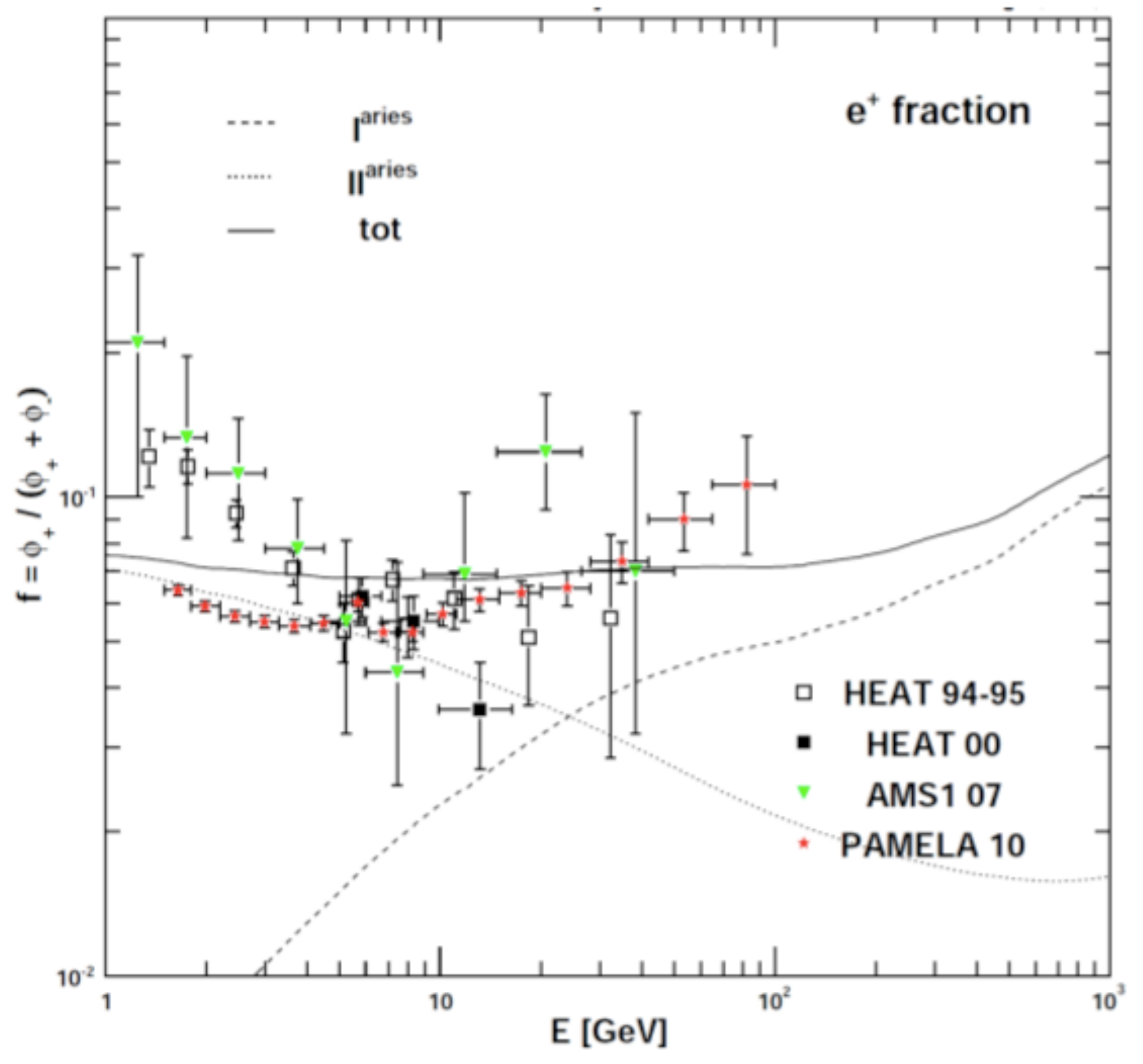}
 \caption{Template calculation for the positron fraction 
 including all primary (disk rete local 
  and smooth distant) and secondary electrons and positrons 
 \citep{2010arXiv1002.1910D}.}  
 \label{fi:eplus} 
 \end{minipage}\hfill
 \begin{minipage}{0.47\linewidth}
 \centering
\includegraphics[width=8.cm,height=8cm]{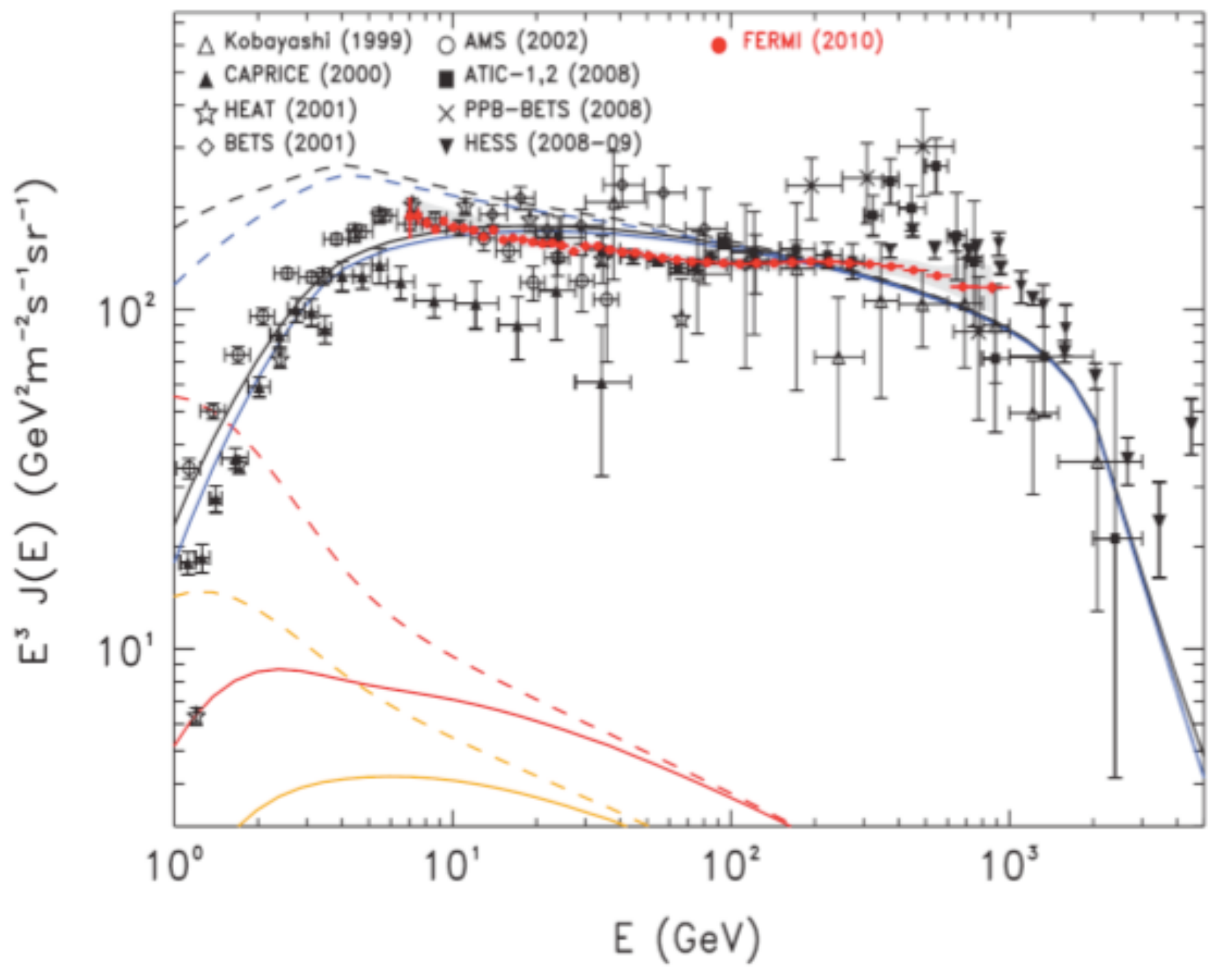}
 \caption{The measured $e^+ + e^-$ spectrum along with 
 a possible estimation of primary and secondary leptonic components. 
 Figure taken from \citep{Ackermann:2010ij}.
 }  
 \label{fi:eplus_tot}
 \end{minipage}
\end{figure}
\vspace{1.2cm}

\centerline{\it Positrons}

Secondary positrons are produced -- like antiprotons and antideuterons --  
by the spallation of the ISM when impinging high--energy particles
\citep{1998ApJ...493..694M,2009A&A...501..821D}.
The main production channel is the collision of protons with hydrogen atoms at rest 
producing
charged pions $\pi^{\pm}$ which decay into muons $\mu^{\pm}$. 
The latter are also unstable
and eventually lead to electrons and positrons. 
Positrons may also be produced through kaons although this channel is rare.
In the case of positrons and electrons, Eq.\ref{eq:CR}
describing the propagation of CRs throughout the diffusive halo is dominated by
space diffusion and energy losses. Above a few GeV, synchrotron radiation in the
galactic magnetic fields as well as inverse Compton scattering on stellar light
and on CMB photons dominate.\\
Data on the absolute \eplus ~flux are less precise than 
on \pbar  ~\citep{2000ApJ...532..653B,1997ApJ...482L.191B,Alcaraz:2000bf,2011PhRvL.106t1101A,2010APh....34....1A}. 
Nevertheless, the data are well described by the contribution from spallation 
reactions within experimental error bars. 
The positron fraction \efrac  ~has been measured by the PAMELA 
satellite experiment \citep{2009Natur.458..607A,2010APh....34....1A} and confirmed 
by the Fermi-LAT Collaboration disk riminating 
the lepton charge using the Earth magnetic field \citep{2011arXiv1109.0521T}. It increases with energies, at variance with 
the predictions from pure secondary production of cosmic positrons. 
A viable explanation of the experimental result resides in the additional 
contribution of astrophysical sources accelerating leptons in their sites
\citep{2009PhRvL.103h1103B,2010arXiv1002.1910D}.
This result is illustrated in Fig. \ref{fi:eplus}, where the \efrac  
~has been calculated adding to the secondary production (mostly relevant for positrons)
the contributions from standard astrophysical sources, such as supernova
remnants and pulsars. As shown in \cite{2010arXiv1002.1910D} (and refs. therein), 
the cosmic fluxes of positrons and leptons are quite sensitive to the presence of 
sources in the near Galaxy (few kpc), whose physics could be explored also in this
peculiar channel. 
The most recent experimental data on $e^+ + e^-$ spectrum are displayed
 in Fig. \ref{fi:eplus_tot} \citep{Ackermann:2010ij}. 
The $e^+ + e^-$ spectrum has been computed with a GALPROP
  model (shown by solid black line) with  breaks in the acceleration 
power spectrum and a strong cutoff  above 2~TeV. 
  Blue lines show $e^-$ spectrum only.
  The dashed/solid lines show the before modulation/modulated spectra.
  Secondary $e^+$ (red lines) and $e^-$ (orange lines) are also shown.  
Secondary electrons and positrons from CR proton and helium interactions with 
interstellar gas make a significant contribution to the total
lepton flux, especially at low energies. The total leptonic 
flux is characterized by peculiar spectral features which could be 
explained by local strophysical or exotic sources in a few kpc region 
around the Solar System (see \cite{Ackermann:2010ij,2010arXiv1002.1910D} and 
refs. therein for details). 

\section{From 100 TeV/n to 100 PeV/n}

\label{sec:indir}

\subsection{Extensive air Showers}
\label{secEAS}
\begin{figure}[!]
\begin{minipage}{0.47\linewidth}
 \centering
 \includegraphics[width=10.0cm,height=8.0cm]{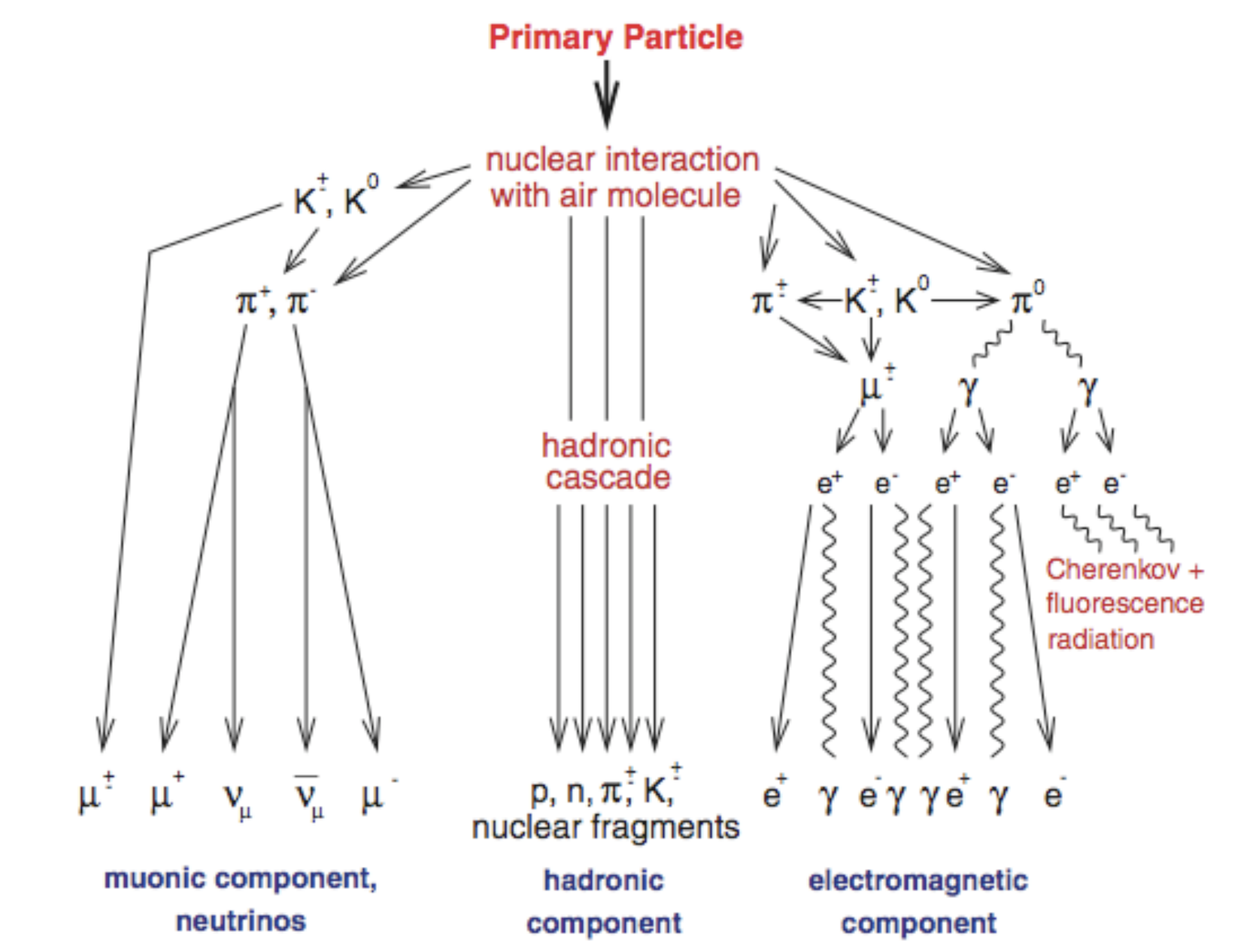} 
 \end{minipage}\hfill
 \begin{minipage}{0.47\linewidth}
 \centering
 \includegraphics[width=6.cm,height=4.cm]{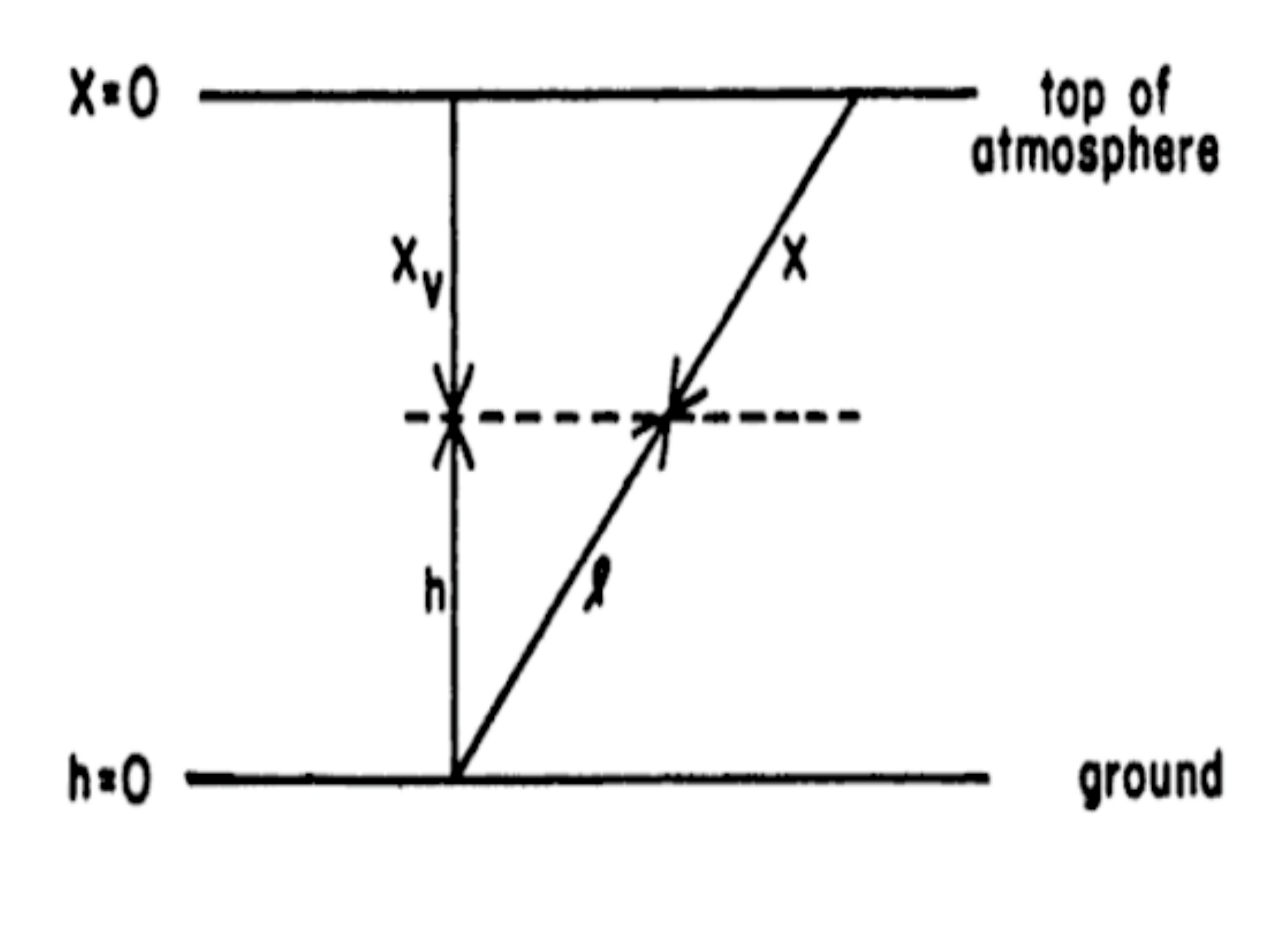}  
 \end{minipage}
\caption{\em {Left: Sketch of an Extensive air Shower \citep{hau03}. Right: the variables used to describe the atmospheric depths \citep{gai90}.}} 
 \label{skEAS}
\end{figure}

Primary CRs above $\simeq 10^{14}$ eV are characterized, as shown in Fig.\ref{fi:fig1}, by a low flux, and their energy, mass and arrival directions can be studied only indirectly by exploiting the particle cascades that they produce in the atmosphere. 
The measured observables are the longitudinal and lateral distributions of the charged components or the Cherenkov and the fluorescence light produced during the propagation of the extensive air shower (EAS) in the Earth atmosphere down to the experimental level. 
Since all observables are interrelated and depend in different ways on both energy and mass of the primaries, multiparametric measurements are to be preferred: modern experimental setups in fact include detectors of many shower components.\\
When a primary CR nucleus interacts with an air nucleus in the upper atmosphere,  a leading nucleon emerges, while a fraction  (the so-called inelasticity {\it k}) of its initial energy goes into production of secondaries\footnote{the definition of "secondaries" applies here to CRs produced in the Earth atmosphere, not to be confused with the secondary particles originating from the spallation of primary CRs in the ISM.}, mainly $\pi$ mesons; due to charge independence, the energy is equally shared among $\pi^{+},\pi^{-}$ and $\pi^{0}$. \\
The electromagnetic component (electrons\footnote{From now on, "electrons" stay for both $e^{-}$ and $e^{+}$} and photons) originates from the fast decay of neutral pions into photons, which initiates  a rapid multiplication of particles  in the shower, mainly through  two production processes: bremmstrahlung by electrons and pair production of electrons by photons. The multiplication continues  until the rate of energy loss by bremmstrahlung equals that of ionization, at a critical energy which in air is $E_{c} \simeq 86$ MeV. 
The hadronic back-bone of the shower continuously feeds the electromagnetic part; the charged pions can either interact or decay. 
The nucleon interaction length in air (with $\langle A \rangle \simeq 14.5$) is $\simeq 80 \ g \ cm^{-2}$.
The transverse momentum of nucleons and pions and the multiple scattering of the shower particles, particularly of the electrons, are responsible for the lateral spread of the particles in the shower.
Finally, charged pions decay into muons (and neutrinos). Since muons lose energy mainly through ionization and excitation,  they are not attenuated very much and  give rise to the most penetrating component of the shower\footnote{This is why, despite being electromagnetic particles, muons are traditionally not included in the electromagnetic shower component, but in the separated muonic one.}.
A sketch of the different components of an Extensive air Shower is shown in Fig.\ref{skEAS} (left).\\
Transport and cascade equations describe the development and propagation of the EAS in the atmosphere. Pathlengths are generally measured in units of $g \ cm^{-2}$ to remove the effect of the density of the medium:  the vertical atmospheric depth is 
\begin{equation}
X_{V}(h) = \int_{h}^{\infty} \rho(h') dh' \ \ {\rm g \ cm^{-2}}
\end{equation}
where $\rho(h)$ is the density of the atmosphere at altitude h. The atmospheric depth measured downward from the top along the direction of the incident  particle is the {\it slant depth} $X$. They are sketched in Fig.\ref{skEAS} (right).
\vspace{1.4cm}

\centerline{\it The electromagnetic component}
The  number of charged particles (the shower {\it size} $N_e$) at each atmospheric depth X can be derived from the solutions of the cascade equations \citep{rg41}.  Most recently, the shower profiles are characterized in data analysis using \citep{gh77}
\begin{equation}
N_{e}(X) = N_{e}^{max} \left (\frac{X-X_{1}}{X_{max}-X_{1}} \right)^{(X_{max}-X_{1})/\lambda} exp\left (\frac{X_{max}-X}{\lambda} \right)
\label{eq:size}
\end{equation}
where $X$ and $X_{1}$ are the depths of observation and  of first interaction and $\lambda$ is the attenuation length.
The depth of maximum development of the shower, $X_{max}$ is obtained by determining the number of lengths needed for each particle to reach the critical energy $E_{c}$:  $X_{max} \simeq X_{R} \left[ ln(\frac{E_{0}}{E_{c}})+\alpha \right]$, 
where $E_{0}$ is the primary energy, the radiation length  $X_{R}$  depends on the composition of the medium ($X_{R} \simeq 37$ g cm$^{-2}$  in air) and $\alpha$ depends on the primary (electron or photon). The rate of variation of $X_{max}$ per decade of primary energy, $\left( dX_{max}/dlog_{10}E \right)$,  is  called the {\it elongation rate}.\\
The density of the $N_{ch}$ charged particles as a function of the distance from the shower core (the intersection of the shower axis with the ground) is described by the Nishimura-Kamata-Greisen function \citep{kg58,gra60}
\begin{equation}
\rho_{ch}(r) = \frac{N_{ch}}{2 \pi r_{M}^{2}} \ C(s)  \left( \frac{r}{r_{M}} \right)^{s-2}  \left( 1+\frac{r}{r_{M}} \right)^{s-4.5}
\label{eq:NKG}
\end{equation}
and depends on the multiple Coulomb scattering of electrons.  Here $s$ is the shower {\it age}, which describes the shape of the distribution ($s=1$ at shower maximum); $C(s)=\Gamma(4.5-s)/\Gamma(s)\Gamma(4.5-2s)$, with $\Gamma$= gamma function.\\
The root mean square scattering (the width of the approximate Gaussian projected angle distribution) undergone by an electron of initial momentum p (MeV/c) as it traverses a thickness $x$ of material is $\theta=(21 MeV/p \beta) \sqrt{x/X_{0}} $; at the critical energy, the scattering angle for an electron crossing one radiation length $X_{0}$ is $\theta \simeq 14^{0}$.  
One Moliere unit, $r_{M}=21 MeV/E_{c}$ is the lateral distance by which an electron with $E=E_{c}$ is scattered as it traverses  1 $X_{0}$ 
($r_{M} \simeq 0.25 X_{0}$ in air). It is the distance encompassing $90 \%$ of the shower energy.\\
At the highest energies, the electromagnetic shower development undergoes modifications due to two competing effects: (a) above $10^{18}$ eV, the Landau-Pomeranchuk-Migdal effect \citep{sta82}, for which particle production is suppressed in certain kinematic regions, leading to an increase of the shower to shower fluctuations (due to the stochastic development of the cascades)
and to deeper maximum; (b) above $10^{19.5}$ eV, the interaction of$\gamma$ rays  with the geomagnetic field of the Earth \citep{erb66}: in this case, magnetic bremmstrahlung causes the shower to behave as a superposition of hundreds of lower energy showers, and  the shower to shower fluctuations are significantly reduced.\\
\vspace{0.4cm}

\centerline{\it The hadronic and muonic components}
In a hadronic shower, the first interaction happens at $X_{0}=\lambda_{I}ln2$, where $\lambda_{I}$ is the interaction length of strongly interacting particles, and about 1/3 of the produced charged particles are $\pi^{0}$. The attenuation length for the hadronic component is larger than $\lambda_{I}$, due to secondary particle production, and is on average $\simeq 120 \ g \ cm^{-2}$.\\
The depth of maximum development of the shower after $X_{0}$ is thus the same as for an electromagnetic shower of energy $E_{0}/3N_{ch}$, so that \citep{mat05}
\begin{equation}
X^{p}_{max} = X_{0}+X_{R} \ ln[E_{0}/3N_{ch}E^{\pi}_{c}]=X^{EM}_{max}+ X_{0}-X_{R} \ ln(3N_{ch})
\label{eq:xp}
\end{equation}
As a consequence, hadronic showers cannot have higher elongation rate than electromagnetic ones. The reduction is due to the increase of both multiplicity of charged particles and cross section for the hadronic showers. For protons, the elongation rate is $\simeq 58$ g cm$^{-2}$ per decade of energy, as estimated from calculations that model the shower development using the best estimates of the relevant features of the hadronic interactions. 
Muons are produced in each of the n generations of a hadronic shower when any of the $N_{ch}$ charged particles has energy equal than some decay energy $E^{\pi}_{c}$. The total number of muons is
\begin{equation}
N_{\mu} \propto (N_{ch})^{n}  = \left( \frac{E_{0}}{E^{\pi}_{c}}\right)^{\beta}
\label{eq:eqnmu}
\end{equation}
where $\beta \simeq 0.85 \div 0.92$ depending on the hadronic interaction model used. The critical pion energy $E^{\pi}_{c} \simeq 20$ GeV in a shower generated by a 1 PeV proton.\\
Due to energy conservation, $E_{0}=E_{EM}+E_{had}$, where $E_{had}=N_{\mu}E^{\pi}_{c}$. The fraction of primary energy going into the electromagnetic component is $E_{EM}/E_{0} = 1- \left( \frac{E_{0}}{E^{\pi}_{c}}\right)^{\beta-1}$, which can be approximated to a power law $E_{EM}/E_{0} \simeq a \ \frac{E_{0}}{E^{\pi}_{c}}^{b}$. \\
Since the electromagnetic size can be expressed as $N_{EM} \propto E_{0}/E^{\pi}_{c}$, by series expansion around $\frac{E_{0}}{E^{\pi}_{c}} \simeq 10^{6}$ (if e.g. $E_{0}=1$ PeV)  
\begin{equation}
N_{EM} \propto E_{0}^{b}   \ \ \ \  {\rm with} \ \ \  b=1+\frac{1-\beta}{10^{6(1-\beta)}-1} \simeq 1.02
\label{eq:eqne}
\end{equation}

\begin{figure}[!]
\begin{minipage}{0.47\linewidth}
 \centering
 \includegraphics[width=8.5cm,height=8.5cm]{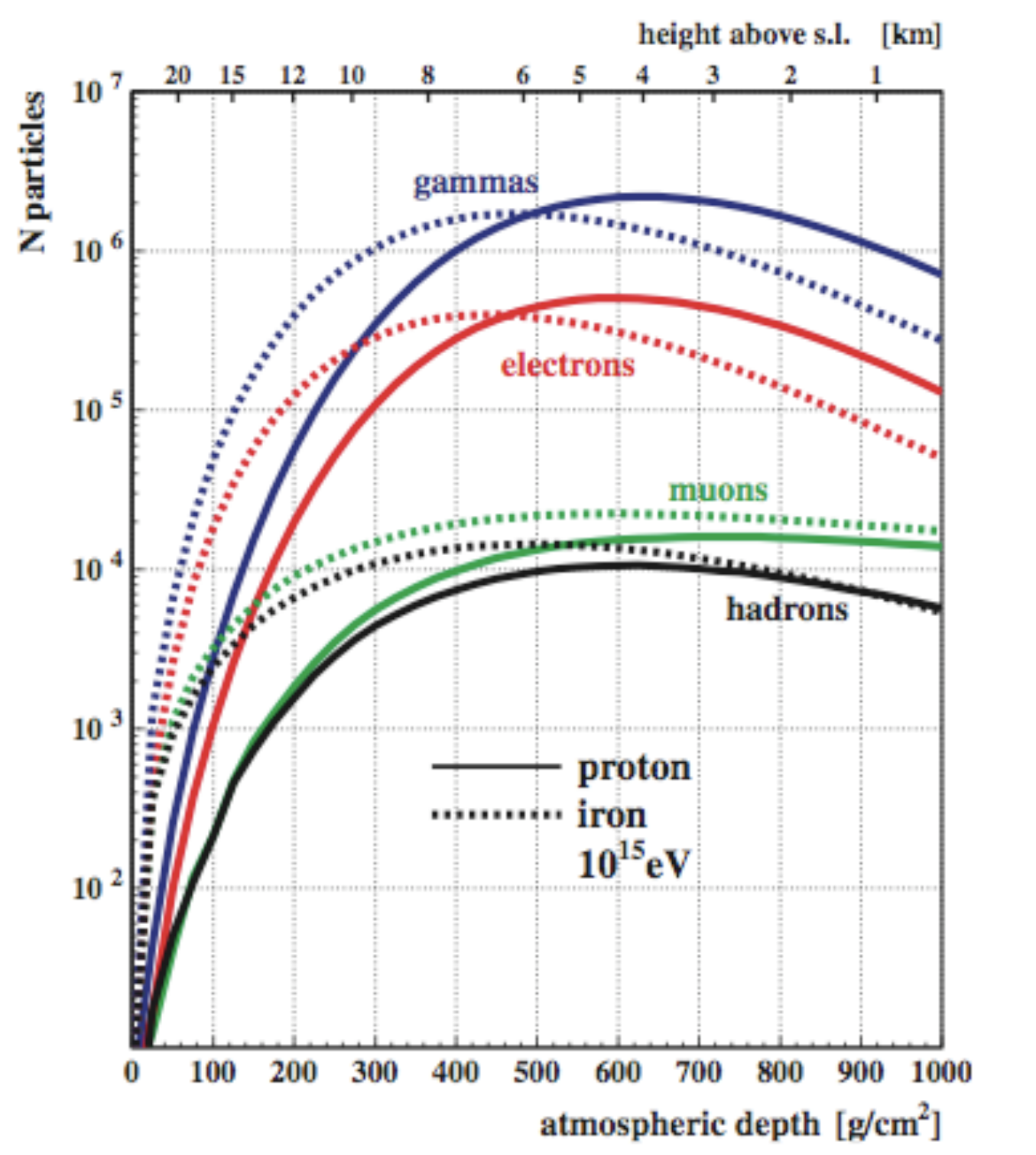}
 \end{minipage}\hfill
 \begin{minipage}{0.47\linewidth}
 \centering
 \includegraphics[width=8.5cm,height=8.5cm]{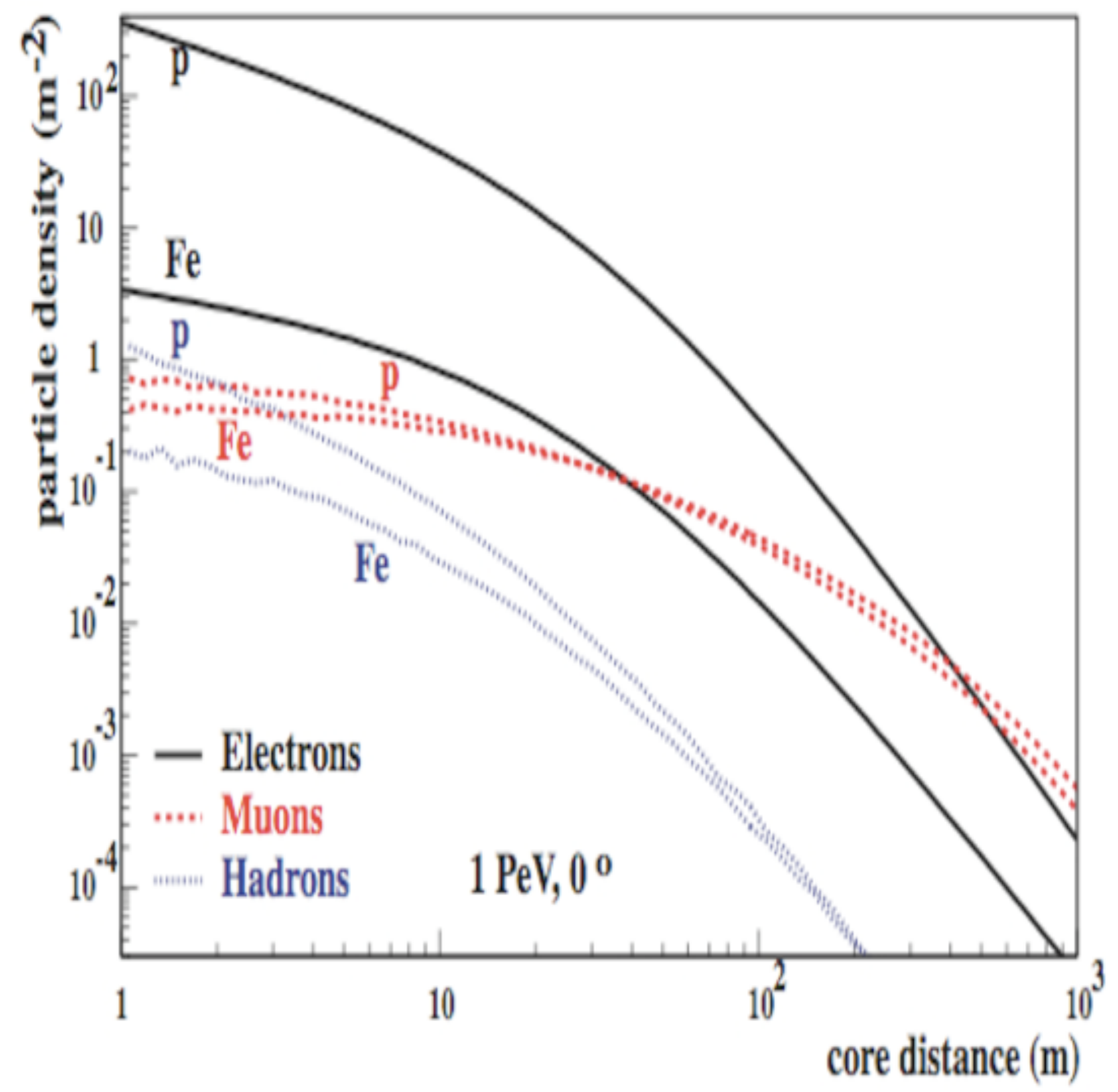}
 \end{minipage}
\caption{\em {Left: longitudinal EAS development of the average total intensities (sizes); right:  lateral distributions of the particle densities at sea level (from CORSIKA simulations \citep{hec98}).}}  
 \label{longlat}
\end{figure}

The {\it superposition model} states that a primary nucleus with mass $A$ and energy $E_{0}$ acts as $A$ independent nucleons of energy $E_{0}/A$; according to it, for the superposition of $A$ nucleon showers Eq.\ref{eq:xp} gives
\begin{equation}
X_{max}^{A} = X^{p}_{max}-X_{R} \ ln~A
\label{eq:eqx}
\end{equation}
\begin{equation}
N_{\mu}^{A} \simeq A \left( \frac{E_{0}/A}{E^{\pi}_{c}} \right)^{\beta} =  A^{1-\beta} N_{\mu}
\label{eq:eqm}
\end{equation}
It follows from (\ref{eq:eqx}) and (\ref{eq:eqm}) that the depth of shower maximum and the number of muons depend on the mass of the primary particle: showers originated by a proton develop lower in atmosphere (higher $X_{max}$), while the higher the primary mass (at a given energy) the more muons are expected. Being the superposition of A nucleon subshowers, heavy nuclei showers will also have smaller shower to shower fluctuations as compared to protons. Gamma ray showers fluctuate much less and are muon-poor, due to the small cross sections for meson production and muon pair creation. However, the superposition assumption is a simplification of the correct treatment of nucleus-nucleus interactions, which does not take into account the fact that in most collisions the number of interacting nucleons is not equal to that of the projectile.
As shown in \citep{iva10}, it could result in an overestimation of the primary energy when evaluated using surface arrays particle density measurements.\\
The muon lateral distribution was parametrized first by \citep{gra60}. Simplified parameterizations, tuned for each particular experiment, can be found in the literature. For example \citep{khris77}:
\begin{equation}
\rho_{\mu}(r) \propto r^{-\alpha} exp  \left( \frac{-r}{r_{M}} \right)
\label{eq:Gre}
\end{equation}
As an example, the longitudinal and lateral distributions are shown in Fig.\ref{longlat} for $10^{15}$ eV proton and iron primaries and different components.\\
\vspace{0.4cm}

\centerline{\it Cherenkov light}
A charged particle of the air shower produces Cherenkov light if its velocity is such that $v/c > n$, where $n(z)$ is the index of refraction as a function of height $z$. 
Most of the Cherenkov light is emitted in air by electrons, for which the threshold energy is about 21 MeV at sea level (to be compared to about 500 MeV for muons), in a cone of emission which, well above threshold, is $\theta_{C} (z)=cos^{-1}\left(1/\beta n(z) \right)$, a decreasing function of the atmospheric height. 
The fraction of light $F_{C}$ reaching the ground if emitted at an atmospheric depth $x$ with zenith angle $\theta$ is \citep{hill82}
\begin{equation}
F_{C} = exp [ -(1020-x)sec \theta / \Lambda ]
\end{equation}
where $\Lambda$ is the absorption length in $g \ cm^{-2}$. This light is in fact attenuated (in UV/blue) by scattering on molecules (Rayleigh scattering) and aerosol particles (Mie scattering), by scattering on water vapor in clouds or by the absorption by ozone molecules.  \\
The  light emitted by a typical$\gamma$ ray shower at 10 km results at ground in a ring focussed at $\simeq 120$ m from the core; the number of emitted photons is about 0.1/cm at sea level. \\
The lateral density distribution of the Cherenkov light at ground is mainly determined by the Cherenkov angle and the Coulomb scattering and can be parametrized as \citep{fow01}
\begin{equation}
  C(r) =
   \cases{
        C_{crit} \ e^{ \ s(r_{crit}-r)}                 & if 30 m $< r \leq r_{crit}$,\cr
        C_{crit} \ (r/r_{crit})^{-\beta}  & if $r_{crit} < r \leq$ 350 m.\cr
     }
     \label{eq:cer}
\end{equation}
Here, $C_{crit}$ is the Cherenkov light  at $r_{crit}$ m from the core; the distance $r_{crit}$ marks a sharp change of slope in the light distribution and is $\simeq 120$ m at sea level. $s$ and $\beta$ are the inner and outer slopes of the distribution. \\
A different contribution comes from the emission of direct Cherenkov light by primary particles with velocities above threshold before their first interaction in atmosphere. Being created higher in atmosphere, it generates on ground a light cone  which is narrower (within $\simeq 100$ m) compared to that produced by the secondaries of the EAS. Since the intensity of this light is almost constant above a given energy, while the EAS one increases almost linearly with energy,  the direct light is outshined by the EAS one above few hundreds TeV.  

\vspace{0.4cm}

\centerline{\it Fluorescence light}
Charged secondary particles of the air showers, mostly electrons and positrons, excite nitrogen molecules in the atmosphere: the de-excitation results in the isotropic emission of a fluorescence spectrum in the near UV region (300-400 nm).  
The fluorescence efficiency of photons, that is the ratio of the energy emitted in fluorescence by the excited gas to the energy deposited by the charged particles, is very low, of the order of $5 \ 10^{-5}$. This is the reason why only at the highest energies, above $\simeq 10^{18}$ eV, the enormous number of particles allows this light to be detected.
Quenching effects cause part of this energy to go to other molecules through collisions; since a shower typically crosses many km of altitude and since the collision rate depends on the average separation distance and velocity of the molecules,  the fluorescence emission will depend on  the gas pressure and temperature.\\
The number of emitted fluorescence photons can be written as
\begin{equation}
\frac{d^{2}N_{\gamma}}{dX d \lambda} = Y(\lambda,P,T,e_{v}) \cdot \frac{dE_{tot}^{dep}}{dX}
\end{equation}
where $dE_{tot}^{dep}/dX$ is the energy deposited in the atmospheric depth $dX$. The fluorescence yield $Y(\lambda,P,T,e)$  depends on the wavelength $\lambda$, the air pressure and temperature $P$ and $T$ and the vapor pressure $e_v$. 
Its value is $(5.05 \pm 0.71)$ photons/MeV of energy deposited, in air at 293 $^{\circ}$K and 1013 hPa in the 337 nm band according to \citep{nag04}.
Exploiting the fact that the fluorescence light is mainly induced by MeV electrons, the emission mechanism and the absolute fluorescence yield have been studied in various laboratory experiments. The most recent results are reviewed by \citep{arq08}. \\

\vspace{0.4cm}
\centerline{\it Radio emission}
Air showers generate coherent radio emission, first observed by \citep{jel65}, due to the cascade electrons emitting synchrotron radiation in the Earth magnetic field. 
In fact these electrons, with $\langle E_{e} \rangle \simeq 30$ MeV, are spread in a thin shower front of $\leq 2$ m, smaller than 1 wavelength for emitted radio pulses of 100 MHz.  Up to this frequency, coherent emission can be expected. The electric field strength is proportional to the primary energy of the CR particle initiating the shower.\\
Radio signals can arise also from coherent radio Cherenkov emission (the Askaryan effect): cascade $e^+$ can annihilate with the $e^-$ of the medium (air), thus generating a $\simeq 20 \%$ electron excess in the air shower, which behaves as a relativistic charge, emitting Cherenkov radiation \citep{ask62}. The emission in this case is coherent up to GHz in sufficiently dense materials like ice, and can be exploited to reach the huge detection acceptance needed to study shower neutrinos.

\subsection{Experimental methods for indirect measurements}
\label{meth}
\begin{figure}[h]
\begin{minipage}{0.47\linewidth}
 \centering
 \includegraphics[width=8.5cm,height=7.0cm]{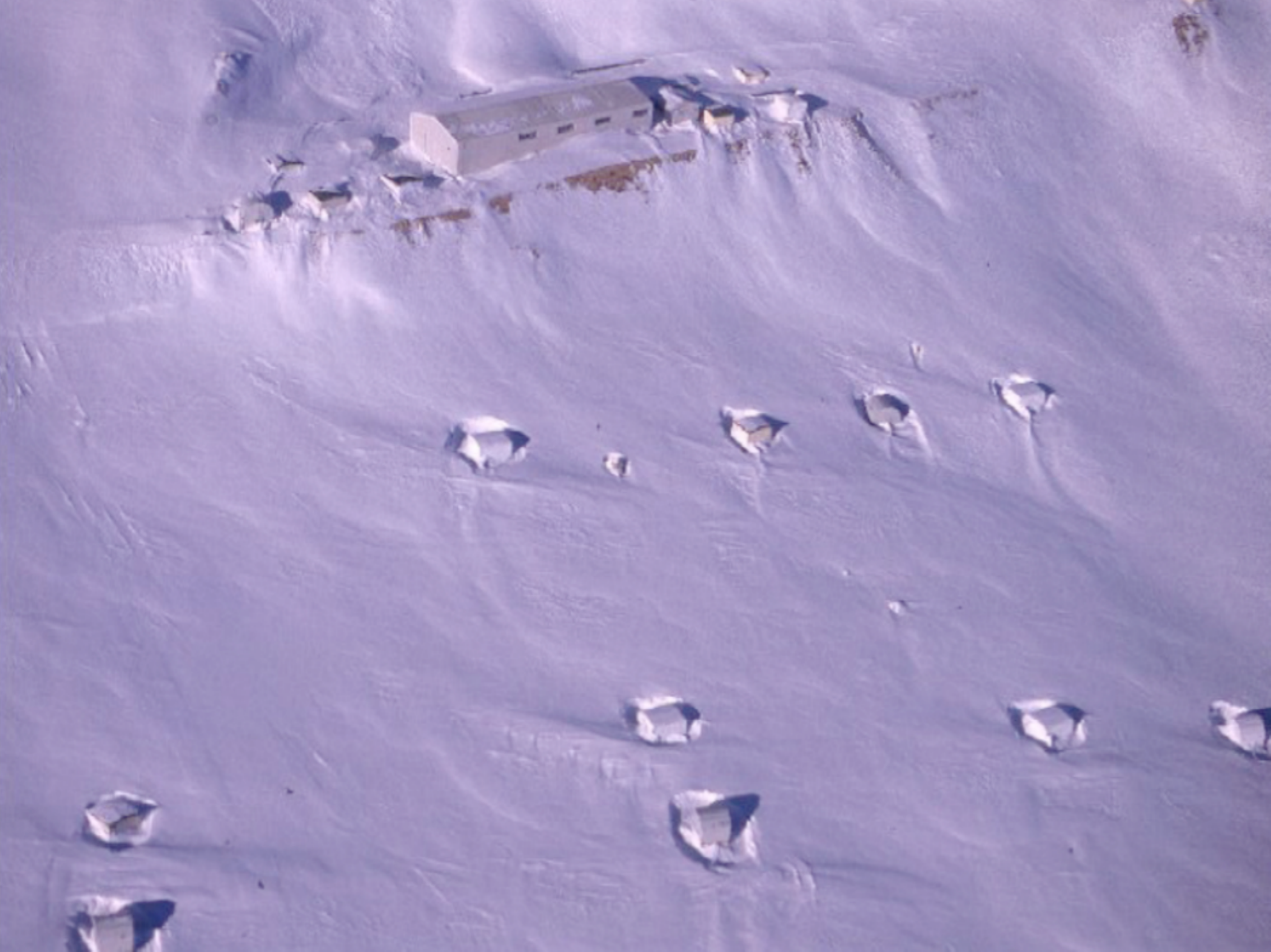}  
 \end{minipage}\hfill
 \begin{minipage}{0.47\linewidth}
 \centering
 \includegraphics[width=8.5cm,height=7.0cm]{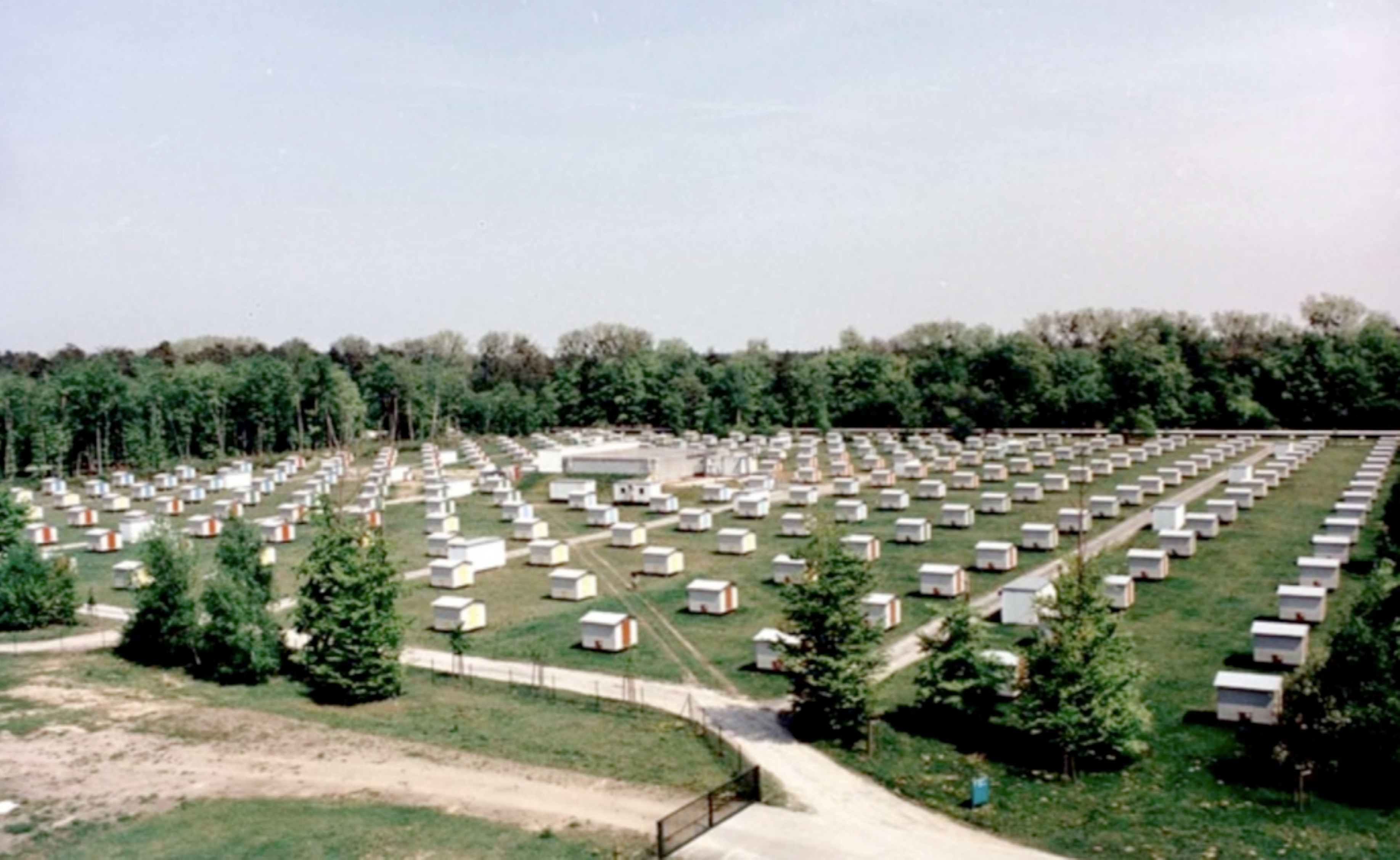}   
 \end{minipage}
\caption{\em {The EAS-TOP, left  (2000 m a.s.l.) and  KASCADE, right (sea level)  air shower arrays  \citep{et89,ka97}.}}
\label{fi:etkfig1}
\end{figure}

\vspace{0.5cm}
\centerline{\it The measure of the charged component}
A classical air shower experiment (two examples are shown in Fig.\ref{fi:etkfig1}) consists of an array of detectors, either scintillator counters or water Cherenkov tanks, distributed over a wide area, which surface is chosen depending on the rate of events to be studied. The separation between the detectors is tuned to match the scale of the shower footprint at the observation level (tens of meters in the PeV region, hundreds of m to km for the arrays studying the extreme energy region).  
At each location, the particle density of one or more charged components is measured with detectors of size suitable for the component under study (few $m^2$ for the electromagnetic one, much larger for muons and hadrons), together with the arrival times of the particles and their time spread. Due to the large number of secondary particles,  the active area to be covered can be much smaller than the total one: sensitive/enclosed area ratio can go from $\simeq 3 \ 10^{-3}$ for the EAS-TOP array (37$\times$10 m$^{2}$ detectors over 10$^5$ m$^2$)  to $\simeq 5 \ 10^{-6}$ for the Pierre Auger Observatory (1600$\times$10 m$^{2}$ detectors over 3000 km$^2$).
This ratio, together with the altitude of the detector (that is thickness of the atmosphere above it) and the ability to detect different components of an EAS, determines the energy thresholds of different detectors. 
To lower the energy threshold, completely different apparatuses have also been designed , where the active area reaches at least $50 \%$ of the total one, thus allowing for the so-called "full coverage" \citep{atk04,aie09}. \\
Besides the aim of lowering the energy threshold, high altitude experiments can be foreseen to study the air showers of higher energies  at an early stage of development, to measure the direct primary spectrum of protons (before their interaction in the atmosphere), to study hadronic production after few interaction lengths, that is in the very forward region.

The more penetrating muon component is generally measured  by shielded detectors, like scintillator slabs: a shielding of thickness some radiation lengths (e.g. 20 $X_{0}$ for KASCADE) can absorb the electromagnetic component without significantly affecting the muon one.     Alternatively, one can use tracking devices (limited streamer or proportional tubes) or measure the muons, together with the electromagnetic component, in water Cherenkov tanks. In this last case, the electromagnetic particles are completely absorbed in water, while the muon signal is proportional to the track length in the detector. Since low energy muons (below 1 GeV) mainly decay before reaching the ground, this component basically consists of  muons with energies of few GeV. 
High energy muons, with energies above few TeV, on the other hand, give information on the first interactions of the primary particle. They can be detected in underground laboratories, shielded by rock, water or ice \citep{macro93,ice06}, either as single muons or bundles.\\
The  core location and the total number of charged particles are obtained by means of a fit to a function describing their measured lateral distribution (e.g. Eq.\ref{eq:NKG} and Eq.\ref{eq:Gre}).  The shower size  (recall Eq.\ref{eq:size}) is typically evaluated above $N_{e}>10^{5}$ with an accuracy  $\sigma_{N_{e}}/N_{e} = 10 \%$;
the core position is determined within few meters \citep{et93}.
The arrival direction of the primary particle is derived from the measure of the arrival  time of particles on the stations  (the shower front). Most detectors have time resolution from 0.5 to few ns, 
with angular resolution typically below $0.5^{\circ}$, which can be evaluated from internal consistency of data. 
An absolute measurement  of the angular resolution for an EAS array is possible by detecting  the reduction in CR intensity due to the "shadow" cast by the Moon and the Sun on the high  energy primary CR flux.  
Being the Sun larger and much further away with respect to the Moon, they have basically the same angular diameter of $\simeq 0.52^{\circ}$: the measure is in principle possible for arrays with angular resolution $\leq 1^{\circ}$ (\citep{et91} and refs. therein).  A very large sample of events is however needed to get a statistically significant result, due to the smallness of the effect.\\
Hadrons can be detected by means of calorimeters, measuring the energy dissipated  by the incoming particles: two big devices have been used e.g. in the EAS-TOP \citep{et99} and KASCADE \citep{ka99} experiments. The observables are in this case the hadron number and their energy sum.\\
\vspace{0.8cm}

\centerline{\it The measure of the Cherenkov light}
To observe the Cherenkov  light emitted by the shower particles in the atmosphere, two techniques can be used:

$\bullet$ {\it Light integrating detectors}  combine a large angular acceptance with the advantages of Cherenkov light detection. They are used to measure the lateral distribution of the Cherenkov light with a grid of photomultipliers distributed over a large area on the ground, each enclosed in a Winston cone\footnote{Non-imaging light-collection devices with a parabolic shape and a reflective inner surface. Winston cones are often used to concentrate light from a large area onto a smaller photodetector or photomultiplier.} to help the light collection. 
This distribution is strongly related to the shower energy 
($C_{120} \propto E^{1.07}$, see also Eq.\ref{eq:cer}). The critical radius (here 120 m)  is far enough from the core, at these energies, to minimize the fluctuations, and close enough to ensure a measurable light density. The dependence on the primary mass is fully included in the inner exponential  slope $s$ of the distribution, which is in fact a function of the depth of the shower maximum. 
After many early attempts in the years 1972-1989, this technique was widely used in the dedicated HEGRA/AIROBICC experiment \citep{lor96,airob00} and subsequently in BLANCA \citep{bla01}, both detectors operating in coincidence with a classical shower  array. Another wide angle Cherenkov array, TUNKA,  is installed near lake Baikal, in Siberia \citep{tunka05}.

$\bullet$ {\it Imaging Detectors} are used to reconstruct air showers generated by$\gamma$ ray primaries in TeV $\gamma$-astronomy, but the technique can also be used to study hadronic showers \citep{dice01}. Basically, an image corresponding to the intensity pattern and direction of Cherenkov light is produced in the focal plane. When the direction of the shower and the distance of its core from the telescope are known, a simple geometrical procedure allows to measure the light emitted at each atmospheric depth. Its integral over the crossed atmospheric depth is used to determine the shower size, while a fit to the shape of the shower in the telescopes allows to derive the depth of maximum development $X_{max}$ in a way which is almost independent on simulations. The interpretation of the results in term of composition is however possible only with the help of Monte Carlo simulations. \\
Imaging detectors can also be used to detect the direct Cherenkov light emitted by the primary particle before interacting. Since this light is emitted in a cone with emission angle between $0.15^\circ$ and $0.3^\circ$, telescopes equipped with cameras using $\leq 0.1^\circ$ pixel size are needed. The light can be seen \citep{aha07} as a single high intensity pixel between the reconstructed shower direction and the center of gravity of the EAS image.

\begin{figure}[h]
\begin{minipage}{0.47\linewidth} 
 \centering
 \includegraphics[width=8.5cm,height=7.0cm]{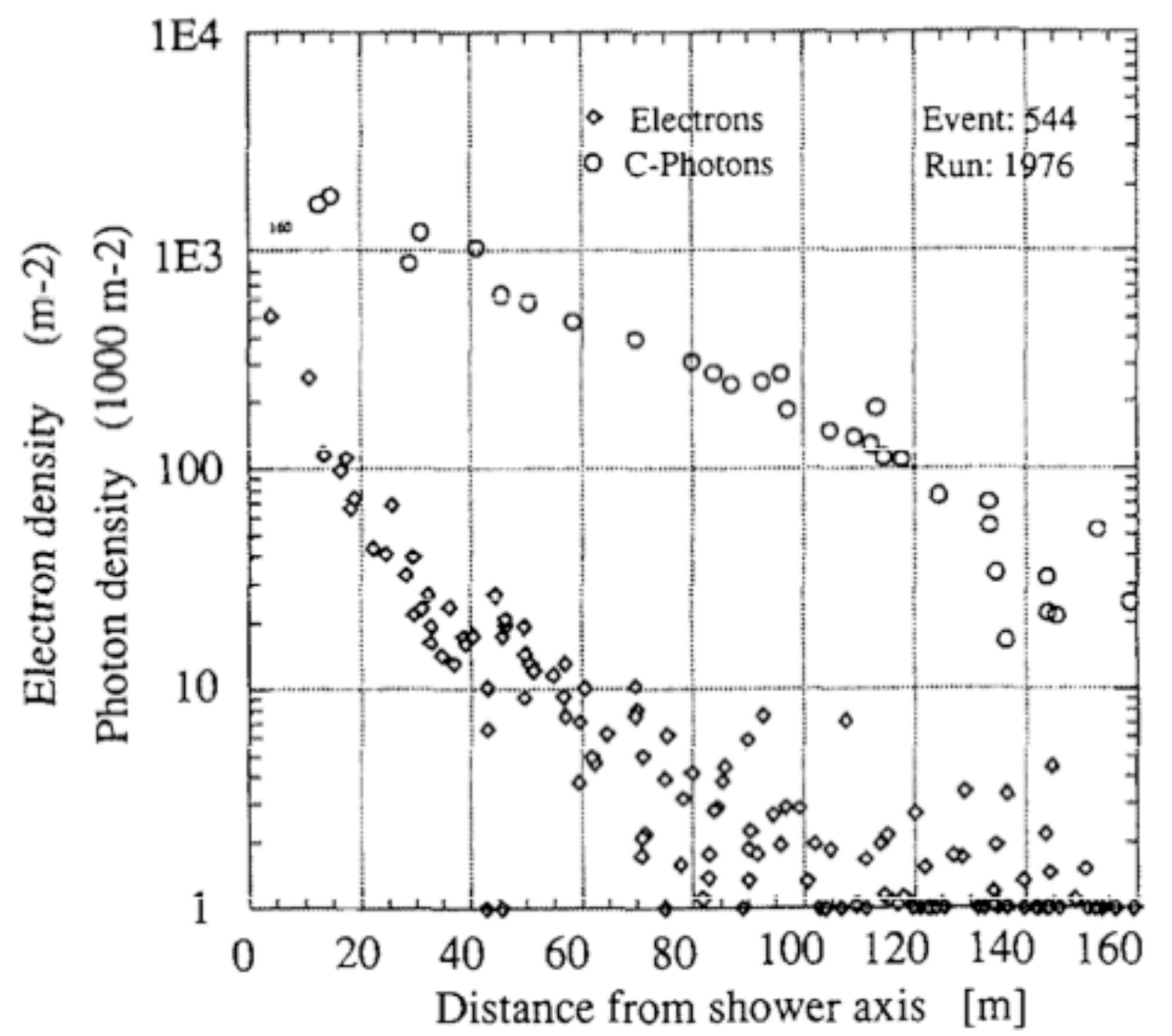}   
\caption{\em {Cherenkov photon density as measured by HEGRA-AIROBICC \citep{kar95}, compared to the electron density at ground level by the scintillator 
array.}}
\label{fi:cer}
 \end{minipage}\hfill
 \begin{minipage}{0.47\linewidth}
 \centering
 \includegraphics[width=8.5cm,height=7.0cm]{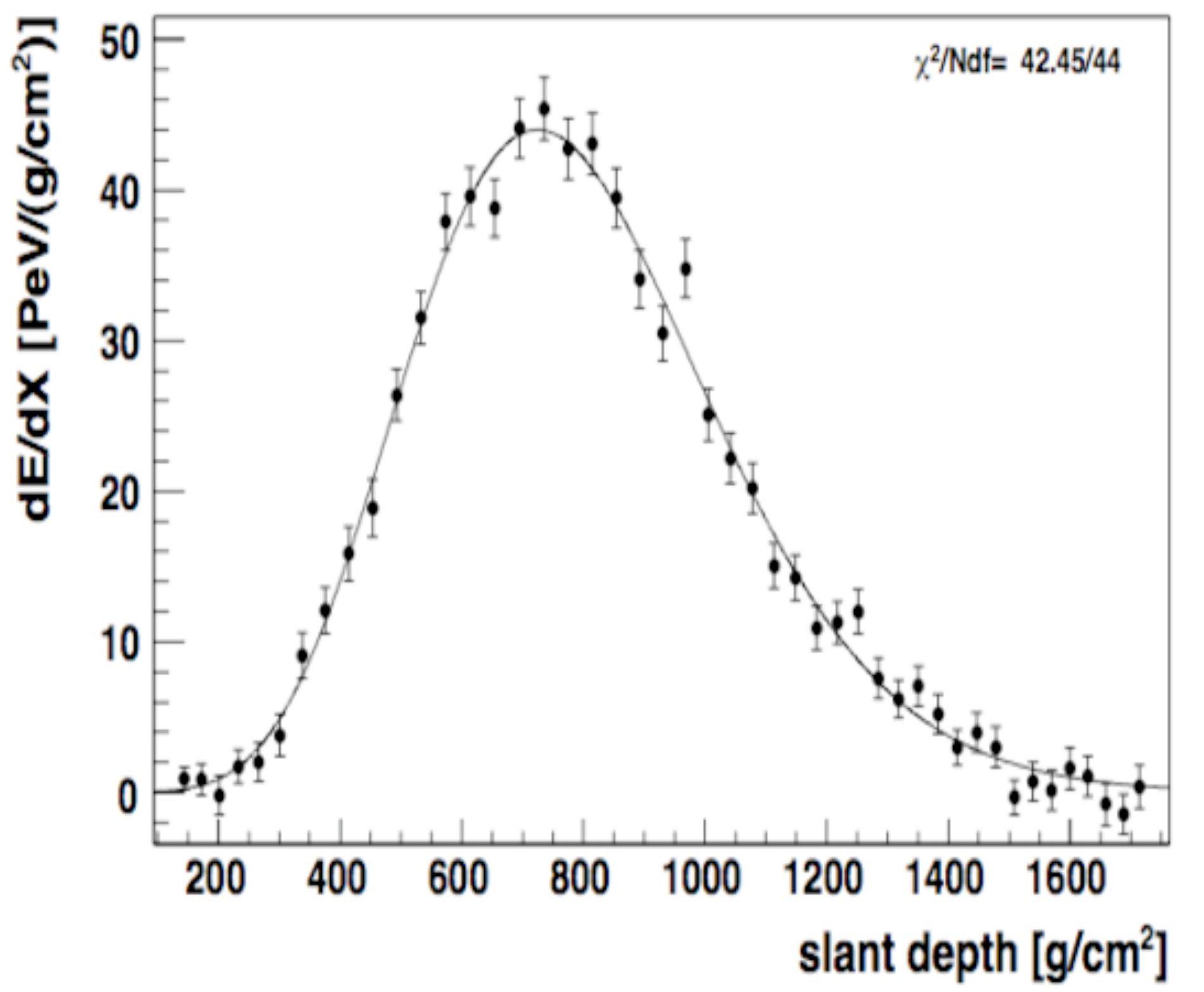}   
\caption{\em {Reconstructed energy deposit profile for an EAS in the Pierre Auger fluorescence detector. The reconstructed energy is $(3.0 \pm 0.2) \ 10^{19}$ eV \citep{augFD}.}}
\label{fi:fluo}
 \end{minipage}
\end{figure}

The broader lateral distribution (see Fig.\ref{fi:cer}), due to the smaller absorption of photons in atmosphere, and the high photon density, that means a better signal-to-noise ratio even for smaller arrays, are the main advantages in using these detectors as compared to classical EAS arrays for charged particle. 
On the other hand, the duty cycle for  Cherenkov observations does not go above 10$\%$, since they  are possible only in clear moonless nights.\\
\vspace{0.4cm}

\centerline{\it The measure of the fluorescence}
The fluorescence light is collected with mirrors and projected onto a camera generally made of a large number of photomultiplier tubes (pixels), which record a time sequence of light. The EAS appears as a trace of illuminated pixels. As simulations show, most of the total energy of the shower is detectable as ionization energy, but a correction must be introduced to take into account the fraction of energy not contributing to the total signal (carried below observation level in the ground, by muons and neutrinos).  This fraction decreases with increasing energy, because at higher energy pions mostly interact, producing $\pi^0$'s of lower energy which in turn go to feed the electromagnetic component. At energies around 1 EeV, $\simeq 15 \%$ and $\simeq 10 \%$ of the energy is missing for iron and proton primaries respectively, but the dependence on the primary mass becomes very small as energy increases. This quantity is also mildly  dependent on the interaction model employed in the Monte Carlo, but being the energy mostly released in the electromagnetic component, this reliance remains at the level of few $\%$.\\
The geometry of the shower, together with the evaluation of the Cherenkov light background and of the atmospheric absorption gives a complete reconstruction of the event. The fluorescence telescopes can be used in monocular (one telescope) or in stereo (two or more telescopes) mode. In the latter case, the angular resolution can reach 0.6$^0$.\\
This technique was firstly used by the Fly's Eye experiment \citep{fly99}; 8 years of data came from its successor HiRes \citep{boy02}.\\
In the Pierre Auger Observatory, 24 fluorescence telescopes are employed \citep{augFD}, located in 4 stations overlooking the surface array. 
The hybrid technique, combining data recorded by both the surface and the fluorescence detectors, has great advantages:  (a) the energy scale is set with the fluorescence telescopes and is thus nearly independent of shower simulations and hadronic interaction models; (b) the shower arrival directions are determined  with very high precision, cross-checking the angular resolution derived from the surface stations; (c)  the two techniques are complementary, different observables are measured in a redundant way and many cross-checks are possible.\\
An example showing the energy deposit in atmosphere by a EAS as reconstructed in one of the Pierre Auger fluorescence detectors is shown in Fig.\ref{fi:fluo}.
\vspace{0.4cm}

\centerline{\it Energy and composition estimators}

The experimental observables which are measured in order to extract information about the energy spectrum are the charged components of showers as measured by ground based detectors with scintillator counters, muon and hadron detectors, the Cherenkov light produced by shower particles as they propagate through the atmosphere and, at higher energy, the fluorescence emission.
Mass and energy of the primary particles are strictly correlated and, in general, the EAS development depends on the interplay of the two.\\
In most cases a single observable, like for example the electromagnetic or muonic shower size (that is the total number of electromagnetic charged particles $N_{e}$ or the total number of muons $N_{\mu}$)  is used to derive the all-particle energy spectrum. The relation of the size to the energy is obtained from simulations with some hypotheses on the mean primary mass. As an example, \citep{et99a} find the relation between shower size and primary energy 
\begin{equation}
N_{e} (E_{0} , A) = \alpha(A_{eff} )E^{\beta(A_{eff})}
\end{equation}
where  the normalization $\alpha$ and the slope $\beta$ depend on  the effective mass $A_{eff}$, calculated from the extrapolation of the spectra $\Phi_{i}(N_{e})$ of the individual elements (each with atomic mass $A_{i}$) measured at low energies by direct measurements:\\

\centerline{$A_{eff}(N_{e}) = \frac{\sum_{i} A_{i}\Phi_{i}(N_{e})}{\sum_{i} \Phi_{i}(N_{e})}$}

\noindent
Above the knee a rigidity dependent cutoff is used. \\
Combining the measure of the electromagnetic and muonic sizes, it is possible to obtain an energy determination almost independently of the primary mass; this method was exploited e.g. in  the CASA-MIA experiment \citep{casa99}, where 
\begin{equation}
E_{0}  =  0.8 \ GeV (N_{e} + 25 \ N_{\mu})
\end{equation}
for any primary mass within an uncertainty of $\simeq 5 \%$.\\
An estimate of energy can be derived from a combination of the amount of Cherenkov light and the location of shower maximum; the lateral distribution and intensity of Cherenkov light at a given total energy depends in fact both on the primary particle 
mass, hence on the mean $X_{max}$, and the distance of the measurement from the shower.\\
For giant arrays, like those measuring EAS above the knee region up to the highest energies, the energy is generally determined by measuring the particle densities at a specific distance from the core. In fact, the effects of intrinsic shower to shower fluctuations are minimized if the signal  is measured at an optimal core distance, which depends almost only on the detector spacing in the array \citep{hil71}. Following this work, the particle density S(600) at 600 m distance was used as energy estimator in the Haverah Park experiment  and later on in the AGASA array \citep{aga03}, while in the Pierre Auger experiment \citep{aug08}, where detectors are spaced on a 1.5 km grid, this distance is increased to 1 km. Finally, for fluorescence telescopes, the energy is measured almost calorimetrically by integrating the fluorescence light along the shower path.\\
The deconvolution of the primary energy and mass can be successfully performed in surface detectors by correlating different observables, like e.g. $N_{e}$ and $N_{\mu}$. The procedure consists basically of an unfolding of the two dimensional $N_{e} \ - \ N_{\mu}$ distribution of  the electromagnetic and muonic sizes of the EAS into the energy spectra of the primary mass groups, their correlation thus being taken into account. 
The number of events in each cell ($N_{e},N_{\mu})^{j}$ can be considered as the superposition of contributions from different primary particles of mass A and energy E:
\begin{equation}
N_{j} = S_{s} T_{m} \sum_{A=1}^{N_{A}} \int_{\Omega} \int_{-\infty}^{+\infty} \frac{dJ_{A}}{d logE} \times p_{A}\ d logE \ d\Omega
\end{equation}
where $dJ_{A}/d log E$ is the differential flux of an element with mass number A and the summation is carried out for all elements present in the primary cosmic radiation. $T_{m}$ represents the measurement time over a sampling area $S_{s}$; $d\Omega = sin\theta d\theta d\phi$ is the differential solid angle.
The  probability $p_{A}$ to measure at ground the sizes $N_{e}, N_{\mu}$  from a shower of primary energy E and primary mass A is evaluated using Monte Carlo simulations and includes the shower fluctuations, the detection and reconstruction efficiencies \citep{ka05}.\\
Techniques able to extract an almost pure light component (p+He) without the need for deconvolution have been used in \citep{etma04}, exploiting the space correlation of the EAS-TOP surface detector and the MACRO underground apparatus. 
The two experiments were separated by a rock thickness ranging from 1100 up to 1300 m depending on the angle  and located at a respective zenith angle of about $30^{\circ}$.  The muon energy threshold at the surface for muons reaching the MACRO depth ranged between 1.3 and 1.8 TeV within the effective area of EAS-TOP.
Light primaries can be selected based on their energy/nucleon by means of high energy muons, while the associated Cherenkov light detected on surface is proportional to the primary energy. \\
Air shower arrays with  large area muon detectors can study different regions of the muon multiplicity distribution as a function of the electron component of showers, with high sensitivity to different nuclear groups present in the primary flux \citep{gra09}. \\
Detectors measuring the Cherenkov or  fluorescence light produced by EAS in the atmosphere can derive the primary mass by measuring the depth of maximum development of the showers, $X_{max}$  (see Eq.\ref{eq:eqx}).\\
\vspace{0.4cm}

\centerline{\it The link to particle physics}
The interpretation of the ground level observations in terms of primary particle characteristics is far from straightforward, because of the significant (mass dependent) fluctuations of EAS observables and of the  strong dependence on hadronic interaction models used in the simulations of the production and propagation of particles through the atmosphere. At present, the latter constitutes the dominant source of systematic uncertainty. \\
Most of the observables that are relevant for the shower development (the total inelastic cross sections, the multiplicities of the final states and the inclusive energy spectra) must be obtained from extrapolations of the measurements in accelerator experiments, which are performed in a much lower energy range. The new measurements at the CERN LHC collider will reach $\sqrt{s}=14$ TeV, that is $E_{lab}\simeq 10^{17}$ eV, well below the maximum energy observed in CRs, and  only for  p-p interactions, while Pb-Pb collisions at  $\sqrt{s}=5.5$ TeV/nucleon are foreseen. Furthermore, the kinematic region of interest for CR interactions is the projectile fragmentation region, while in collider experiments the central region of the interaction is best explored. 
A couple of experiments starting at LHC will be of particular interest for CR physicists: LHCf \citep{adr08} will explore the very forward region of the interaction at $|\eta| > 8.5$\footnote{The pseudorapidity $\eta=-ln[tg(\theta/2)]$ measures the angle of the particle respect to the beam direction. The forward region is characterized by $|\eta| > 4$}, TOTEM \citep{ant10}  will derive the p-p cross section  in the range $3.1 \leq |\eta| \leq 6.5$.\\
On the other hand,  EAS measurements are the only way to give information about very high energy hadronic interactions, and can perform precise tests to constrain the models. The data from KASCADE, for example, allowed to compare the measured correlations between different hadronic observables (such as number of hadrons, their energy sum, their lateral distribution and energy spectra) with the number of muons and electrons at ground level with the results of the CORSIKA simulation code for various models (\citep{ape07} and refs. therein). \\
The measurement of the inelastic proton-air cross section  can be performed using EAS arrays from the mean depth of the first interaction and its fluctuations (\citep{ulr09} and refs. therein). It can be determined indirectly  either investigating the unaccompanied hadrons, at low energies, or basing on the analysis of the muon and electromagnetic shower sizes using simulations \citep{agl09cross} . 
Most recently, the Pierre Auger Collaboration used their hybrid data to analyze  the shape of the depth of shower maximum distribution: the choice of showers more deeply penetrating in the atmosphere allows to obtain a proton rich sample of events \citep{ulr11}.  The data favour  a moderately slow rise of the cross- section towards higher energies, a result that seems to be confirmed in the first data from LHC \citep{aad11} at lower energy.

\subsection{The knee region}

\begin{figure}[!]
\begin{minipage}{0.47\linewidth}
 \centering
 \includegraphics[width=8.5cm,height=8.5cm]{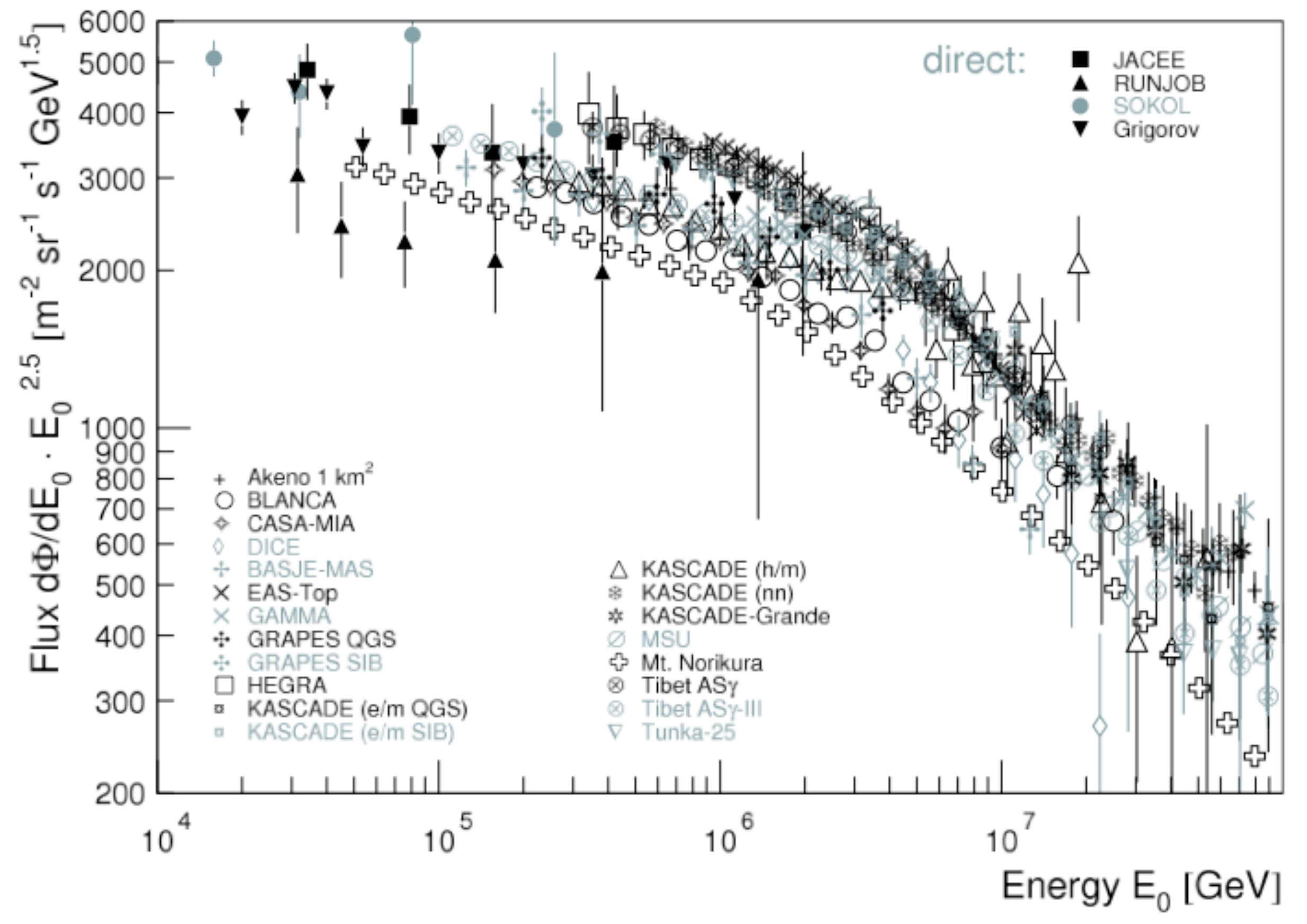}  
\caption{\em {The all-particle energy spectrum in the knee region  \citep{blu09}.}}  
 \label{fi:spek2}
 \end{minipage}\hfill
 \begin{minipage}{0.47\linewidth}
 \centering
 \includegraphics[width=8.5cm,height=8.5cm]{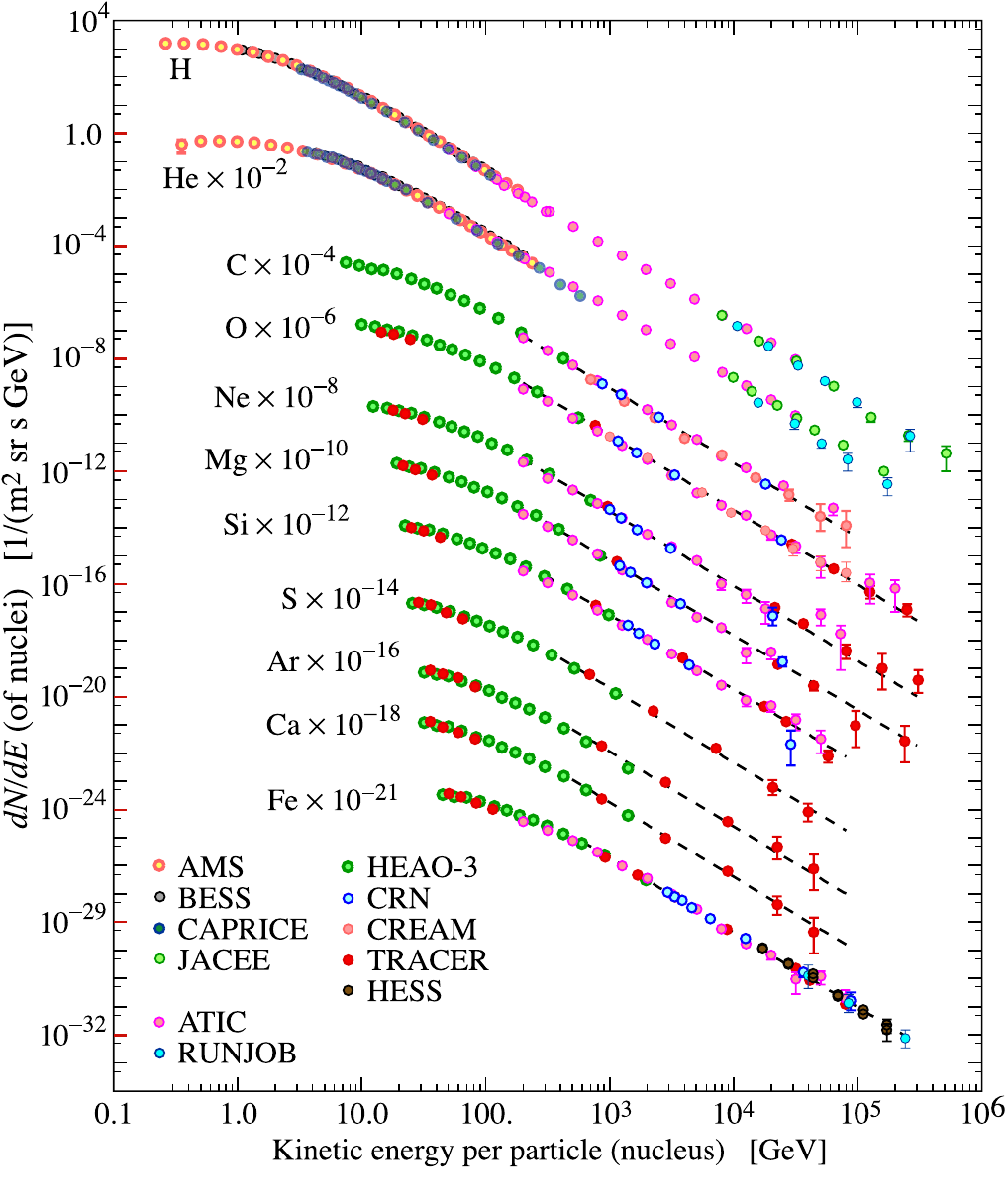}   
\caption{\em {Single elements spectra from direct measurements \citep{rpp09}.}}
 \label{fi:spedir}
 \end{minipage}
\end{figure}

\begin{figure}[!]
\begin{minipage}{0.47\linewidth}
 \centering
 \includegraphics[width=8.5cm,height=8.5cm]{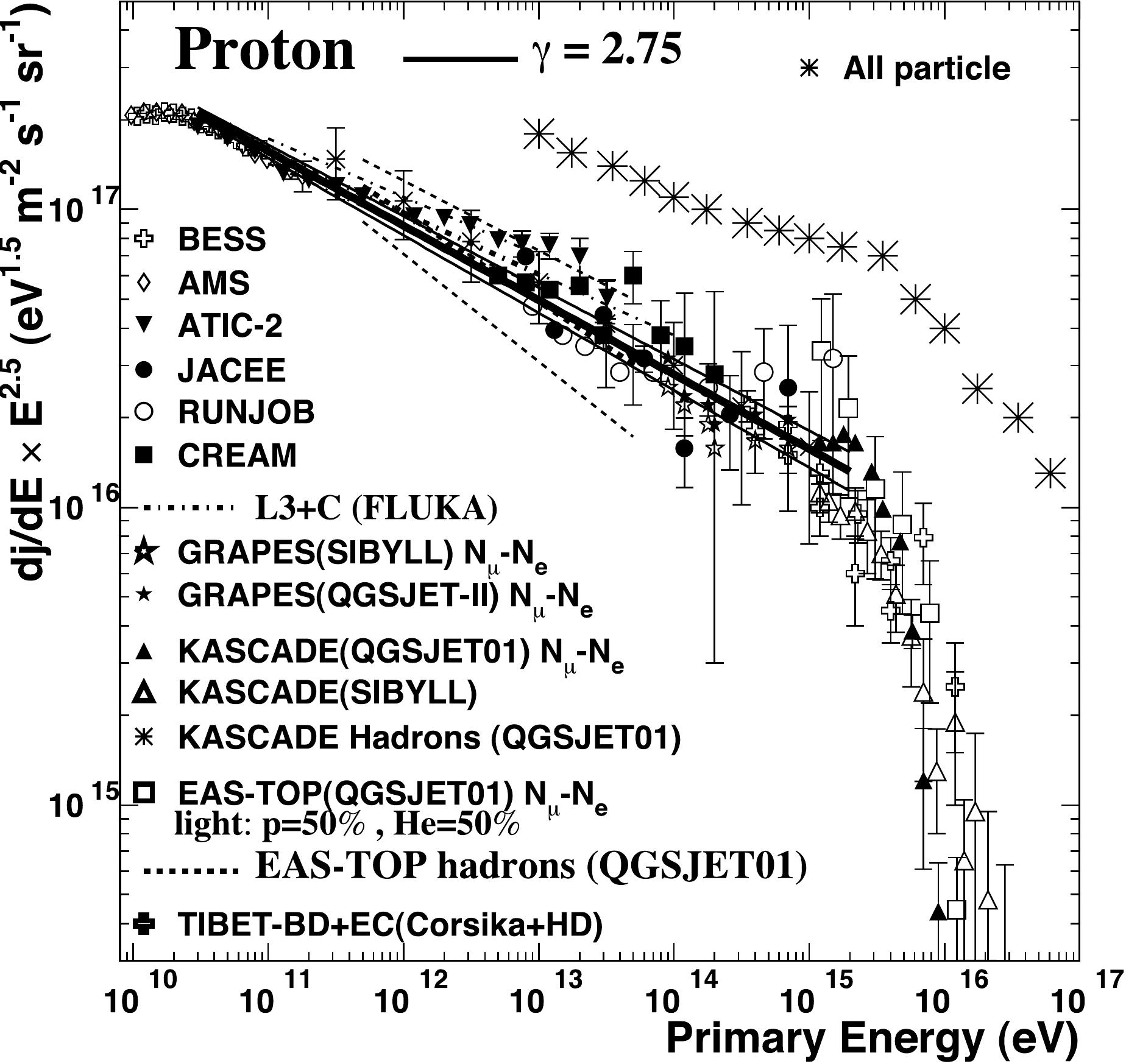} 
 \end{minipage}\hfill
 \begin{minipage}{0.47\linewidth}
 \centering
 \includegraphics[width=8.5cm,height=8.5cm]{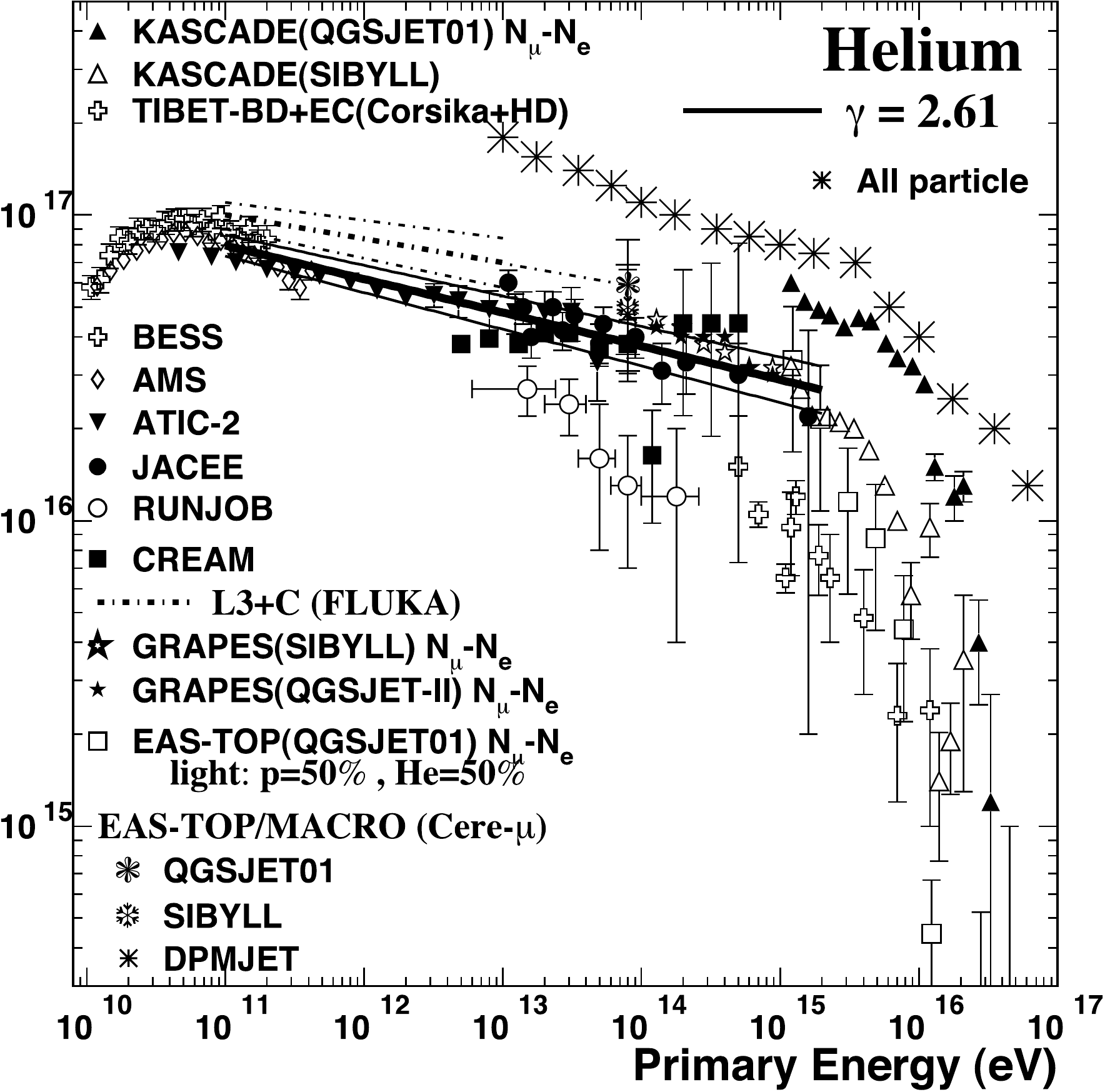}  
 \end{minipage}
\caption{\em {Compilation of the proton and Helium spectra \citep{bert08}. The all-particle spectrum is shown by asterisks for reference.}}  
 \label{fi:spettri1}
\end{figure}

\begin{figure}[!htbp]
\begin{minipage}{0.47\linewidth}
 \centering
 \includegraphics[width=8.5cm,height=8.5cm]{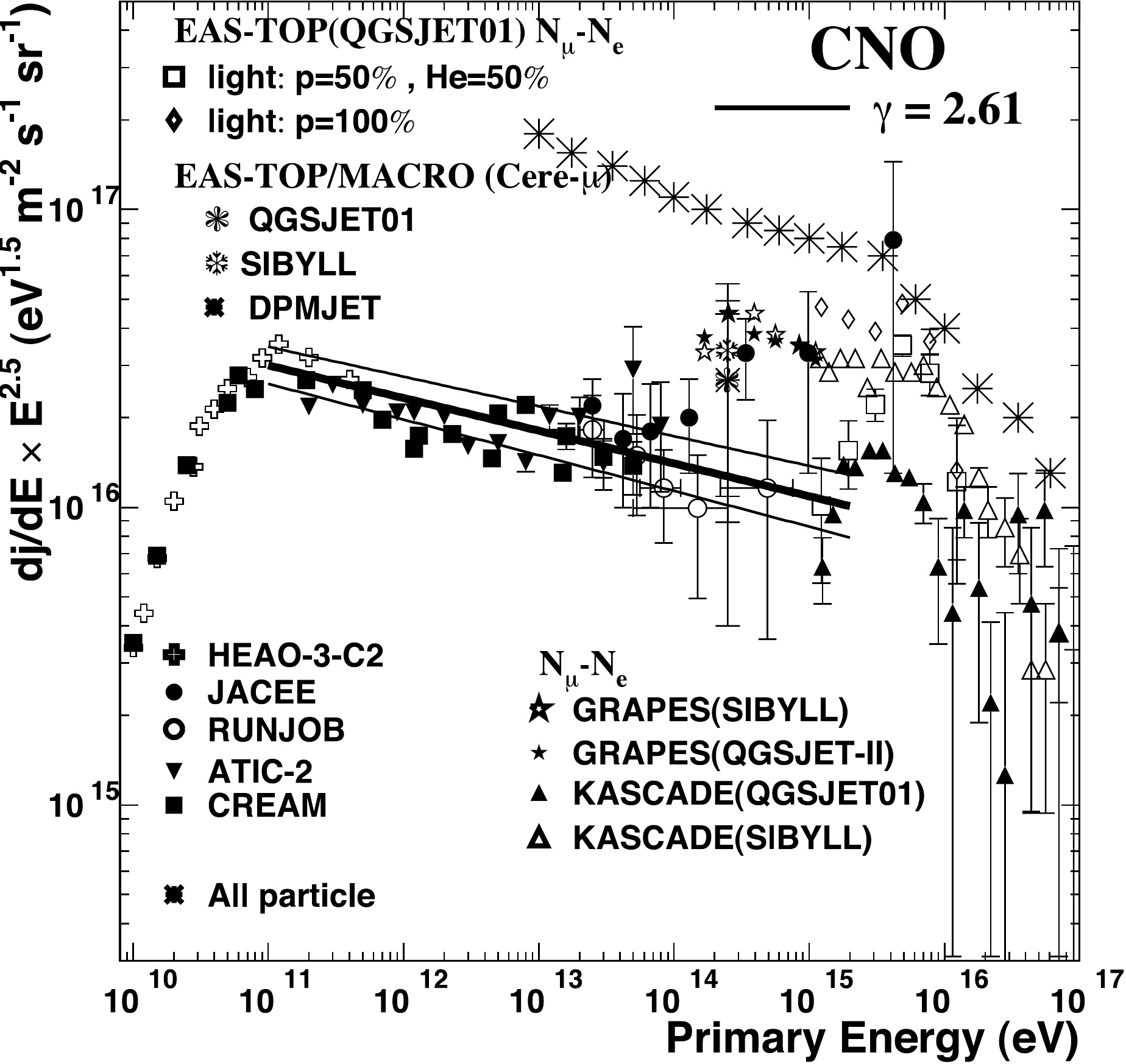}  
 \end{minipage}\hfill
 \begin{minipage}{0.47\linewidth}
 \centering
 \includegraphics[width=8.5cm,height=8.5cm]{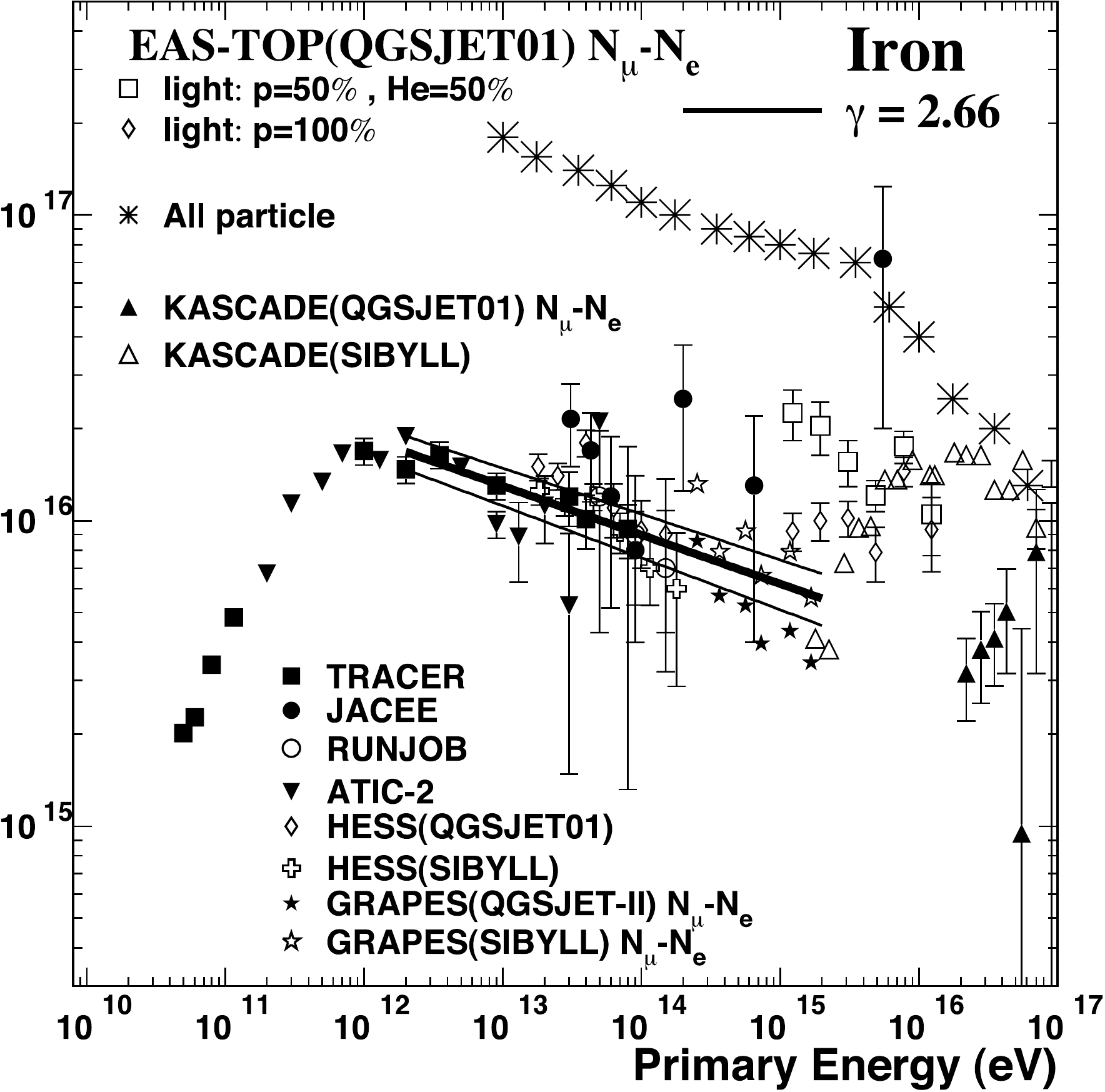}   
 \end{minipage}
\caption{\em {Compilation of the CNO and Iron spectra \citep{bert08}. The all-particle spectrum is shown by asterisks for reference.}}  
 \label{fi:spettri2}
\end{figure}

\vspace{0.4cm}
\centerline{\it The energy spectrum}
A huge number of experiments contributed to the measure of the all-particle energy spectrum, shown in Fig.\ref{fi:fig1}. In the lowest energy region, the fluxes are obtained by means of direct measurements above the atmosphere, the RunJob \citep{run05} and JACEE \citep{jac98} being the only two balloon experiments extending above 100 TeV. 
The all particle spectrum above $10^{13}$ eV is shown in Fig.\ref{fi:spek2}. All experiments agree on a spectrum described by a power law with a slope changing from $\langle \gamma \rangle = (2.66 \pm 0.06)$ below the knee to $\langle \gamma \rangle =(3.10 \pm 0.09)$ above. The mean value of the knee energy is $E_{knee}=(3.9 \pm 0.7)$ PeV.  \\
In the lower energy region, the results from balloons  and EAS experiments overlap, showing a good agreement considering that indirect data  below about $10^{15}$ eV are dominated by systematic uncertainties while direct measurements suffer from statistical ones.\\
The systematic uncertainties depend on the fact that the all-particle energy spectrum has been obtained using different observables, dissimilar assumptions on composition or in some case no assumption at all, in apparatuses located at quite different altitudes (from underground laboratories to the sea level, to high mountains). Each experiment employs  a specific  "particle" unit to define the signal, and the observables are measured with quite dissimilar accuracies. Furthermore, the methods of conversion from the observables to the energy are often based on different simulation codes which in turn exploit various hadronic interaction codes. A recent discussion of the differencies among the experiments can be found in \citep{nag09}.\\
More information can be gathered by measuring the spectra of single elements: a recent compilation of their spectra is shown in Fig.\ref{fi:spedir} from the results of direct measurements. 
Updated data can be found in \citep{2010ApJ...715.1400A,2011arXiv1108.4838O,2010ApJ...723L...1M,rau09}.\\
In addition to the balloon and satellite results, direct measurements of single elements can be obtained by measuring the Cherenkov light emitted directly by the primary particle in the $\simeq 8-15$ g/cm$^2$  of atmosphere above the first interaction with an atmospheric nucleus.  Exploiting the different emission angle and time of arrival of the direct and shower Cherenkov contributions (see Sect.\ref{meth}), it is possible to separate the two signals.
In this way, the energy spectrum of iron nuclei was determined between 13 and 200 TeV employing the HESS Imaging Atmospheric Cherenkov Telescopes \citep{aha07}. \\
The knowledge of the energy spectra of single (or groups) of elements from direct measurements has been extended to higher energies by exploiting the ability of the modern ground experiments to detect the different components of the EAS. A compilation of the world results is shown in Fig.\ref{fi:spettri1} and \ref{fi:spettri2} for the single (or groups of) elements spectra \citep{bert08}.\\
The spectra of the different elements can be fitted using a power law up to the corresponding knee. For protons, for which there is a remarkable agreement of all techniques in the whole energy range:
\begin{equation}
\frac{dN}{dE}=(8.8 \pm 0.5) \ E^{-2.75 \pm 0.01}  \ \ m^{-2} \ s^{-1} \ sr^{-1} \ TeV^{-1}
\end{equation}
while for all the other elements the slope is close to 2.61. Most experiments confirm a difference between the proton and Helium slopes, which could be interpreted with  two different types of sources/acceleration mechanisms for the two elements \citep{bie93}. The only contradicting results come from the RunJob collaboration; however, a very recent reanalysis of their data \citep{kop09} attributes the differences as being due to to the very low statistics and systematic uncertainties, especially in the high energy region. 
The larger spread among the results for the CNO can be partly explained by the slightly different definitions of "CNO" group in the various experiments; some of the data shown in the Fe plot refer to a more general group of heavy elements around iron.\\
The measured spectra show that the knee in the all-particle spectrum is mainly due to the light elements suppression. The cut-offs in the different spectra seem to show up at energies proportional to the nuclear charge; the most recent results \citep{ape09} confirm this conclusion, underlining once more the limiting factor due to the uncertainty in the hadronic interaction models.\\
Dividing their air shower events in electron-rich and electron-poor groups,  based on the measure of the charged particle and muonic components, \citep{KG11} found the first experimental evidence for a knee-like break in the  CR spectrum of heavy primaries at about $ 9 \times 10^{16}$ eV, in agreement with models where the components of the primary beam bend at subsequent knees proportional to the charge of the primary nuclei.\\
A different conclusion is reached in the Tibet experiment \citep{ame11}, where data indicate a heavy component dominance around the knee, with light nuclei bending below the all particle knee and contributing to not more than 30$\%$ at the knee. According to the authors, the disagreement with other data (e.g. KASCADE) can be attributed to the different kinematic region explored and to model dependence. The result could be explained either assuming CRs produced in  nearby sources with source composition dominated by heavy nuclei or  nonlinear effects in the diffusive shock acceleration mechanism, but the limited statistics could affect the conclusion.
\vspace{0.4cm}

\centerline{\it The composition in the knee region}

Above  about 100 TeV,  the particle fluxes are very low and the measurements are subject to large fluctuations; that is why the results obtained with the different techniques on the primary composition are generally given in terms of the mean logarithmic mass, defined as
\begin{equation}
\langle ln \ A \rangle = \sum_{i} r_{i} \ ln A_{i}
\end{equation} 
where $r_{i}$ is the relative fraction of nuclei of mass $A_i$.
The experimental measurement of this quantity is performed exploiting either the proportionality of  $\langle ln \ A \rangle$  to the mean shower maximum or that of A to the ratio of the charged components of the shower. In fact, in the frame of a simple description of the shower development in atmosphere (see Sect.\ref{secEAS}) and using the superposition model, Eq.\ref{eq:eqx} gives the dependence of $<X^{A}_{max}> $ on $ln~A$, while recalling Eqs.\ref{eq:eqne} and \ref{eq:eqm},  the ratio of the electromagnetic and muonic particle sizes
\begin{equation}
N_{e}/N_{\mu} \propto \left( \frac{E_{0}}{A} \right)^{0.15}
\end{equation}

\begin{figure}[h]
\begin{minipage}{0.47\linewidth}
 \includegraphics[width=8.5cm,height=7.5cm]{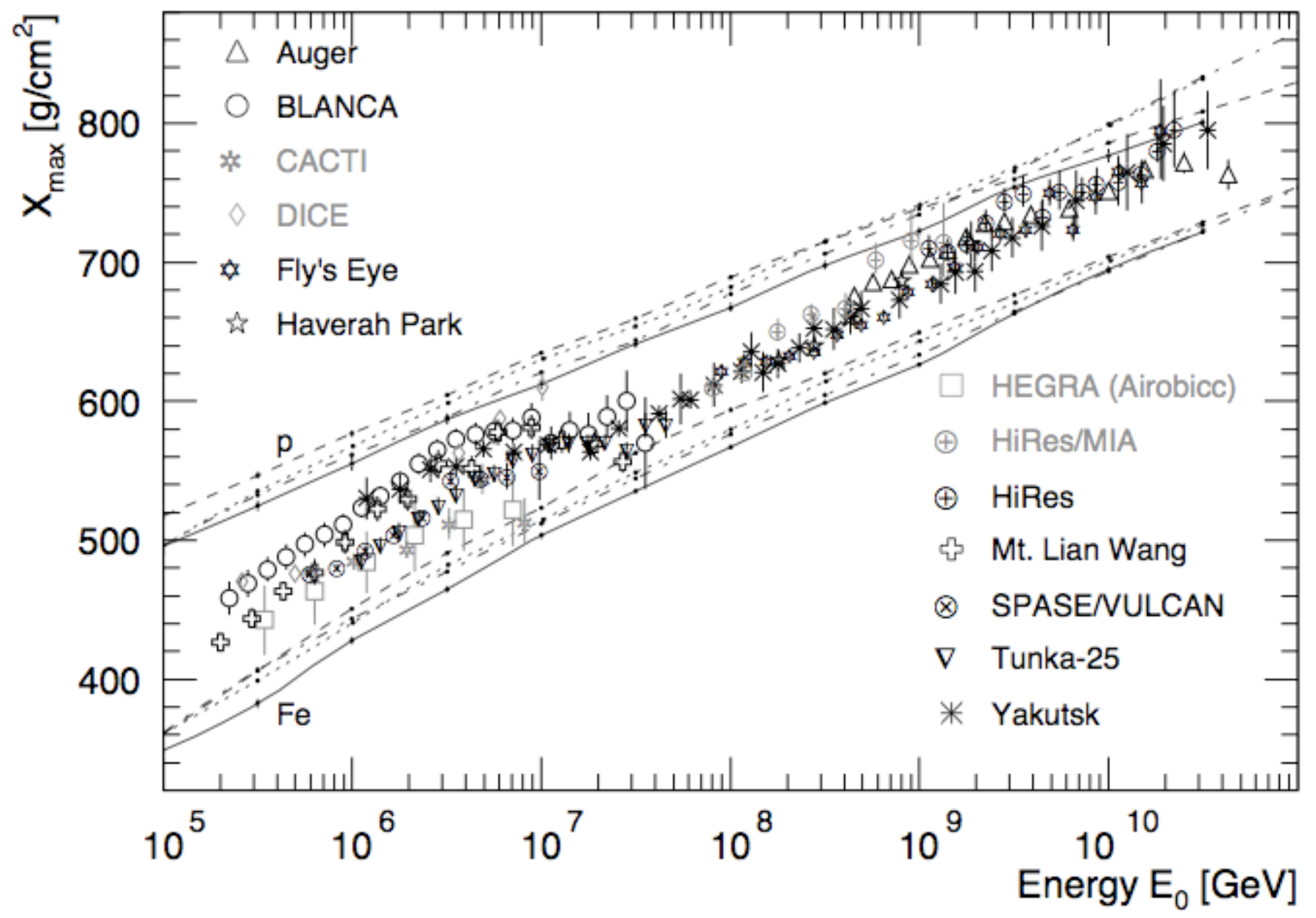}   
\caption{\em {Average depth of the shower maximum as a function of energy  as derived from Cherenkov or fluorescence detectors.}}  
 \label{fi:logaX}
 \centering
 \end{minipage}\hfill
 \begin{minipage}{0.47\linewidth}
 \centering
 \includegraphics[width=8.5cm,height=7.5cm]{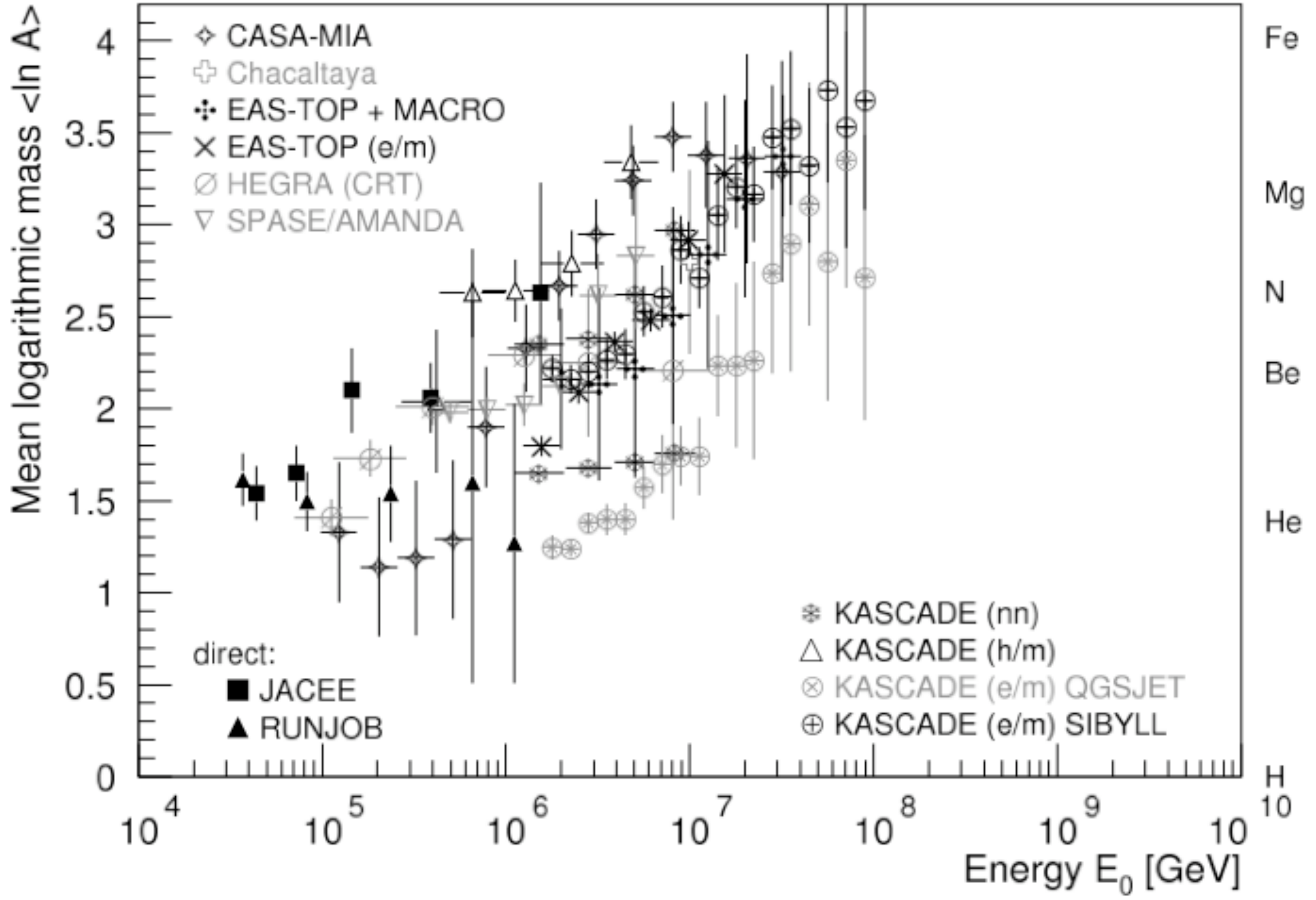}   
\caption{\em {The mean logarithmic mass from the measurement of charged components of EAS at ground as a function of energy.}}  
 \label{fi:logaCh}
 \end{minipage}
 \end{figure}

The results on the measure of  $<X_{max}>$ are shown in Fig.\ref{fi:logaX}. Monte Carlo simulations with specific hadronic interaction model choice are used to get the expected average depths of maximum for protons and iron, $X_{max}^{p,sim}$ and $X_{max}^{Fe,sim}$, so that
\begin{equation}
\langle ln \ A \rangle = \frac{<X_{max}>-X_{max}^{p,sim}}{(X_{max}^{Fe}-X_{max}^{p})^{sim}} \cdot lnA_{Fe}
\end{equation}
The composition is getting lighter towards the knee, while  the opposite happens above. From $10^7$ GeV up to the transition region, it remains almost constant. 
 
The mean logarithmic mass from the measurements of the charged components of the showers is shown in Fig.\ref{fi:logaCh}, obtained using the QGSJet-01 hadronic interaction model. The increase in $<lnA>$ across the knee region is clear, but we do not see it to become lighter towards the knee, as it was observed by means of the measurement of $X_{max}$. A better agreement among the different methods can be obtained modifying some of the features of the models; e.g., a decrease in the inelastic cross section and an increase in the elasticity turn into a deeper depth of shower maximum, which reflects into the number of particles produced at ground \citep{hor09}.\\

\subsection{Models for the knee}

\begin{figure}[!]
\centering
 \includegraphics[width=18.0cm,height=16cm]{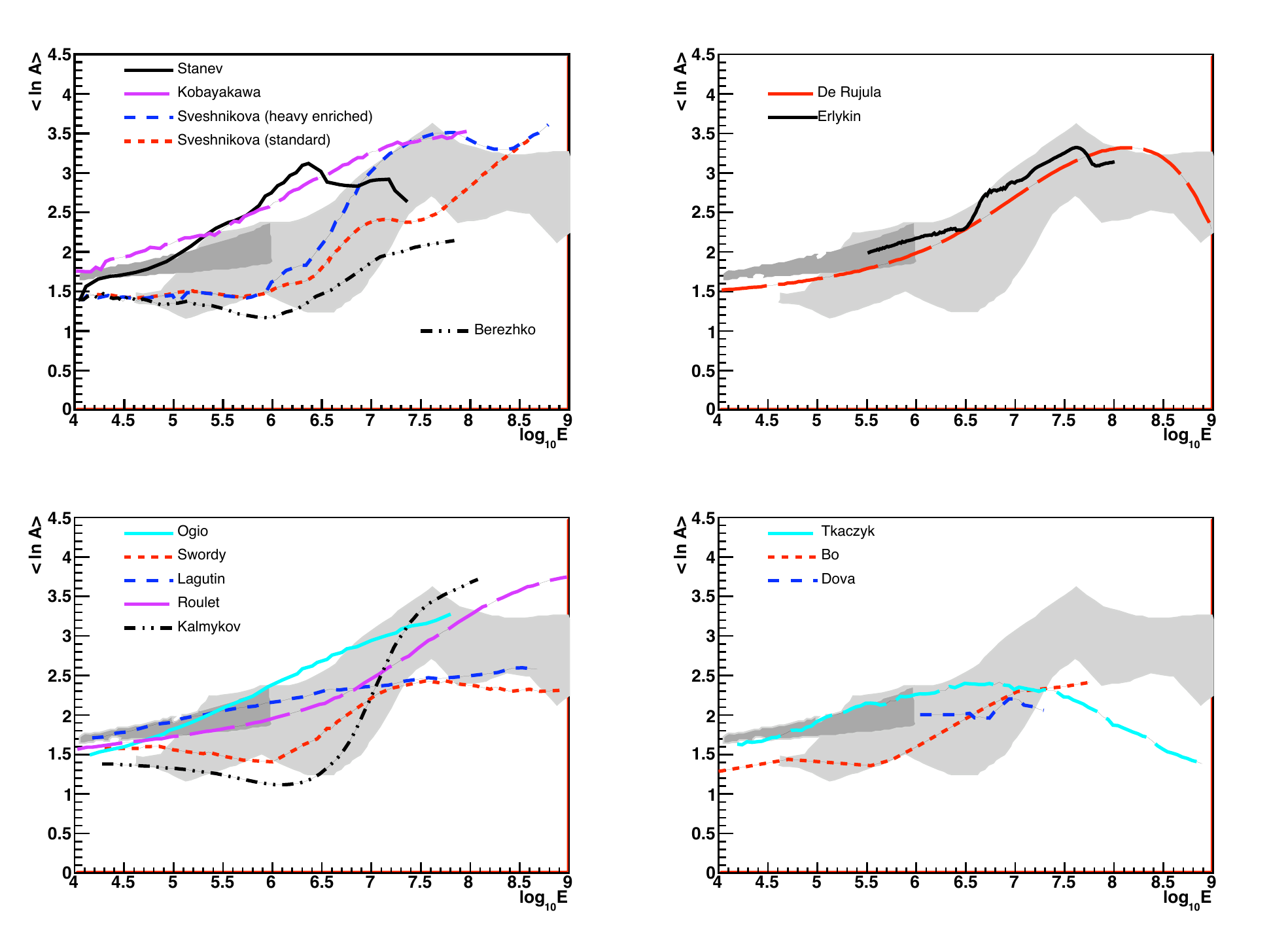}  
\caption{\em {The mean logarithmic mass from direct (dark gray) and indirect (light gray) measurements  as a function of the logarithm of energy, compared with different models. Top left: acceleration models, top right:   cannonball and single source models. Bottom line: propagation in the Galaxy (left) and interaction with background particles (right)  (see text for details). Adapted and modified from \citep{hor04}.}}
 \label{fi:lnA}
 \end{figure}

Fig.\ref{fi:spettri1} and \ref{fi:spettri2} show a compilation \citep{bert08} of the data on the single (or groups of) elements spectra; together with those from composition and anisotropy, these results can give some hints about the origin of the spectral feature of the  knee.  \\
The most popular scenario attributes the knee to the maximum energy attainable by a relativistic particle of charge Ze and energy E(PeV) during the acceleration process   in a site with magnetic field B and size L (see Eq.\ref{eq:eq1}). \\
The actual value of the break depends on the properties of the source and of the medium, on the strength of the magnetic field and its orientation with respect to the particles motion. While breaks around $E_{max} \propto Z \times 10^{14}$ eV are predicted in \citep{sta93,bie93,kob02}, higher energies can be reached if  $B$ is increased, e.g. taking into account preacceleration in the precursor wind \citep{ber99} or on the contrary increasing the shock parameters and summing the contributions of different kinds of Supernovae \citep{sve03}.\\
In Fig.\ref{fi:lnA} (top left), the prediction of these models are compared to the experimental results obtained by means of direct (dark gray area) and  indirect (light gray area)  measurements. \\
Pertaining to the same group, the model of \citep{ew01,ew09} assumes that, out of a background of many undefined sources,  a single close-by SNR be responsible for the knee.
From a recent comparison among the spectra measured by 10 different groups  \citep{erl11}, the knee appears sharper\footnote{The "sharpness" of the knee is here defined as $S = -d^{2}logI/d(logE)^{2}$, where I(E) is the primary CR energy spectrum} than that predicted in diffusion models, Helium is the dominant nucleus around the knee and a second peak around 50-80 PeV (maybe the iron peak) is found.
The mean logarithmic mass derived from this model is shown in Fig.\ref{fi:lnA} (top right); viable candidates for this single source are suggested to be  B0656+14 in the Monogem ring or J0833-45 in the Vela SNR.\\
A single type of source,  the "cannon ball"\footnote{jet of plasmoids of ordinary matter emitted in the explosion of a core-collapse SN, in addition to the ejection of a non relativistic spherical shell},
is supposed to be the origin of all non solar CRs, at all energies in  \citep{dar08}: a sequence of magnetic collisions in the turbulent field of the cannonball accelerate CRs to higher and higher energies. The model is developed as a generalization of the one used for$\gamma$ ray bursts \citep{dar04}. The prediction is shown in the same figure, and it seems to be in good agreement with data. However,  further work is needed in this model to understand the mechanisms of confinement and acceleration and the  diffusion to Earth must be proven negligible.
The knee can also be generated by propagation effects: in this case, as the energy of the CRs increases, their confinement in the Galaxy becomes more difficult and  their escape easier.  Diffusive shock acceleration with energy dependent propagation pathlength has been used by \citep{swo95}, with the introduction of a "residual pathlength" $\lambda_{res} \simeq 0.013$ g/cm$^2$ which is supposed to provide a minimum pathlength at high energy: $\lambda_{e}=\lambda_{0} \left( \frac{R}{r_{M}} \right)^{-\delta} + \lambda_{res}$. The propagation process has been described in terms of anomalous diffusion in \citep{lag01}, where the knee appears to be due to  the fractal structure of the galactic magnetic fields. \\
As shown in Fig.\ref{fi:lnA} (bottom left), the propagation process has been described with different assumptions on the galactic magnetic fields. In \citep{ptu93}, the drift in the regular magnetic field, effective at high energy, is responsible for the knee, which appears to be sharp, while leakage through diffusive motion along the turbulent galactic magnetic fields is described in \citep{ogi03}. Diffusion and drift in a 3 component magnetic field is modeled in \citep{rou04}.  \\
A combination of both acceleration and propagation effects could give an acceptable description of the data up to $10^{16}$ eV.\\
According to a  third group of models, the knee is a threshold effect, due to the interactions of charged CRs with background particles.
Massive neutrinos, as proposed in \citep{dov01,wig03} could in principle explain the spectral features. However, besides producing a too light composition above the knee, they appear to be excluded by the results of WMAP and 2dFGRS \citep{han04}.  Photodisintegration of the CR nuclei by interactions with optical and soft UV photons in the source region  has also been considered \citep{kar93,can02}, but  it does not succeed in describing the composition in the direct measurements realm and at the highest energies, as shown in Fig.\ref{fi:lnA} (bottom right). \\
A more recent proposal relates the knee to the $e^{+}e^{-}$ pair production  by CR nuclei interacting with  background photons \citep{wa10}. \\
Finally, new hadronic interactions could take place in the atmosphere at very high energies. The knee would be in this case just due to an underestimation of energies of the shower particles.  The  missing energy could be transferred to unobservable particles, e.g. gravitons \citep{kaz01}; a test of the model will come from LHC results. 
The production of PeV muons  from the leptonic decay of hypothetical heavy short lived particles was proposed in \citep{pet03}. They would not be detected by EAS arrays (which generally count muons but do not measure their energy).  An exotic component or high energy atmospheric muons seems however to be already excluded by the measurements of the Baikal $\nu$ telescope \citep{wis05}.

\section{Above 100 PeV/n: the onset of EGCRs}
\label{sec5}

\subsection{Energy spectrum and composition}

\begin{figure}[!h]
\begin{minipage}{0.47\linewidth}
 \centering
 \includegraphics[width=8.5cm,height=8.5cm]{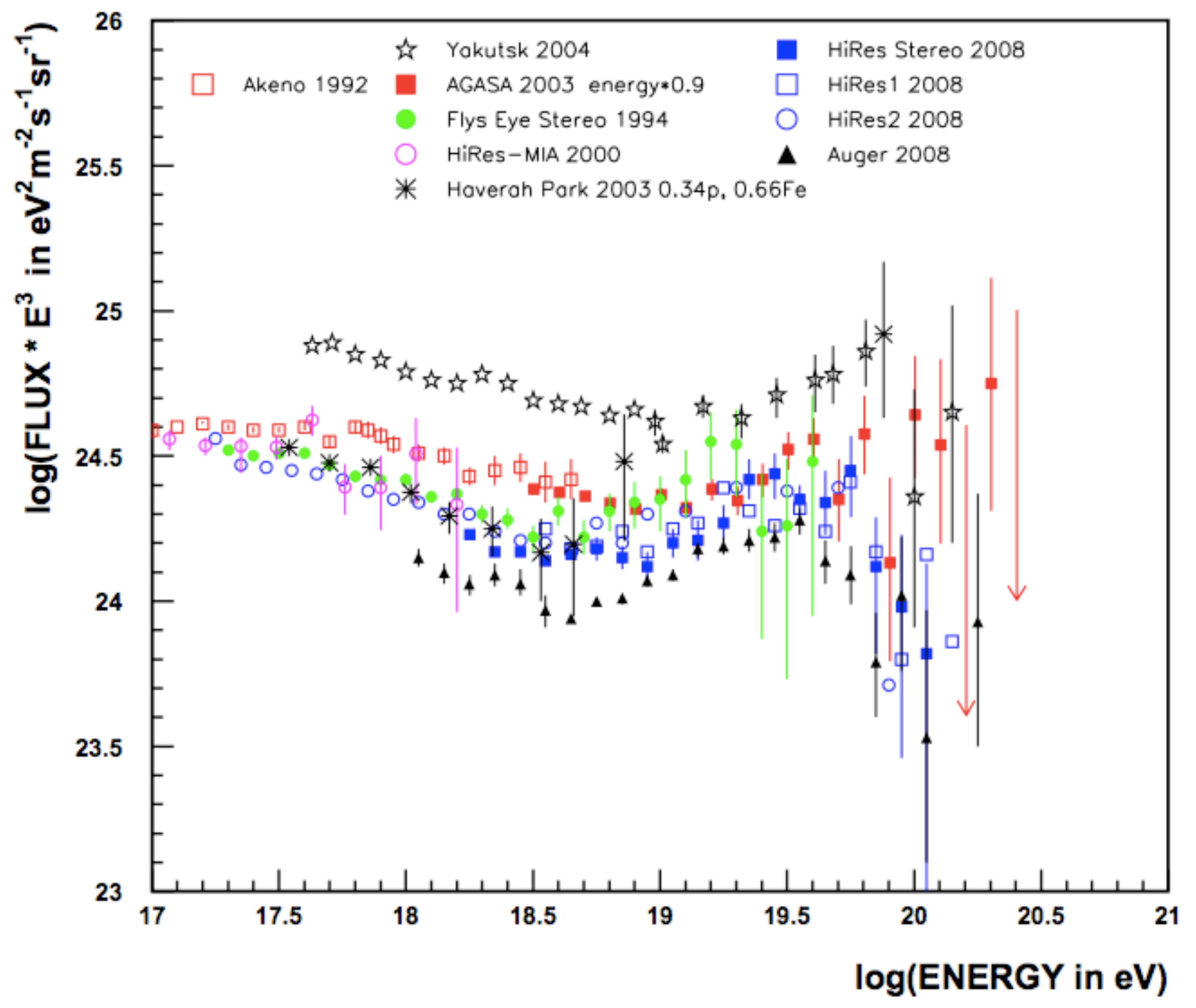}   
\caption{\em {Primary energy spectrum in the transition region and above. The 
differential flux in each bin is multiplied by an energy-dependent power $E^3$ \citep{nag09}.}}  
 \label{fi:nagspe}
\end{minipage}\hfill
 \begin{minipage}{0.47\linewidth}
\centering
\includegraphics[width=8.5cm,height=8.5cm]{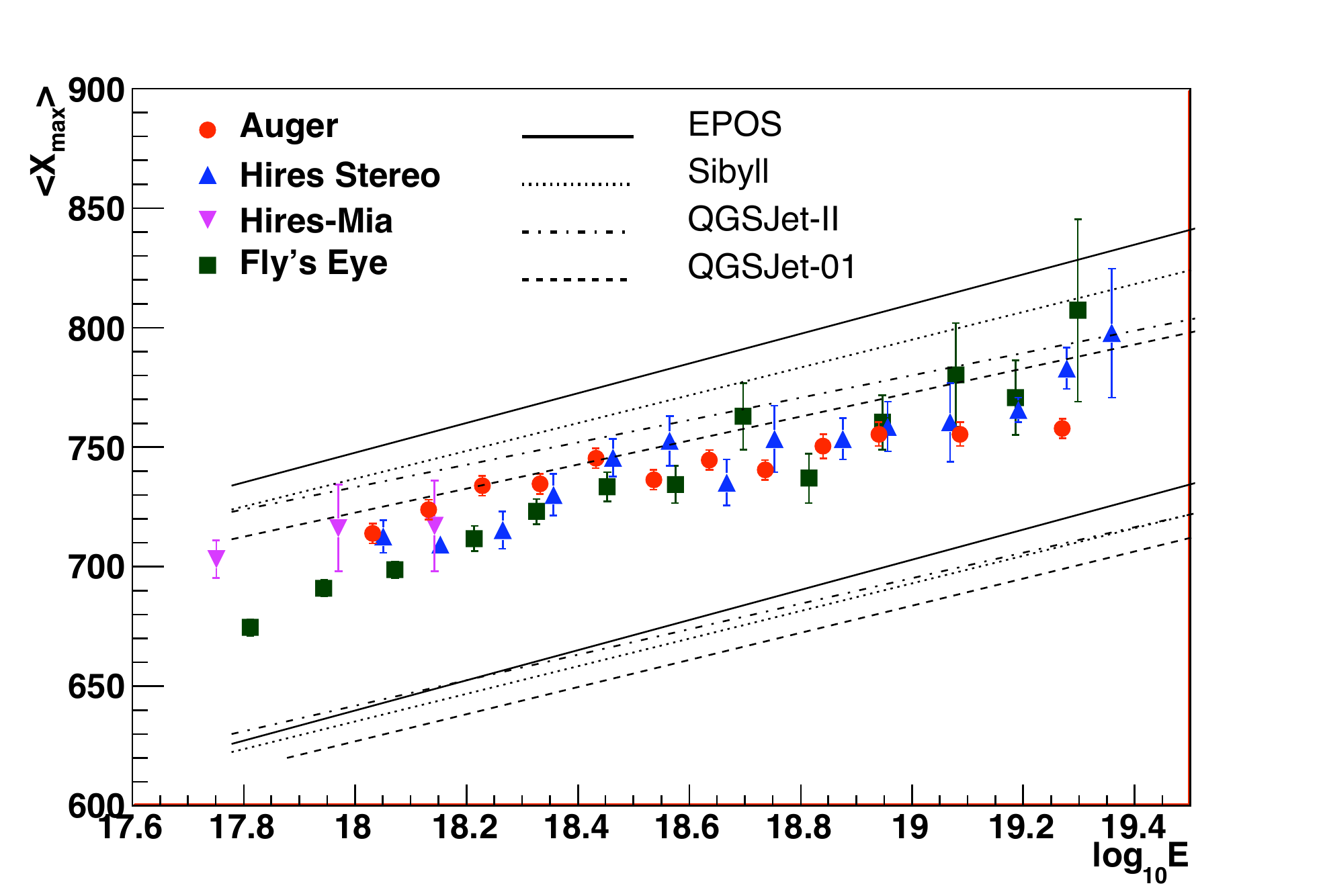}   
\caption{\em {Mean depth of shower maximum  vs primary energy, compared to different hadronic interaction models for protons and iron (modified from \citep{aug10}).}}  
\label{fi:xmx}
 \end{minipage}
\end{figure}

Different experiments have collected data in the galactic/extragalactic  transition region, from the second knee to the ankle: Haverah Park \citep{law91,ave03}, Akeno \citep{ake92}, Fly's Eye \citep{bir94}, Yakutsk \citep{ego01}, HiRes-MIA \citep{abu01},  AGASA \citep{tak03}, Auger \citep{aug10a}. 
Some of them are (or were) surface detectors: water  Cherenkov detectors  over an area of $\simeq 12$  km$^{2}$ at Haverah Park, scintillators spread over 100 km$^2$ at AGASA. The Yakutsk array in Siberia covered an area of $\simeq 18$ km$^2$ with surface scintillators, underground detectors and a photomultiplier system to measure the Cherenkov light in air.  Fly's Eye and its successor HiRes are based on the fluorescence technique. The latter  used a system of two detector stations 12.6 km apart, with 22 and 42 telescopes respectively, each with 3.7 m diameter spherical mirrors. The Pierre Auger Observatory exploits both the surface array (1600 water detectors over 3000 km$^2$) and the 24 fluorescence telescopes  overlooking the apparatus from 4 sites to measure EAS in hybrid mode.\\
The primary energy spectrum is shown in Fig.\ref{fi:nagspe}, multiplied by a factor $E^{3}$ to better show the deviations from a pure power law; the "ankle" feature is clearly visible in all data, at an energy corresponding to $log_{10}(E/eV)=18.60 \div 18.65$. \\
The systematic uncertainties play a leading role here: they are entangled with the statistical ones in the spectrum representation of the figure (due to the $E^{3}$ multiplication), but it can be shown that a good agreement both in the flux normalization and in the ankle position can be reached if they are properly taken into account.
For a surface detector like AGASA, they depend mainly on the simulations used to relate the signal measured at 600 m, $S(600)$,  to energy. In the case of HiRes, the overall systematic uncertainty on the energy scale is $\simeq 17\%$ \citep{abb04}; in this apparatus, the aperture is rapidly growing with energy, since at higher energy  the showers are brighter and can thus be detected at larger distances, and it is determined by simulations.\\
In the Pierre Auger Observatory, the energy dependent exposure includes the trigger, reconstruction and selection efficiencies and the evolution of the detector in time (through the construction phase). Its total systematic uncertainty  is estimated to be $10 \%$ at 1 EeV, decreasing to $6 \%$ above 10 EeV. The uncertainties in the evaluation of the energy scale depend on  the fluorescence yield, the absolute calibration, the reconstruction method and amount to $\simeq 22 \%$ \citep{aug08}. The hybrid technique exploited in the Auger experiment allows to directly correlate the  signal measured by the ground array at 1000 m from the shower axis, $S(1000)$,  to the calorimetric energy measured by the fluorescence detector. In such a way, the energy assignment is almost independent of simulations, which enters only in the evaluation of the energy which can be missed because carried  away by neutrinos and muons and thus not contributing to the fluorescence signal.

Our knowledge of the mass of primary CRs in the transition region and above is quite poor. The conclusions reached by the different groups depend on the methods used to measure the mass and, most important, they  are dependent, to a greater or lesser extent, upon the interaction model that is 
assumed in the simulations needed to interpret the data. \\
 The composition can be studied by measuring the depth of maximum development of showers, $X_{max}$.
 the results from Fly's Eye \citep{bir94}, HiRes-MIA \citep{abu01}, Hires-stereo \citep{abb05} and Auger \citep{aug10} are shown in Fig.\ref{fi:xmx}, together with the prediction of different models for proton or iron primaries. \\
As it is clear from the figure, the differencies among the predictions are quite big; the model ambiguity is even higher when using the observables from surface arrays, since the predictions for the number of muons at ground differ as much as $30 \%$. However, if their description of hadronic interactions at these energies is realistic, then a trend from heavier to lighter composition is observed in data from Fly's Eye and HiRes, while a more mixed composition is suggested by the Auger data. Above 10 EeV, the latter suggest a gradual increase of the primary CR mass, a conclusion also confirmed by the analysis of the  magnitude of fluctuation of the position of $X_{max}$, which, for fixed primary energy, is expected to be larger for protons than for iron nuclei. \\

\subsection{Models for the transition}
\begin{figure}[!h]
 \centering
 \includegraphics[width=16.5cm,height=8.cm]{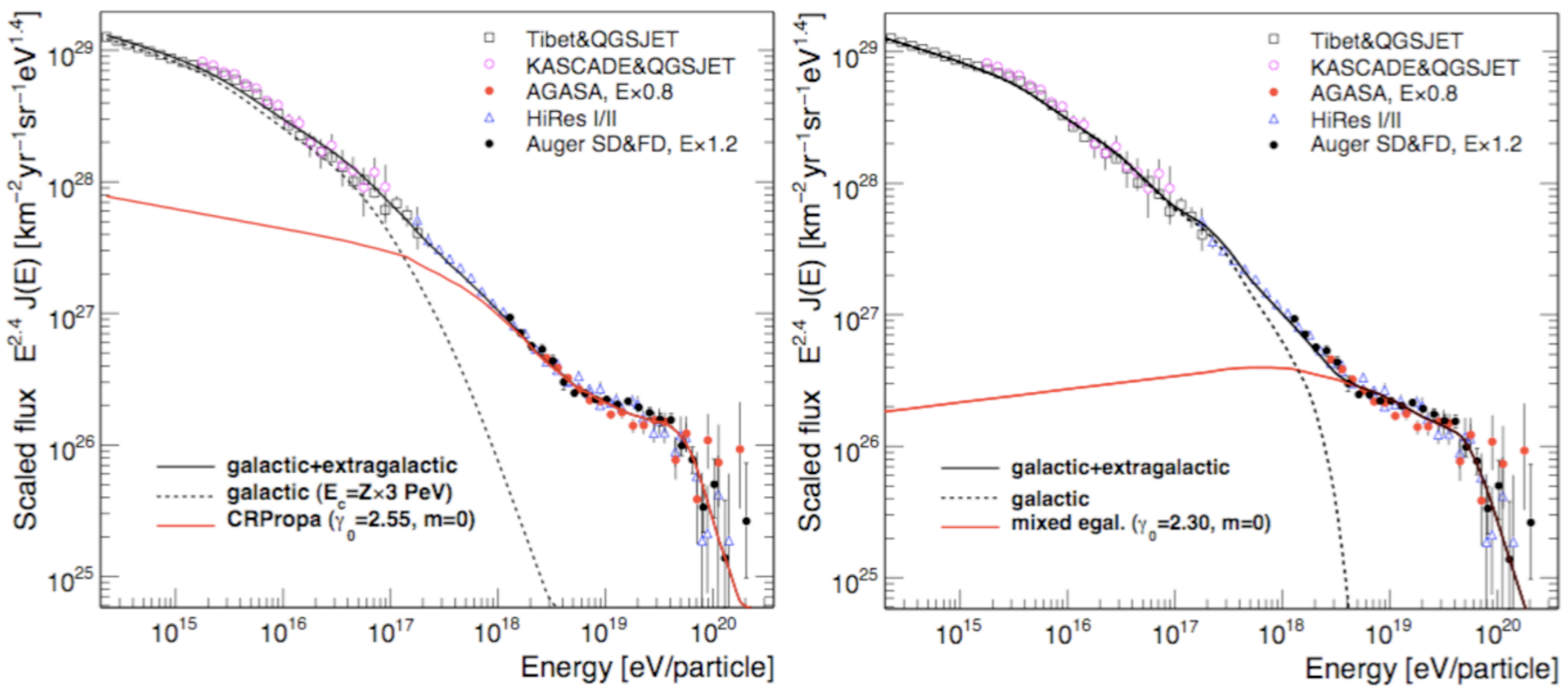}   
\caption{\em {The measured all-particle flux compared to the dip and the mixed composition models described in the text \citep{ung08}.}}  
 \label{fi:speTrMod}
\end{figure}

\begin{figure}[!h]
\begin{minipage}{0.47\linewidth}
 \centering
 \includegraphics[width=8.5cm,height=8.5cm]{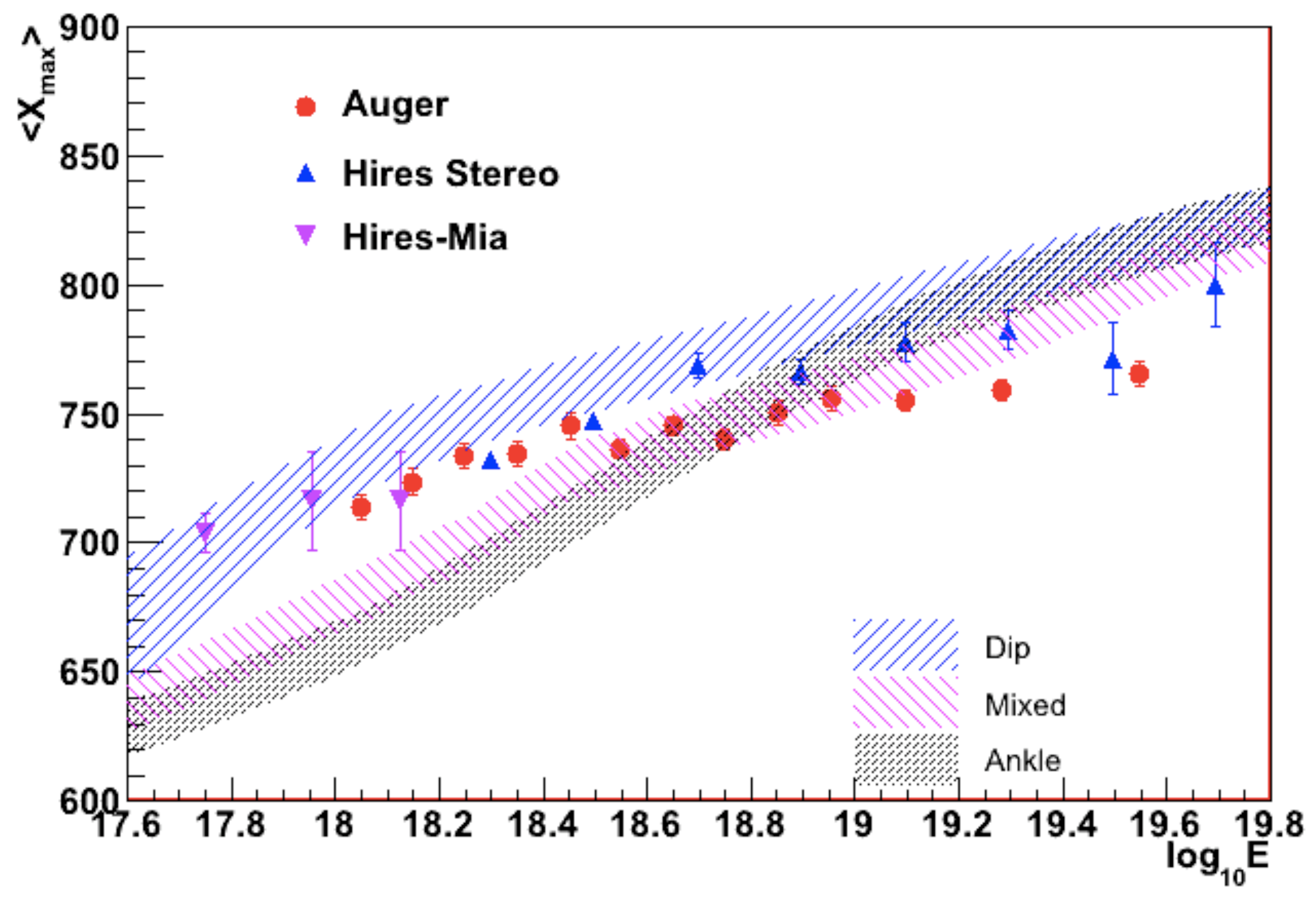}   
\caption{\em {$\langle X_{max} \rangle$ vs energy compared to model predictions  (adapted from \citep{ung08},  Auger data from \citep{aug10}).}}  
 \label{fi:transComp}
 \centering
 \end{minipage}\hfill
 \begin{minipage}{0.47\linewidth}
 \centering
\includegraphics[width=8.5cm,height=8.0cm]{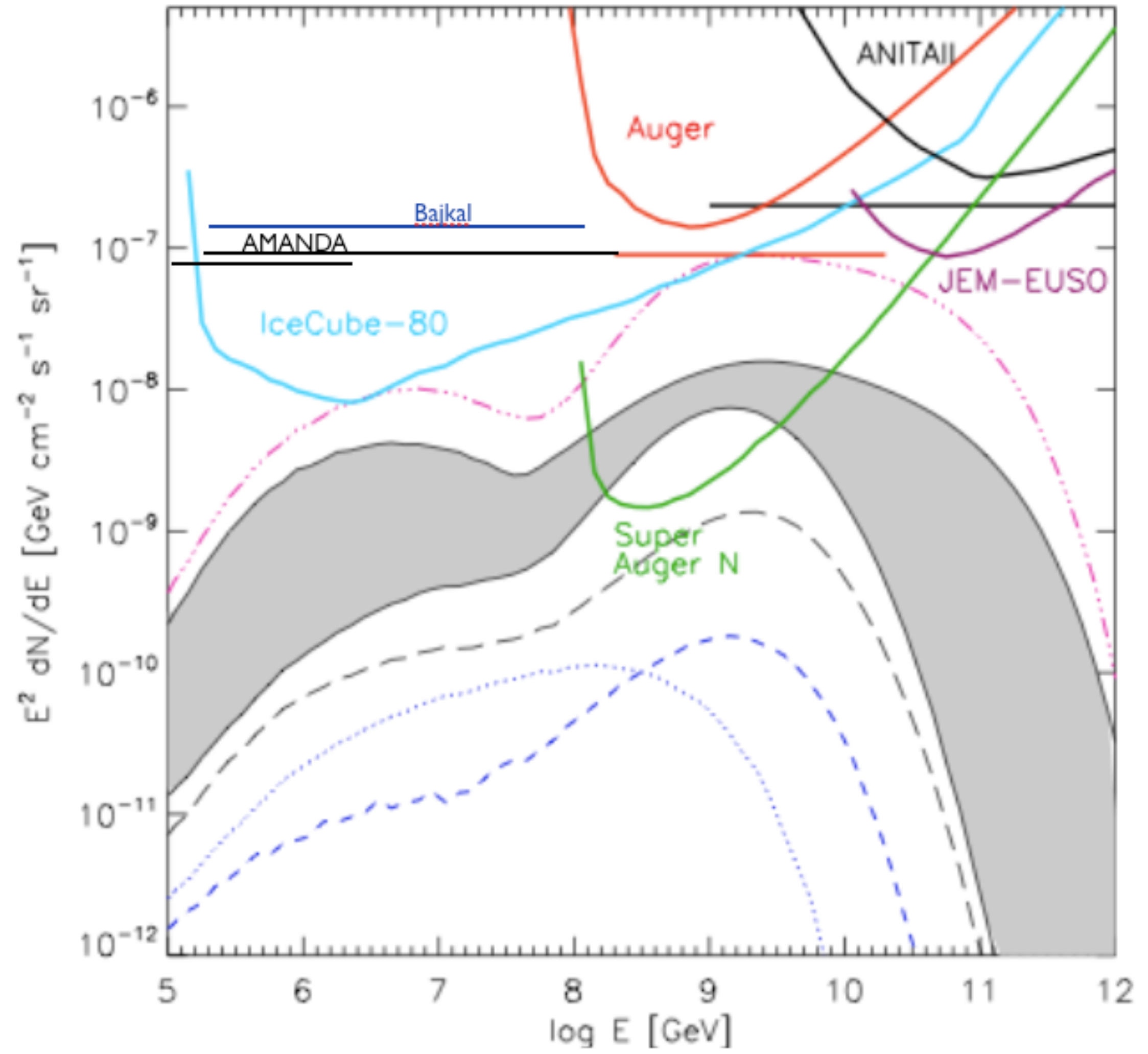}    
\caption{\em {Predicted cosmogenic neutrino flux (modified from \citep{2010JCAP...10..013K}, see legend therein).}}  
 \label{fi:transNeu}
 \end{minipage}
 \end{figure}

As outlined in the previous section, in the context of the standard model of CR origin in supernova remnants subsequent "knees" or steepenings of the spectra are predicted at $E_{max} \propto Z \times 10^{15}$ eV, reaching $\simeq 8 \ 10^{16}$ eV and above for the iron group. Above several $10^{18}$ eV  the detected cosmic particles must be of extra-galactic origin. This assumption also explains the lack of observation of a strong anisotropy that would be expected for charged particles with a large gyro-radius at this energy.  \\
The simplest and most natural  way of producing a flattening (an ankle) in the CR spectrum is that of intersecting the steep galactic spectrum with a flatter extragalactic one. This is the basic idea behind the {\it ankle model} \citep{hil05}: the transition from galactic to EGCRs appears around $10^{19}$ eV, at the crossing of the two spectra, producing also the observed "dip", a concavity in the spectrum  at $10^{18} \lesssim E \lesssim 4 \ 10^{19}$ eV . The generation spectrum of the extragalactic component has a slope between 2.2 and 2.5, as predicted by Fermi acceleration both at non relativistic and ultra relativistic shocks, even if these slopes are rather model dependent \citep{lem06}.  However, it is clear that  additional mechanisms able to accelerate  the galactic component beyond the iron knee have to be introduced \citep{bell01} in the ankle model, in order to fill the gap between the iron knee and the onset of the EGCRs.\\
n the {\it dip model} \citep{ber06}, the dip is supposed to be produced by the $e^{+}e^{-}$ pair production in the interactions between extragalactic protons (after their escape from the sources \citep{sig05}) and CMB photons. The transition takes place at the energy at which the adiabatic energy losses due to the expansion of the Universe equal the pair production ones, that is around the so-called second knee ($E \simeq 5 \ 10^{17}$ eV); the propagated spectrum and the dip do not depend on the hypotheses on  source evolution, although both the beginning of the dip region  and the energy at which the transition is completed do. The model requires an almost pure proton composition, with a maximum allowed contamination from He of $\simeq 10 \%$; at these energies, in fact, the pair production cross section is relevant only for protons, being proportional to the energy/nucleon.  A broken power-law generation spectrum is needed to avoid a too large emissivity when extrapolated at lower energy \citep{ber04}.\\
The {\it mixed composition model} \citep{all07} assumes that the  EGCR source composition is mixed and similar to that of the GCRs. The observed spectrum can be reproduced by assuming a source spectrum $E^{(-2.2 \div 2.3)}$; the transition region covers energies up to the ankle, while the galactic component extends up to more than $10^{18}$ eV.  Interestingly, the same source spectrum describes low energy CR data, in line with the ideas of holistic models \citep{par05}, which propose that CRs of any energies be produced by the same sources. A possible difficulty of the mixed model is the fact that the single element spectra cut off at energies proportional to their mass A.   The composition could be dominated by protons below $10^{18}$ eV, unlike in the ankle model case. \\
Energy spectrum, composition and anisotropy are used to disk riminate among the different models. The first one is the best measured, but  (as shown in Fig.\ref{fi:speTrMod} for the dip and mixed composition cases) all models give a good description of the all-particle energy spectrum. In particular, there is an impressive agreement of the dip predictions with the spectrum in the ankle region.  A very small level of anisotropy below 1 EeV  and isotropy when all CRs become extragalactic can be eventually expected in the transition region.\\
The study of composition seems to be more efficient in disk riminating among the different hypotheses. According to the dip model, the $\langle X_{max} \rangle$ evolution with energy is steep till the transition ends, becoming then flatter and corresponding to extragalactic protons. In the mixed model, the transition is wider and the evolution of $\langle X_{max} \rangle$ less steep, going from the heavy galactic part to the mixed but light extragalactic composition. 
The experimental data are shown in Fig.\ref{fi:transComp}, together with the expectations for proton and iron primaries from two different hadronic interaction models and the predictions described above.
None of the three models can satisfactorily reproduce the data; in the case of the mixed composition, the higher number of free parameters allows an easier adjustment.\\
Another possible key to explore the transition region is that of cosmogenic neutrinos. They originate from the decay of charged pions and neutrons produced in photo-pion and photo-nuclear interactions of ultra high energy CRs on  the cosmic background photon fields \citep{ber69}. 
The neutrino fluxes strongly depend on the source density evolution, CR composition and maximum energy of the accelerated CRs.  
In Fig.\ref{fi:transNeu}, they are calculated by \citep{2010JCAP...10..013K} for different model parameters, distinguishing three possible regions: strong evolution for sources and pure proton composition (pink dot-dashed curve), uniform evolution with pure iron injection and iron rich composition (blue lines) and intermediate models (grey area). In the same figure, the existing limits from Auger \citep{aug09} and ANITA-II \citep{gor10} and the upcoming experimental neutrino sensitivities are also shown. The model predictions spread a wide range of possible fluxes, with an estimated uncertainty $\simeq 50 \%$, and can be even further affected by the inclusion of the effects of extragalactic magnetic fields or of galactic strongly magnetized regions.  As pointed out in \citep{ber10}, at ultra high energy some of these scenarios could be excluded by Fermi/LAT results on the diffuse extragalactic$\gamma$ ray background \citep{abd10}; if so, the detection of the UHE neutrinos would require an increase of at least a factor $\simeq$ 10 in the current experiments sensitivity.
On the other hand, a positive detection of a neutrino flux in the UHE domain, together with PeV measurements would give information about the galactic to extragalactic transition and the source composition models.

\section{The measure of the anisotropy}

The study of CR anisotropies and their evolution with energy is a powerful tool to study propagation properties and sources. \\
At energies below $10^{17}$ eV, the CR trajectories  are bent and isotropized  by the galactic magnetic field. Large scale anisotropies can arise  due to the density gradients produced by the propagation of CRs in the galactic magnetic field, described in Sec.\ref{sec:propa} and the high statistics would allow to detect even a small degree of anisotropy.
For increasing energy, when the particle gyroradius becomes comparable to the galactic disk thickness, we can expect anisotropy  if the sources are distributed in the galactic disk. At the highest energies, no large scale modulation would be expected, but  we could observe events coming from nearby sources (within the GZK horizon), since the magnetic deflections of the particle trajectories are small.

\subsection{Large scale anisotropy}
\label{anis}
\begin{figure}[!h]
\begin{minipage}{0.47\linewidth}
 \centering
\includegraphics[width=8.5cm,height=8.5cm]{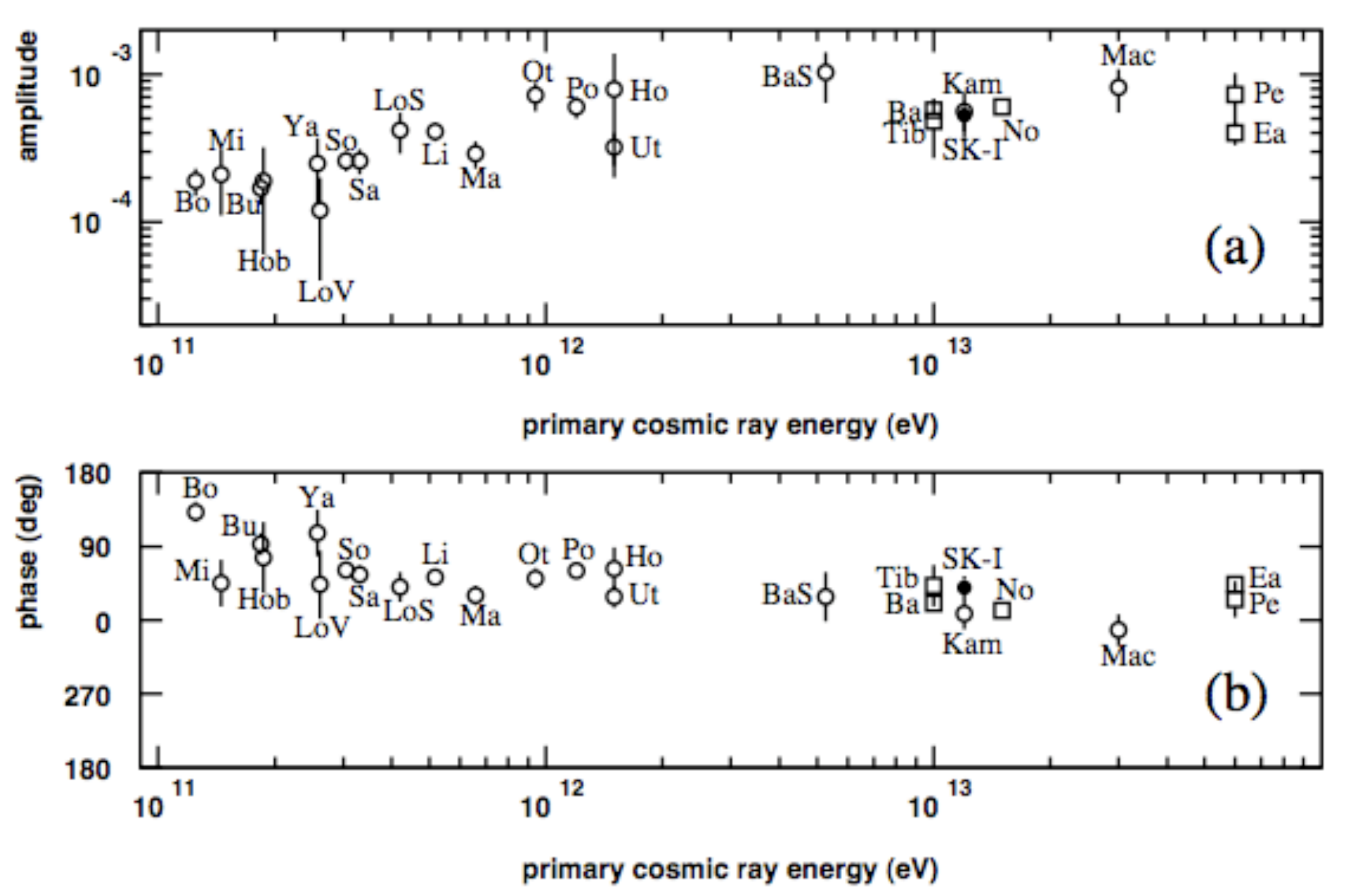}  
 \caption{\em {Amplitude and phase of anisotropy below 100 TeV (from \citep{gui07}, see legend therein).}}  
 \label{fi:anisLE}
 \end{minipage}\hfill
 \begin{minipage}{0.47\linewidth}
 \centering
 \includegraphics[width=8.5cm,height=8.5cm]{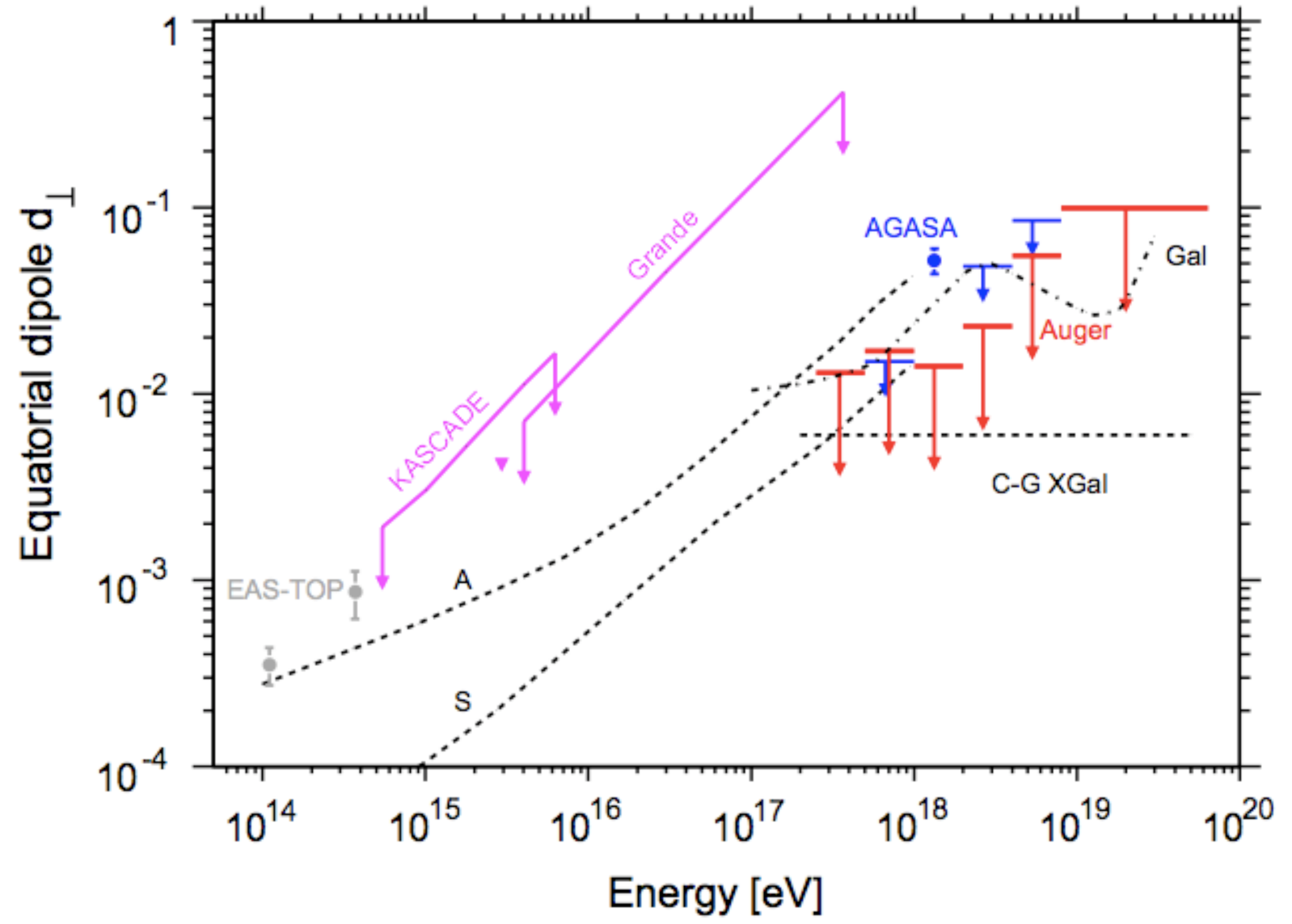}   
 \caption{\em {Upper limit on the anisotropy amplitude as a function of energy \citep{bon09}.}}  
 \label{fi:anisHE}
\end{minipage}
\end{figure}

The simplest anisotropy signal that can be looked for is a dipole in a given direction $\vec{j}$, which gives rise to an intensity 
$I(\vec{u})=I_{0}+I_{1} \cdot \vec{j}$, with amplitude $ (I_{max}-I_{min})/(I_{max}+I_{min}) = I_{1}/I_{0}$. \\
EAS arrays, due to the Earth rotation, operate uniformly with respect to the sidereal time, so that the shower detection and reconstruction are dependent only on declination. Data are generally analyzed in right ascension only, fixing the declination band, due to the difficulty to define the dependence of the detector exposure on declination.\\
The standard technique to look for large scale anisotropies is the Rayleigh method \citep{lin75}, which, performing an harmonic analysis, allows the  amplitude  $A_{k}$ and the phase $\phi_{k}$  of the k-th harmonic to be extracted.   If ($\alpha_{i},\delta_{i}$) are the galactic coordinates of the i-th of $N$ events
\begin{equation}
A_{k}=\sqrt{a_{k}^{2}+b_{k}^{2}} \ \ \ \ \ \ \ \  \phi_{k}={\rm arctan}(b_{k}/a_{k}) 
\end{equation}
where $a_{k}= \frac{2}{N} \sum_{i=1}^{N} {\rm cos}(k \alpha_{i})$ and  $b_{k}= \frac{2}{N} \sum_{i=1}^{N} {\rm sin}(k \alpha_{i})$
and the probability of detecting a spurious amplitude by chance is $P(\geq A_{k})=exp(-NA_{k}^{2}/4)$. \\
From the experimental point of view, the measure is feasible only with large collecting areas and long term observations. The detectors must be uniform in time and area, and operate continuously. Systematic effects linked to the atmospheric temperature and pressure variations have to be carefully taken into account.\\
Anisotropy measurements are obtained either by underground muon observatories \citep{amb03,gui07}, below $10^{13}$ eV, or by EAS arrays above this energy. 
A compilation of the results is shown in Fig.\ref{fi:anisLE}: the measured amplitudes from $10^{11}$ to $10^{13}$ eV amount to few times $10^{-4}$. \\
The EAS-TOP experiment demonstrated that the main features of anisotropy are similar at $10^{14}$ eV to those measured at lower energies and extended the measurements up to $4 \times 10^{14}$ eV where $A_{sid}= (2.6 \pm 0.8) \times 10^{-4}$ and $\phi_{sid}= (0.4 \pm 1.2)$ h \citep{agl09}. The analysis was performed  adopting a different method, based on the counting rate differences between eastward and westward directions, so removing variations of atmospheric origin.\\
A large scale anisotropy signal is also expected at all energies due to the motion of the observer with velocity $\vec{V}$ with respect to a locally isotropic flux of CRs, the Compton-Getting effect \citep{com35}. For a  CR particle spectrum $I(E) \propto E^{-\gamma}$, the amplitude is  $\Delta I /I = \frac{V}{c} (\gamma +2) cos \theta$, where $\theta$ is the angle between the arrival direction of 
CRs and the moving direction of the observer.
For example, the orbital motion of the Earth with velocity $\vec{V} \simeq 30$ km/s leads to a dipolar anisotropy which can be as large as $\simeq 5 \times 10^{-4}$ (assuming a CR spectrum slope $\gamma$=3).
The expected rate modulation in solar time has in fact been measured at 10 TeV by EAS-TOP \citep{agl96} and Tibet \citep{ame04}. \\
A similar effect could be expected from the motion of the Solar System around the galactic center at $\vec{V} \simeq 220$ km/s, which would produce an anisotropy of few $10^{-3}$ if the CR plasma were at rest with respect to the Galaxy. The measured sidereal time modulation is an order of magnitude lower, thus demonstrating that the CR plasma corotates with the Galaxy \citep{ame06}.\\
At higher energy, the anisotropy would be expected to increase if the knee be due to an 
increasing efficiency of higher and higher energy CRs to escape the Galaxy; the level of anisotropy could then help in choosing among different propagation models. On the other hand, a knee due to the existence of a maximum acceleration energy in the galactic sources would result in a decrease of the anisotropy amplitude with increasing energy, because of the increasing contribution of the extragalactic isotropic CRs component.
At these energies, many groups applied the Rayleigh formalism to study the large scale anisotropy, but only upper limits could be derived \citep{ant04,ove07}.  The current status of the large scale anisotropy measurements in the knee region and above is shown in Fig.\ref{fi:anisHE}. The results are plotted taking into account the sky coverage of each apparatus.\\
As for the higher energy region, different scenarios for the transition bring to different predictions for the anisotropy: if the transition takes place at the ankle energy, the dominant component at 1 EeV is galactic, and a modulation at few percent can be expected due to their escape driven by diffusion and drift. If on the contrary EGCRs appear at lower energies,  the large scale distribution of isotropic CRs  would be influenced by the relative motion of the observer with respect to the frame of the sources. The cosmological Compton-Getting effect, evaluated in the frame in which the CR distribution is isotropic coincides with the CMB rest frame, is shown at about $0.6\%$. Three times the present statistics would be needed to be sensitive to such amplitude. \\
The AGASA group reported a large scale anisotropy with dipole-like modulation in right ascension of amplitude $\simeq 4 \%$  for a dipole oriented near the galactic center at $E \simeq 1-2$ EeV, using 11 years of data  \citep{aga99}.
The Pierre Auger Observatory  \citep{bon09}  studied the sidereal anisotropies using both the Rayleigh  and the East-West differential methods, deriving bounds on the first harmonic amplitude at the 1$\%$ level at EeV energies which did not confirm the AGASA claim.  As an example of comparison with models,  the expectation derived for the dipole amplitude from diffusion depending on different galactic magnetic field models and source distribution \citep{can03} is also shown in Fig.\ref{fi:anisHE} (calculated assuming transition at the ankle).
The Pierre Auger data are already excluding  models where an antisymmetric halo magnetic field is used (indicated with "A"), while nothing can at the moment be said about  the "S" model, where the field is symmetric.

\subsection{Point sources}
Clustering at small and intermediate scale can also be searched for in data, to derive information about the number of sources and their distribution. 
The standard technique is that of comparing the measured arrival direction distribution in equatorial coordinates with an isotropic background distribution, and different methods can be used to derive  the signal strength and significance \citep{mol08}.\\
Single sources have also been searched for by many groups: no excess of particles from point sources has been detected by KASCADE  \citep{ant04} above 0.3 PeV. In particular, the possible signal from the  Monogem ring (a supernova remnant suggested to be the possible single source of GCRs \citep{ew04})  has not been confirmed, either by KASCADE or Tibet \citep{ame05}.\\
Due to the high density of stars and to the presence of a super massive black hole, the galactic center is considered a possible site of acceleration for GCRs.  
Reports from  AGASA indicate a 4.5 $\sigma$ excess of CRs  in the EeV energy range; this result was confirmed by the Fly's Eye and SUGAR detectors \citep{bir99,bel01}, which both claimed for an excess in the direction of the galactic center (the latter with an offset with respect to AGASA). The claim was not supported by the Auger collaboration from the analysis of all  data above $10^{18.5}$ eV in the same region of the sky, with statistics much greater than those of previous experiments \citep{aug07}.\\
The study of disk rete sources of CRs, if any, can be best performed exploiting the highest CR energies, since the isotropic background of distant sources would in this case be eliminated by the GZK cut-off and the ultra high energy would guarantee sufficient magnetic rigidity for the particles to trace back to their sources. 
The extremely large exposures and small angular accuracy are the challenge of the on-going and future experiments  aiming to perform this study. Being out of the scope of this review, the present results and discussion pertaining to the anisotropy of EGCRs will not be discussed here and we address the reader to the review of \citep{som09}.

\section{The future of cosmic ray astrophysics}
The astrophysics of GCRs is living a flourishing
era, thanks to the wealth of experiments of different conception:
ground based, ballon-borne, satellites, or installed on the International
Space Station.\\
Present-day results indicate how powerful and complex is this field in the
understanding of the physics and astrophysics of the Milky Way.
From one side, we dispose of phenomenological models able to reproduce on
a qualitative ground  -- and in some cases also quantitatively -- many
different data on cosmic nuclei. On the other side, we still lack a
precise modelization of the magnetic Galaxy and of the interactions
occurring to CRs in their wandering from their sources to our Planet.
The understanding of CR data founds on the physics and astrophysics of
their sources: mostly SNRs, with some contribution from pulsars.
We need a careful -- morphological, energetic -- study of SNRs in
connection with $\gamma$ ray data from ground-based as well as space-based
telescopes.\\
By means of direct measurements, both on satellites and balloons, we understood that GCRs are produced accelerating seeds mainly composed by refractory elements, producing a composition at source very similar to the Solar System one and that  $\simeq 20 \%$ of the sources of GCRs are to be found in OB associations. \\
The diffusion coefficient, which is in principle connected with the
inhomogeneities of the galactic magnetic field, can be determined only
from the measured spectra of CRs, the best indicator being the boron--to--carbon ratio.
Despite the recent experimental achievements, it remains of the utmost importance to dispose of more accurate
data on B/C from hundreds of MeV/n to at least the TeV/n region and to increase the statistics in the highest energy region. \\
Crucial tests of diffusive models are some $\beta$-decaying isotopes, the
most relevant being $^{10}Be/^{9}Be$. Data are still scarce and very
limited in energy. It would be desirable having a measured spectrum
extending from around GeV/n up to hundreds (tens at least) GeV/n.\\
The case of antimatter has received an enormous improvement in the very
last years. Antiprotons have been shown to be powerful tests of galactic
propagation. Their high agreement with data and the small theoretical
uncertainties make this rare species a gauge for astrophysical,
non-thermal contributions or exotic contributions from dark matter
particles in the galactic halo. 
It is amazing to notice that in some cases the strongest constraints to particle dark
matter models come from the cosmic antiproton data.\\
The recent measurements of the positron fraction and electron absolute
flux clearly indicate that the interpretation of the data requires a
careful understanding of the astrophysics of sources located few kpc
around the Solar System. Leptons from SNRs and/or pulsars could have
registered feeble footprints
on the cosmic flux arriving at Earth.\\
Future direct measurements of CRs will also need to extend the investigation to the high energy region, towards to knee,  in order to 
establish a benchmark for the cosmic-ray composition and so reduce the free parameters in the evaluation of the composition at higher energies. \\
The requirement of higher statistics, at least 10 events per particle type above $10^{15}$ eV, implies the use of instruments with larger apertures and longer operation time.
Part of this goal will be achieved with the new technology at the base of the Ultra Long Duration Balloon flight project, which is currently extensively tested with the aim of producing balloons able to fly for 100 days. However, their limitations in weight and exposure will not allow to reach energies $\gsim$ few $10^{14}$ eV, which can be studied only by means of Extensive Air Shower detectors. \\
In general, the all particle flux is measured with good agreement among the experiments up to the highest energies, with some evidence of a proportionality of the cut-off energies for each element  to the rigidity, even if a dependence on the mass A cannot be excluded. The data show that the proton spectrum is steeper compared to the helium and 
CNO groups and that there is a dominance of the helium component in the knee region. 
The results on the composition of single or groups of elements show clearly that the actual limitations are not statistical, but depend mainly on the insufficient knowledge of the characteristics of hadronic interactions, thus underlining the importance of a strict and mutually benefitting link between astroparticle and particle physics. The future results from LHC at CERN will be most helpful in benchmarking the extrapolations to the highest energies.\\
The investigation of the region above the knee towards the transition to EGCRs requires detectors with large areas but  with smaller spacing compared to the arrays studying UHECRs. As for the lower knee region, the requirement is that of employing complementary techniques, so to detect as many components of showers as possible and cross check their systematics.
Numerous new projects have been designed and are now taking data or are under construction. KASCADE-Grande \citep{KG11} most recently claimed evidence of a knee in the heavy component spectrum; Tunka-133 \citep{tunk} and  Ice-Top \citep{ice} aim at exploring the end of the galactic spectrum. \\
Two enhancements in the Pierre Auger Observatory, the Auger Muon and Infill for the Ground Array (AMIGA) \citep{amiga} and the High Elevation Auger Telescopes (HEAT) \citep{heat} will allow to explore the transition region from $10^{17}$ up to $10^{19}$ eV. They will give us a powerful tool to clarify the problem of the transition from Galactic to extragalctic CRs  and complement the measurements at the highest energies.\\
Intense research and development activity is going on to detect the radio emission from EAS, that could enable both an increase of the statistics and a reduction of systematic uncertainties on the determination of the air shower properties.  Progress has been made in recent years by the LOPES  and CODALEMA  groups \citep{lop11,cod11}, while radio detection at higher energies is under scrutiny in Auger \citep{kel09}.

\vspace{0.8cm}

\centerline{\it \bf Acknowledgements}

The authors would like to thank many of their colleagues for interesting comments and fruitful discussions. Special thanks are due to A.A.Watson for the many suggestions.

\bibliography{CaDo_v6}

\end{document}